\documentclass[12pt]{article}
\usepackage{amsmath, amssymb, amsfonts, amsthm, epsfig}

\newtheorem{theorem}{Theorem}[section]
\newtheorem{proposition}[theorem]{Proposition}

\newtheorem{lemma}[theorem]{Lemma}
\newtheorem{definition}[theorem]{Definition}

\newcommand{\tri}{| \! | \! |}
\newcommand{\rd}{{\rm d}}
\newcommand{\be}{\begin{equation}}
\newcommand{\ee}{\end{equation}}
\newcommand{\bey}{\begin{eqnarray}}
\newcommand{\eey}{\end{eqnarray}}
\newcommand{\bef}{\begin{figure}}
\newcommand{\eef}{\end{figure}}
\newcommand{\bec}{\begin{center}}
\newcommand{\eec}{\end{center}}

\newcommand{\sfrac}[2]{{\textstyle \frac{#1}{#2}}}

\newcommand{\bp}{{\bf p}}

\newcommand{\tsi}{{\widetilde\sigma}}
\newcommand{\tpi}{{\widetilde\pi}}

\renewcommand{\a}{\alpha}

\newcommand{\e}{\varepsilon}
\newcommand{\s}{\sigma}
\newcommand{\om}{{\omega}}

\newcommand{\bE}{{\bf E}}

\newcommand{\cX}{{\cal X}}
\newcommand{\cT}{{\cal T}}

\newcommand{\bR}{{\mathbb R}}
\newcommand{\bC}{{\mathbb C}}
\newcommand{\bN}{{\mathbb N}}
\newcommand{\bZ}{{\mathbb Z}}

\newcommand{\wt}{\widetilde}
\newcommand{\wh}{\widehat}
\newcommand{\ov}{\overline}

\newcommand{\cS}{{\cal S}}

\newcommand{\cA}{{\cal A}}

\newcommand{\cE}{{\cal E}}

\newcommand{\cV}{{\cal V}}

\newcommand{\cI}{{\cal I}}

\newcommand{\cL}{{\cal L}}

\newcommand{\cO}{{\cal O}}

\newcommand{\us}{\underline{s}}

\newcommand{\no}{\nonumber}
\newcommand{\non}{\nonumber}

\input epsf

\newcommand{\donothing}[1]{}

\begin{document}

\title{Lecture Notes on Quantum Brownian Motion}
\author{L\'aszl\'o Erd\H os\thanks{Partially supported
by SFB-TR 12 Grant of the German Research Council}\\
Institute of Mathematics, University of Munich, \\
Theresienstr. 39, D-80333 Munich, Germany
\\}

\date{Sep 4, 2010}

\maketitle

\input epsf

\begin{abstract}
Einstein's kinetic theory of the Brownian motion, based
upon light water molecules continuously bombarding a heavy
pollen, provided an explanation of diffusion from the Newtonian 
mechanics. Since the discovery of quantum mechanics it has been
a challenge to verify the emergence of diffusion from
the Schr\"odinger equation.

The first step in this program is to verify the linear
Boltzmann equation as a certain scaling limit of a
Schr\"odinger equation with random potential. In the second
step, one considers a longer time scale that corresponds to
infinitely many Boltzmann collisions. The intuition is that
the Boltzmann equation then converges to a diffusive equation
similarly to the central limit theorem for Markov processes
with sufficient mixing. In these lecture notes (prepared
for the Les Houches summer school in 2010 August) we  present
the mathematical tools to rigorously justify this intuition.
The new material relies on joint papers with H.-T. Yau and
M. Salmhofer.

\end{abstract}


\tableofcontents

\section{Overview of the rigorous derivations  of diffusions}\label{sec:ov}

The fundamental equations governing the basic laws of physics,
the Newton and the Schr\"odinger
equations,  are
time reversible and have no dissipation.
It is remarkable that dissipation is nevertheless ubiquitous
in nature,
so that almost all macroscopic phenomenological equations
are dissipative.
The oldest such example is perhaps the
equation of heat conductance found by
Fourier. This investigation has led to the heat equation,
the simplest type of diffusion equations:
\be
    \partial_t u = \Delta_x u,
\label{heat}
\ee
where $u(x,t)$ denotes the temperature at position $x\in\bR^d$ and time $t$.
One key feature of the diffusion equations is their inhomogeneous scaling;
the time scales as the square of the spatial distance:
$$
  t\sim x^2; \qquad \mbox{time} \sim (\mbox{distance})^2.
$$

In these lectures we will explore how diffusion equations emerge
from first principle physical theories such as the classical 
Newtonian dynamics and the quantum Schr\"odinger dynamics.
In Section~\ref{sec:ov} we give an overview of existing mathematical
results on the derivation of diffusion (and related) equations.
In Sections~\ref{sec:recap}--\ref{sec:qm} we discuss
the basic formalism and present a few well-known 
preliminary facts on stochastic, classical and quantum dynamics. 
An experienced reader can skip these sections.
In Section~\ref{sec:rs} we introduce our main
model, the random Schr\"odinger equation or the quantum
Lorentz gas  and its lattice version,
the Anderson model. In Section~\ref{sec:main} we formulate
our main theorems that state that the random Schr\"odinger 
equation exhibits diffusion and after a certain scaling
and limiting procedure it can be described by a heat equation.

The remaining sections contain the sketch of the proofs. 
Since these proofs are quite long and complicated,
we will not only have to omit 
 many technical details, but even several essential ideas
can  only be mentioned very shortly. We will primarily
focus on the most important aspect: the classification and
the estimates of the so-called non-repetition Feynman graphs.
Estimates of other Feynman graphs, such as recollision
graphs and graphs with higher order pairings will only
be discussed very superficially.

Our new results presented in this review were obtained in collaboration
with H.-T. Yau and  M. Salmhofer. 

\medskip

{\it Acknowledgement.} 
The author is grateful to M. Salmhofer for many suggestions
to improve this presentation and for making his own lecture notes
available from which, in particular, the material
of Section~\ref{sec:l2} and \ref{sec:lowerorder} was borrowed.

\subsection{The founding fathers: Brown, Einstein, Boltzmann}

The story of diffusion starts with  R. Brown in 1827
 who observed almost two centuries ago
that the motion of a wildflower pollen suspended
in water was erratic \cite{Br}. He saw a
picture similar to Fig.~\ref{fig:brown} under his microscope
(the abrupt velocity changes in this picture are exaggerated
to highlight the effect, see later for more precise explanation). 
He ``observed many of them very evidently in motion''
and argued that this motion was not due to water currents
but ``belonged to the particle itself''. First he thought
that this activity characterizes only ``living'' objects like pollens,
but later he found that small rock or glass particles follow
the same pattern.
He also noted that this apparently chaotic motion never seems to stop.
\bef\bec
\epsfig{file=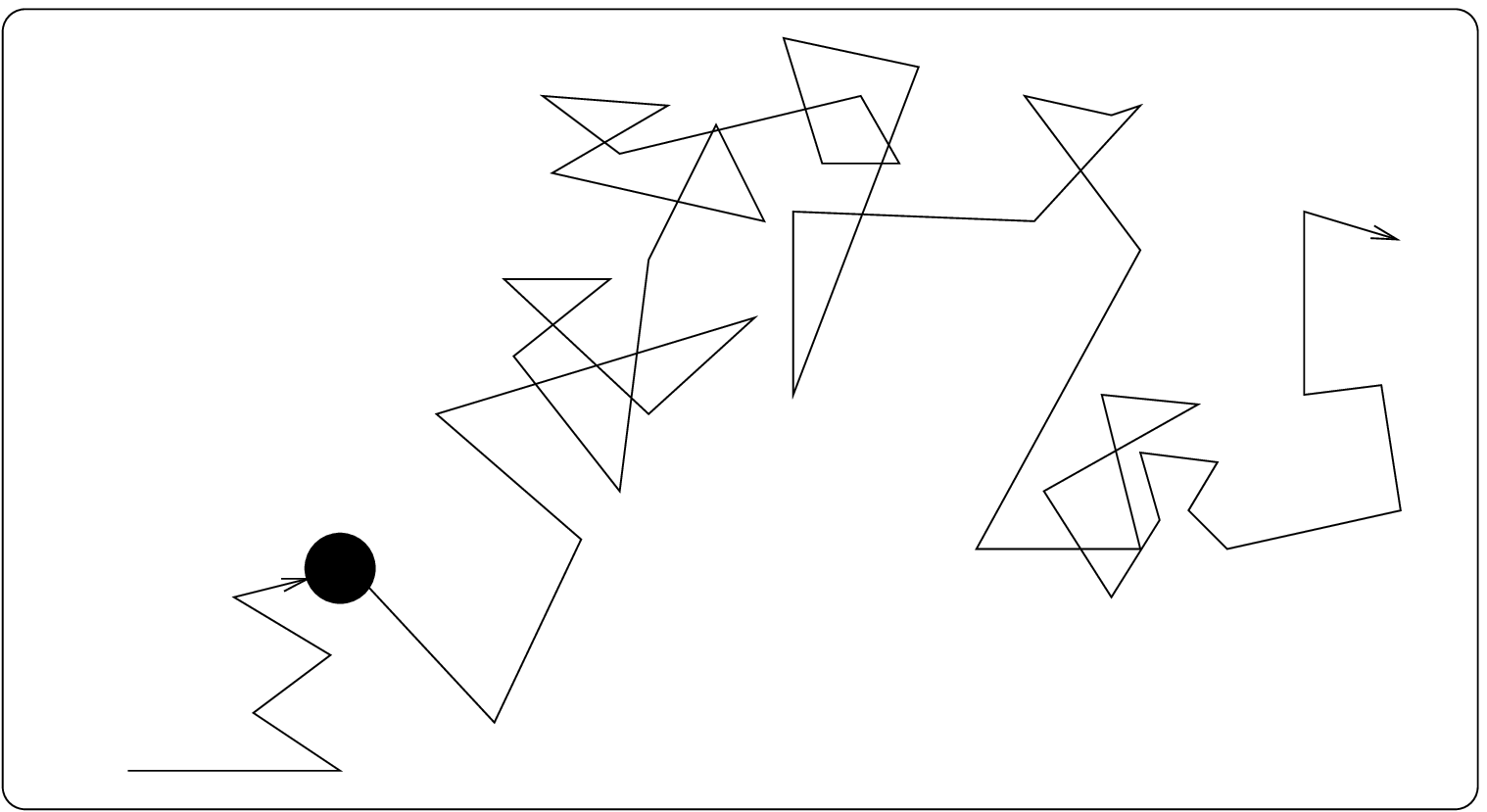, scale=0.50}
\eec
\caption{Brown's picture under the microscope}
\label{fig:brown}
\eef
\bef\bec
\epsfig{file=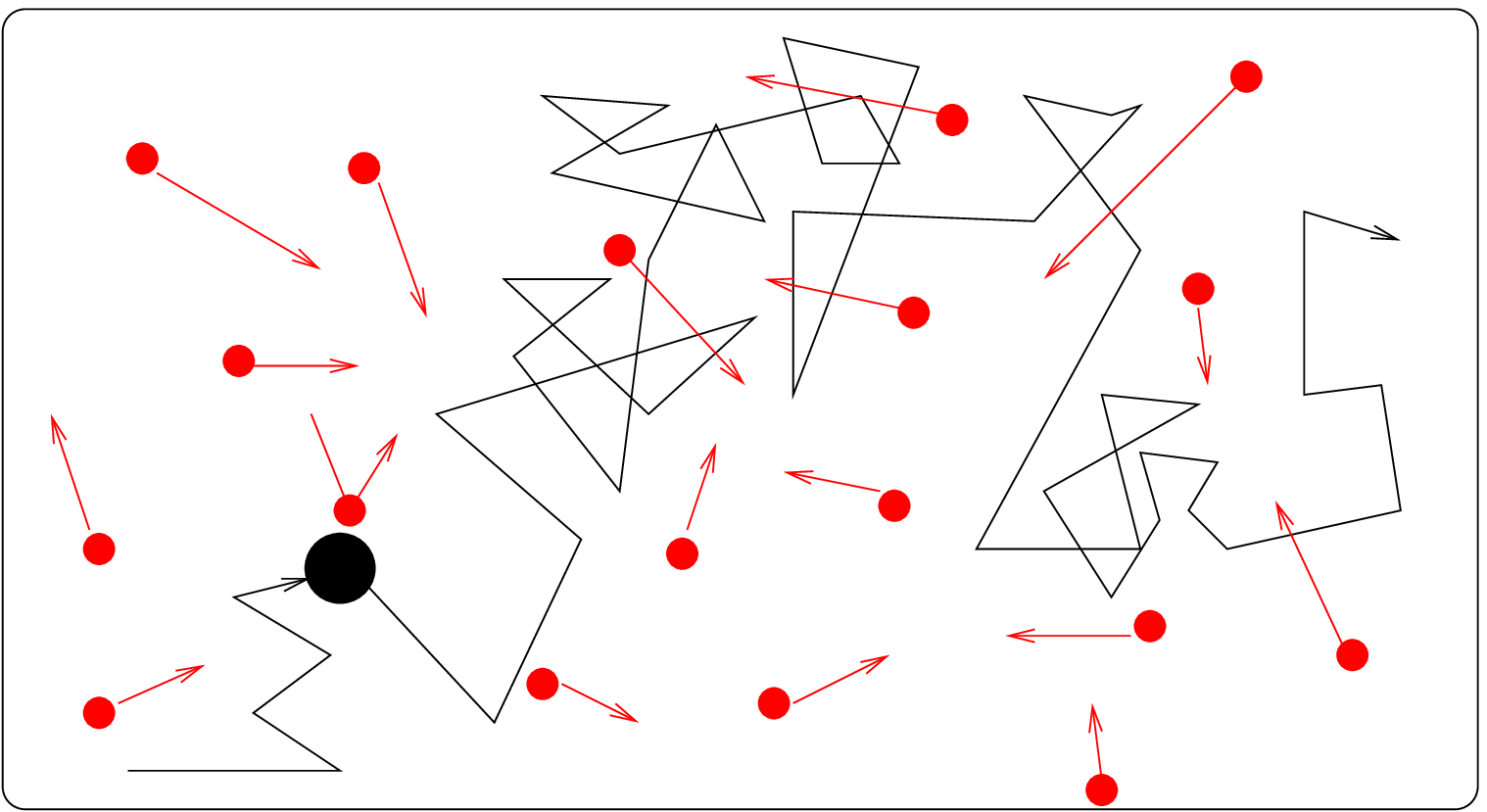, scale=0.50}
\eec
\caption{Einstein's explanation to Brown's picture}
\label{fig:einstein}
\eef

This picture led to the kinetic explanation by
A. Einstein in 1905 \cite{Ei} that such a motion was created by the
constant kicks on the relatively heavy pollen  by the
light water molecules. Independently, M. Smoluchowski \cite{Sm}
arrived at a similar theory.
 It should be noted that
at that time even the atomic-molecular structure of
matter was not universally accepted.
Einstein's theory was strongly supported by
Boltzmann's kinetic theory, which, however,
was phenomenological and seriously debated at the time.
Finally in 1908 Perrin \cite{Per} (awarded the Nobel prize in 1926) experimentally
verified Einstein's theory and used it, among others,  to
give a  precise estimate on the Avogadro number.
These experiments gave the strongest evidence
for atoms and molecules at that time.

\bigskip

We should make a remark on Boltzmann's theory although this
will not be the subject of this lecture. Boltzmann's kinetic
theory postulates that in a gas of interacting particles at relatively
low density the collisions between the particles are
statistically independent (Ansatz of {\it molecular chaos}).
 This enabled him to write up
the celebrated (nonlinear) Boltzmann equation
for the time evolution of the single particle phase space density
$f_t(x,v)$. The main assumption is that
the collision rate is simply the product of the particle
densities with the appropriate velocities, thus he arrived at
the following equation:
\be
   \partial_t f_t(x,v) + v\cdot \nabla_x f_t(x,v) = \int \sigma(v,v_1)
   \big[ f_t(x,v')f_t(x,v_1') - f_t(x,v)f_t(x,v_1)\big].
\label{nlb}
\ee
Here $v,v_1$ is the pair of incoming velocities
of the colliding particles  and $v',v_1'$ are the outgoing velocities.
In case of elastic hard balls,
the pair $(v',v_1')$ is uniquely determined by $(v,v_1)$
(plus a randomly chosen contact vector) due
to energy and momentum conservation, for other type
of collisions the kernel $\sigma(v,v_1)$ describes
microscopic details of the collision mechanism. 
The integration is for the random contact vector
and all other variables apart from $x,v$,
subject to constraints (momentum and
energy conservation) encoded in $\sigma(v,v_1)$.

In addition
to the highly non-trivial Ansatz of independence, this 
theory also assumes the molecular nature of gases and fluids
in contrast to their more intuitive continuous description.
Nowadays we can easily accept that gases and fluids on
large scales (starting from 
micrometer scales and above -- these are called {\it macroscopic scales})
can be characterized by continuous density functions,
while on much smaller scales (nanometers and below, these are
called {\it microscopic scales}) the particle
nature dominates. But at Boltzmann's time the particle
picture was strongly debated. The ingenuity of Boltzmann's
theory is that he could combine these two pictures. The Boltzmann equation
is an equation of continuous nature (as it operates with
density functions) but its justification (especially the determination 
of the collision kernel)  follows an argument
using particles. No wonder that it gave rise to so much controversy
especially before experiments were available.
Moreover, Boltzmann theory implies irreversibility
(entropy production) that was for long thought to be more
a philosophical than a physical question.

\bigskip

After this detour on Boltzmann we return to the diffusion models.
Before we discuss individual models, let me mention
the 
{\it key conceptual difficulty} behind all rigorous derivations
of diffusive equations. Note that
the Hamiltonian dynamics (either classical or quantum)
is reversible and deterministic. The diffusion equation \eqref{heat},
and also the Boltzmann equation, is
irreversible: there is permanent loss of information along
their evolution (recall that the heat equation is
usually not solvable backward in time).
Moreover, these equations exhibit a certain 
 chaotic nature (Brown's picture). How can these
two contradicting facts be reconciled?

The short answer is that the loss of information is due
to a {\it scale separation} and the {\it integration of
microscopic degrees of freedom}. On the microscopic (particle)
level the dynamics remains reversible. The 
continuous (fluid) equations live on the macroscopic (fluid) scale:
they  are obtained by neglecting (more precisely, integrating out)
many  degrees of freedom on short scales.
 Once we switch to the fluid description,
the information in these degrees of freedom is lost forever.

This two-level explanation foreshadows that for a rigorous mathematical
description one would need a scale separation parameter,
usually called $\e$ that describes the ratio between
the typical microscopic and macroscopic scales. In practice
this ratio is of order
$$
   \e= \frac{1 \;\;\mbox{Angstrom}}{1 \;\;\mbox{cm}} = 10^{-8}\;,
$$
but mathematically we will consider the $\e\to 0$ so-called
{\it scaling limit}.

Once the scale separation is put on a solid mathematical ground,
we should note another key property of the macroscopic
evolution equations we will derive, namely their Markovian character.
Both the heat equation \eqref{heat} and the nonlinear Boltzmann
equation \eqref{nlb} give the time derivative of the macroscopic
density at any fixed time $t$ in terms of the density
at the  {\it same time} only. I.e. these evolution equations express
a process where the future state depends only on the present state;
the dependence of the future
on the past is only indirect through the present state.
We will call this feature {\it Markovian property} or 
in short {\it Markovity}. 
Note that colliding particles do build up a memory
along their evolution: the
state of the recolliding particles will remember
their previous collisions. This effect will have to be
supressed to obtain a Markovian evolution
equation,  as it was already recognized by Boltzmann in his Ansatz.
There are essentially
two ways to reduce the rate of recollisions: either by going to a low density situation
or by introducing a small coupling constant.

Apart from the rigorous formulation of the scaling 
limit, we thus will have to cope with the main
technical difficulty: {\bf controlling memory (recollision)
effects.} Furthermore, specifically in quantum models,
we will have to {\bf control interference effects} as well.

\bigskip

Finally, we should remark that
 Einstein's model is simpler than Boltzmann's,
as light particles do not interact
(roughly speaking, Boltzmann's model is 
similar to  Fig.~\ref{fig:einstein} but the light particles
can also collide with each other and not only with the heavy
particle). Thus the verification of
 the Ansatz of molecular chaos from first principle Hamiltonian
dynamics
is technically easier in Einstein's model.
 Still, Einstein's model already addresses the key issue:
how does diffusion emerge from Hamiltonian mechanics?

\subsection{Mathematical models of diffusion}

The first step to setup a correct mathematical model is
to recognize that some stochasticity has to be added.
Although we expect that for a {\bf ``typical''} initial 
configuration the diffusion equations are correct,
this certainly will {\bf not hold for all} initial configuration.
We expect that after opening the door between a warm
and a cold room, the temperature will equilibrate (thermalize)
but certainly there exists a ``bad'' initial configuration of
the participating $N\sim 10^{23}$ particles such that all ``warm'' particles
will, maybe after some time, head towards the cold room and vice versa; i.e. the 
two room temperatures will be exchanged instead of thermalization.
Such configuration are extremely rare; their measure
in all reasonable sense  goes to zero very fast as $N\to\infty$,
but nevertheless they are there and prevent us from
deriving diffusion equations for all initial configurations.
It is, however, practically hopeless to describe
all ``bad'' initial configurations; they may not be recognizable
by glancing at the initial state. The stochastic approach
circumvents this problem by precisely  formulating
what we mean by saying that  the ``bad events are rare''
without identifying them.

The stochasticity can be added in several different ways and
this largely determines the technical complications
involved in the derivation of the diffusion. In general,
more stochasticity is added, the easier the derivation is.

\bigskip

The easiest is if the {\bf dynamics itself is stochastic}; the typical
example being the classical random walk and its scaling limit,
the Wiener process (theorized by Wiener \cite{W} in 1923).
In the typical examples, the random process governing the
dynamics of the system has no correlation.  The whole microscopic
dynamics is Markovian (in some generalizations  the dynamics of the one particle
densities is not fully Markovian, although
 it typically has  strong statistical mixing properties).
 Of course this dynamics is not Hamiltonian
(no energy conservation and no reversibility), and thus
it is not particularly surprising that after some scaling
limit one obtains a diffusion equation. The quantum counterpart of
the classical random walk
 is a Schr\"odinger equation with a Markovian time dependent
random potential, see \cite{Pi} and the recent proof of
quantum diffusion in this model by Kang and Schenker \cite{KS}.

\bigskip

The next level of difficulty is a Hamiltonian
 system with a {\bf random Hamiltonian}. Some physical data
in the Hamiltonian (e.g. potential or magnetic field)
is randomly selected (describing a disordered system),
but then they are frozen forever and then the Hamiltonian
dynamics starts (this is in sharp contrast 
to the stochastic dynamics, where the system
is subject to fresh random inputs along its evolution).
A typical model is the {\it Lorentz gas},
where a single Hamiltonian particle is moving among
fixed random scatterers. Recollisions with the same
scatterer are possible and this effect potentially
ruins the Markovity. Thus
this model is usually considered in the weak coupling
limit, i.e. the coupling parameter $\lambda$ between the
test-particle and the random scatterers is set to converge
to zero, $\lambda\to 0$, and simultaneously a long time
limit is considered. For a weak coupling, the recollision 
becomes a higher order effect and may be neglected.
It turns out that if
\be\label{vh}    t\sim \lambda^{-2}
\ee
then one can prove a nontrivial Markovian diffusive limit
dynamics (see Kesten and Papanicolaou \cite{KP}, extended more
recently to a bit longer times by
Komorowski and Ryzhik \cite{KR}).
The relation \eqref{vh} between the coupling and the time scale
is called the {\it van Hove limit}.
Similar supressing of the recollisions can be achieved
by the {\it low density limit} (see later).

The corresponding quantum  model ({\it Anderson model})
is the main focus of the current lecture notes.
We will show that in the van Hove limit the linear Boltzmann
equation arises after an appropriate rescaling and for longer
time scales the heat equation emerges. In particular, we
describe the behavior of the quantum evolution in the
presumed {\it extended states regime}
 of the Anderson model up
to a certain time scale. We will use a mathematically rigorous perturbative
approach. We mention that supersymmetric methods \cite{Efe} offer 
a very attractive alternative approach to study quantum diffusion,
although the mathematical control of the resulting functional integrals
is difficult. Effective models that are reminiscent to the saddle point
approximations of the supersymmetric approach 
 are more accessible to rigorous mathematics.
Recently Disertori, Spencer and Zirnbauer have proved
a diffusive estimate for the two-point correlation functions in
a three dimensional supersymmetric hyperbolic sigma model
\cite{DSZ} at low temperatures and localization
was also established in the same model at high temperatures \cite{DS}.

\bigskip

The following level of difficulty is a {\bf deterministic
 Hamiltonian with random initial data of 
many degrees of freedom}. The typical example is Einstein's model,
where the test-particle is immersed in a heat bath.
The heat bath is called {\it ideal gas}, if it is characterized by a Hamiltonian $H_{bath}$ of
 non-interacting particles. The initial state is
typically  a temperature
equilibrium state, $\exp{(-\beta H_{bath})}$ i.e. the initial
data of the heat-bath particles are given by this equilibrium measure
at inverse temperature $\beta$.
Again, some scaling limit is necessary to reduce the
chance of recollisions, one can for  example consider
the limit $m/M\to \infty$, where $M$ and $m$ are the
mass of the test-particle and the heat-bath particles,
respectively. In this case, a collision with a single
 light particle does not have sizable effect on the motion
of the heavy particle (this is why the abrupt changes in
velocity in Fig.~\ref{fig:brown} are exaggeration).
Therefore, the rate of collisions has to be increased in parallel with
$m/M\to 0$ to have a sizeable total collision effect. Similarly to the van Hove limit,
there is a natural scaling limit, where nontrivial
limiting dynamics was proven
by D\"urr, Goldstein and Lebowitz \cite{DGL}.
On the quantum side, this model corresponds to a
quantum particle immersed in a phonon (or possibly photon)
bath. On the kinetic scale the Boltzmann equation
 was proven in \cite{E}. 
Recently De Roeck and Fr\"ohlich \cite{DF}
 (see also \cite{DFP} for an earlier toy model) have proved quantum diffusion
for a related model in $d\ge 4$ dimensions, where the mass
of the quantum particle was large and an additional
internal degree of freedom (``spin'') was coupled
to the heat bath to enhance the decay of time correlations.

\medskip

We remark that the problem becomes enormously more complicated
if the {\bf heat-bath particles can interact among each other},
i.e. if we are facing a truly interacting many-body dynamics.
In this case there is even no need to distinguish between
the tracer particle and the heat-bath particles, in
the simplest model one just considers identical 
particles interacting via a two-body interaction
and one investigates the single-particle density function $f_t(x,v)$.
In a certain scaling limit regime, the model should
lead to the nonlinear Boltzmann equation \eqref{nlb}.
This has only been  proven in the classical model for short time 
by Lanford \cite{L}. His work appeared in 1975
and since then nobody could extend the proof to include longer time
scales. We mention that the complications are partly due to the
fact that the nonlinear Boltzmann equation itself
does not have a satisfactory existence theory for long times.
The corresponding quantum model is an unsolved problem;
although via an order by order expansion \cite{BCEP} there is no
doubt on the validity of the nonlinear Boltzmann equation
starting from a weakly coupled interacting many-body 
model, the expansion cannot be controlled up to date.
Lukkarinen and Spohn \cite{LS2} have studied a weakly nonlinear cubic Schr\"odinger
equation with a random initial data (drawn from a phonon bath) near
equilibrium. They proved that the space-time covariance of the
evolved random wave function satisfies a nonlinear Boltzmann-type
equation in the kinetic limit, for short times.
While this is still a one-particle model and only fluctuations
around equilibrium is studied,
 its nonlinear character
could be interpreted as a first step towards understanding the
truly many-body Boltzmann equation.

\bigskip

Finally, the most challenging (classical) model is
the {\bf deterministic Hamiltonian with a random initial data
of a few degrees of freedom}, the typical example being
the various mathematical billiards.
The simplest billiard is the hard-core periodic
Lorentz gas  (also called Sinai's billiard \cite{BS}), where
the scatterers are arranged in a periodic lattice
and a single point particle (``billiard ball'')
moves among these scatterers according
to the usual rules of specular reflections.
The methods of proofs here originate more in dynamical system
than in statistical mechanics and they have a strong
geometric nature (e.g. convexity of the scatterers is
heavily used).

\bigskip

All these models have natural quantum mechanical analogues;
for the Hamiltonian systems they simply follow from
standard quantization of the classical Hamiltonian.
These models are  summarized  in the table below
and they are also illustrated
in Fig.~\ref{fig:mic}. We note that the quantum analogue
of the periodic Lorentz gas is ballistic due to the
Bloch waves (a certain diagonalization procedure
similar to the Fourier transform), thus in this
case the classical and  quantum models behave differently;
the quantum case being relatively trivial.

\bigskip\bigskip

\footnotesize{

\begin{tabular}{l|l|l} 

& CLASSICAL MECHANICS & QUANTUM MECHANICS \\ \hline  && \\
Stochastic dynamics  & Random walk (Wiener) & Random kick
model with \\ 
(no memory) & & zero time corr. potential  \\
&& (Pillet, Schenker-Kang)

\\&& \\  \hline && \\

Hamiltonian particle in a & Lorentz gas: particle &  Anderson model \\

random environment  & in random scatterers & or quantum Lorentz gas \\ 

(one body) &  (Kesten-Papanicolaou) &   (Spohn, Erd{\H o}s-Yau, Erd{\H o}s-Salmhofer-Yau  \\
&  (Komorowski-Ryzhik) & Disertori-Spencer-Zirnbauer) \\ && \\ \hline && \\

Hamiltonian particle & Einstein's kinetic model & Electron in phonon \\
in a heat bath  & (D\"urr-Goldstein-Lebowitz)  & or photon bath \\
(randomness in the & & (Erd{\H o}s, Erd{\H o}s-Adami,  Spohn-Lukkarinen \\

many-body data) & & De Roeck--Fr\"ohlich)\\ && \\ \hline && \\
 
Periodic Lorentz gas & Sinai billiard   & Ballistic\\

(randomness in the & (Bunimovich-Sinai) & (Bloch waves, easy)\\

one-body initial data) &&
\\ && \\ \hline && \\

Many-body interacting &   Nonlinear Boltzmann eq&  Quantum NL Boltzmann \\

Hamiltonian &  (short time: Lanford) &  (unsolved) \\

\end{tabular}

}

\bigskip\bigskip

\normalsize

\bef\bec
\epsfig{file=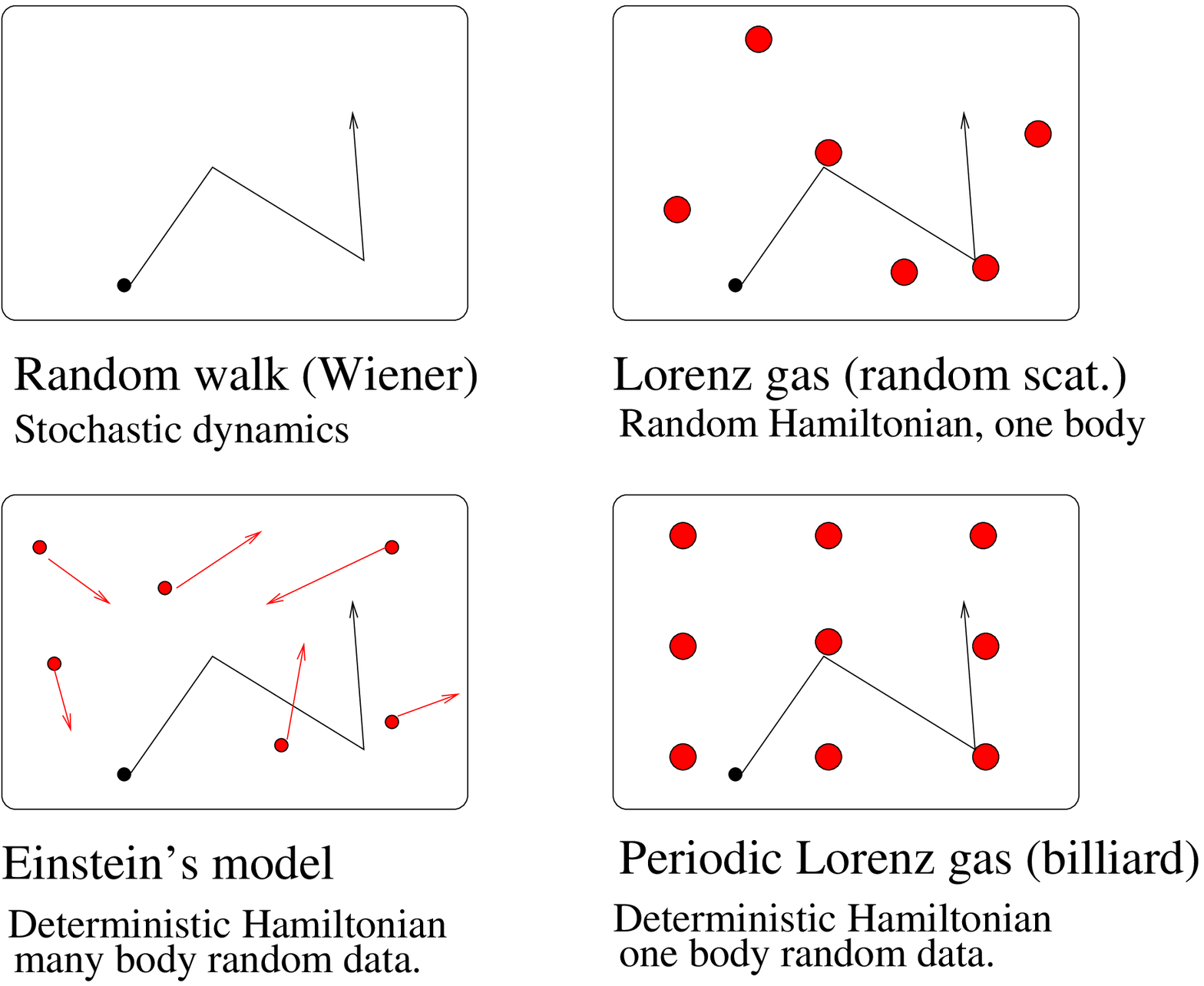, scale=0.55}
\eec
\caption{Microscopic models for diffusion}
\label{fig:mic}
\eef

\section{Some facts  on stochastic processes}\label{sec:recap}
\setcounter{equation}{0}

Our goal is to understand the stochastic features
of quantum dynamics. A natural way to do it is
to compare quantum dynamics with a
much simpler stochastic dynamics whose properties
are well-known or much easier to establish.  
In this section the most important ingredients  of 
elementary stochastic processes are collected
and we will introduce the Wiener process.
Basic knowledge of probability theory
(random variables, expectation, variance, independence,
Gaussian normal random variable,
characteristic function) is assumed.

\subsection{The central limit theorem}

Let $v_i$, $i=1,2,\ldots $,
 be a sequence of independent, identically distributed (denoted
by i.i.d.
for brevity) random variables with zero expectation,
$\bE\, v_i =0$ and finite variance, $\bE\, v_i^2 = \sigma^2$.
The  values of $v_i$ are either in $\bR^d$ or $\bZ^d$. In case
of $d>1$, the variance $\sigma^2$ is actually a matrix (called
{\it covariance matrix}),  defined
as $\sigma^2_{ab}= \bE\, v^{(a)}v^{(b)}$ where $v=(v^{(1)}, v^{(2)}, 
\ldots , v^{(d)})$. 
The $v_i$'s should be thought of as  velocities or steps of a walking particle
at discrete time $i$. 
For simplicity we will work in $d=1$ dimensions, the extension
to the general case is easy.
A natural question:
 Where is the particle after $n$ steps
if $n$ is big?

\medskip

Let 
$$
       S_n:= \sum_{i=1}^n v_i
$$
be the location of the particle after $n$ steps. It is clear
that the expectation is zero,
$   \bE \, S_n =0$, and the
variance is 
\be
\bE\, S_n^2 = \bE \Big(  \sum_{i=1}^n v_i\Big)^2=
 \sum_{i=1}^n \bE\, v_i^2 = n\sigma^2
\label{eq:quadvar}
\ee
(note that independence via 
$\bE \, v_iv_j=0$ for $i\ne j$ has been used).

This suggests to rescale $S_n$ and define
$$
    X_n: = \frac{S_n}{\sqrt{n}} = \frac{ \sum_{i=1}^n v_i}{\sqrt{n}}
$$
then $\bE\, X_n=0$ and $\bE \, X_n^2 =\sigma^2$. 
 The surprising fact is 
that the distribution of $X_n$ tends to a universal one (namely
to the Gaussian normal distribution) as $n\to \infty$,
no matter what the initial distribution of $v_i$ was!
This is the central limit theorem, a cornerstone of
probability theory and one of the fundamental
theorems of Nature. 

\begin{theorem}\label{thm:clt}
Under the conditions $\bE \,v_i=0$ and $\bE\, v_i^2 =\sigma^2<\infty$
the distribution of the rescaled sum $X_n$ of the i.i.d
random variables $v_i$ converges to the normal distribution, i.e
$$
     X_n \Longrightarrow X
$$
in the sense of probability  distributions where 
 $X$ is a random variable distributed according 
to $N(0, \sigma^2)$.
\end{theorem}
We recall that the density of the normal distribution
in $\bR^1$  is given by
$$
   f(x) = \frac{1}{\sqrt{2\pi}\sigma} e^{-\frac{x^2}{2\sigma^2}},
$$
and the convergence is the standard weak convergence
of probability measures (or convergence in distribution
or convergence in law):

\begin{definition}
A sequence of random variables $X_n$ converges
to the random variable $X$ in distribution, $X_n\Longrightarrow X$
if
$$
   \bE \,G(X_n) \to \bE \,G(X) 
$$
for any continuous bounded function $G: \bR \to \bR$
\end{definition}

In particular, $X_n$ converges in distribution to  $N(0, \sigma^2)$
if
$$
   \bE \,G(X_n) \to  \int G(x) f(x) \rd x \; .
$$
Analogous definitions hold in higher dimensions.

\bigskip



Recall that we will have two scales: a microscopic and a macroscopic one.
In this example, the microscopic scale corresponds to one step
of the walking particle and we run the process up to a total of $n$
units of the microscopic time.
 The macroscopic time will be the natural time
scale of the limit process and it will be kept order one even as
$n\to\infty$. Recalling that we introduced $\e\ll 1$
as a scaling parameter, we now reformulate the central limit theorem
in this language.

Let $T>0$ be a macroscopic time (think of it as an unscaled,
order one quantity) and let
$$ 
    n: = [T\e^{-1}],
$$
where $[\cdot]$ denotes the integer part. Let again $\bE \, v_i=0$,
$\bE \, v_i^2 = \sigma^2$. From the central limit theorem it directly follows
that
\be
   \wt X_T^\e:= \frac{1}{\e^{-1/2}}
\sum_{i=1}^{[T\e^{-1}]} v_i 
   \Longrightarrow X_T,\qquad \mbox{as $\e\to 0$},
\label{xtilde}
\ee
where $X_T$ is a normal Gaussian variable with variance $T\sigma^2$
and with density function
\be
   f_T(X) = \frac{1}{\sqrt{2\pi T}\sigma}\exp{\Big(
  -\frac{X^2}{2\sigma^2T}\Big)}.
\label{gaussden}
\ee
Note that the normalization in \eqref{xtilde} is only by $\sqrt{\e^{-1}}$ and not by
$\sqrt{n} \approx \sqrt{T\e^{-1}}$, therefore
the limit depends on $T$ (but  not on $\e$, of course).

Since the Gaussian function $f_T(X)$ gives the probability
of finding $X_T$ at location $x$, we have
$$
   f_T(X)\rd X \approx \mbox{Prob}\{ \mbox{$\wt X^\e_T$ is located at
$X+\rd X$ at time $T$}\}.
$$
Note that we used macroscopic space and time coordinates, $X$ and $T$.
Translating this statement into the original
process in microscopic coordinates we see
that
$$
   f_T(X)\rd X \approx \mbox{Prob}\{ \mbox{finding $S_n$ at 
$\e^{-1/2}(X+\rd X)$ at time $n\approx \e^{-1}T$}\}.
$$
Note that the space and time are rescaled differently; if
$(X,T)$ denote the macroscopic space/time coordinates
and $(x, t)$ denote the microscopic coordinates, then
$$
    x=\e^{-1/2}X, \qquad t = \e^{-1}T.
$$
This is the typical {\it diffusive scaling}, where
\be
   \mbox{time} = (\mbox{distance})^2.
\label{diffscale}
\ee
Finally, we point out that the limiting density function $f_T(X)$
satisfies the heat equation with {\it diffusion coefficient} $\sigma^2$:
\be
    \partial_T f_T(X) =\sigma^2 \partial_X^2 f_T(X).
\label{hh}
\ee
In fact, the Gaussian function \eqref{gaussden} is the fundamental solution to
the heat equation also in higher dimensions.

The emergence of the normal distribution and  the heat equation
is universal; note that apart from the variance (and the
zero expectation) no other information on the distribution
of the microscopic step $v$ was used. The details of the
microscopic dynamics are wiped out, and only the essential
feature, the heat equation (or the normal distribution)
with a certain diffusion coefficient,
remains visible after the $\e\to0$ scaling limit.

\bigskip

We remark that the central limit theorem (with 
essentially the same proof) is valid if $v_i$'s have
some short time correlation. Assume that $v_i$ is a stationary
sequence of random variables with zero mean, $\bE\, v_i =0$,
and let 
$$
  R(i,j):=\bE \, v_iv_j
$$ 
the correlation function. By stationarity, $R(i,j)$ depends
only on the difference, $i-j$, so let us use the notation
$$ R(i-j):=\bE \, v_iv_j
$$
This function is also called the {\it velocity-velocity
autocorrelation function.}

\begin{theorem} \label{thm:vv}
Assume that the velocity-velocity autocorrelation function is summable, i.e.
$$
   \sum_{k=-\infty}^\infty R(k) <\infty.
$$
Then
 the rescaled sums
$$
  \wt X^\e_T : = \e^{1/2}\sum_{i=1}^{[T\e^{-1}]} v_i
$$
converge in distribution to a normal random variable,
$$
   \wt X^\e_T \Longrightarrow X_T, \qquad \e\to0,
$$
whose density function $f_T(X)$ satisfies the heat equation
$$  
    \partial_T f = D\Delta f_T
$$
with a diffusion coefficient
$$
   D:= \sum_{k=-\infty}^\infty R(k),
$$
assuming that $D$ is finite. This relation
between the diffusion coefficient and the sum (or integral)
of the velocity-velocity autocorrelation function is
called the  Green-Kubo formula.
\end{theorem}

\bigskip

Finally we mention that all results remain valid in higher
dimensions, $d>1$. The role of the diffusion
coefficient is taken over by a diffusion matrix $D_{ij}$
defined by the correlation matrix of the single step distribution:
$$
   D_{ij} = \bE \, v^{(i)}v^{(j)}
$$
where $v= (v^{(1)}, v^{(2)}, \ldots , v^{(d)})$.
The limiting heat equation \eqref{hh} in the
uncorrelated case is modified to
$$
    \partial_T f_T(X) =\sum_{i,j=1}^d D_{ij} \partial_{i}\partial_j f_T(X).
$$
In particular, for random vectors $v$ with i.i.d. components the
covariance matrix is constant $D$ times the identity, and we obtain
the usual heat equation
$$
\partial_T f_T(X) = D \Delta_X f_T(X)
$$
where $D$ is called the {\it diffusion coefficient}.

\subsection{Markov processes and their generators}\label{sec:markov}

Let $X_t$, $t\ge 0$, be a continuous-time stochastic process, i.e.
$X_t$ is a one-parameter family of random variables,
the parameter is usually called time.  The {\it state space}
of the process is the space from where $X_t$ takes its values,
in our case $X_t\in \bR^d$ or $\bZ^d$. As usual, $\bE$ will
denote the expectation with respect to the distribution of the
process.

\begin{definition}
$X_t$ is a Markov process if for any $t\ge \tau$
$$
  \mbox{Dist} ( X_t \; | \; \{ X_s\}_{s\in [0,\tau]})
  =\mbox{Dist} ( X_t \; | \; X_\tau)
$$
i.e. if the conditional distribution of $X_t$ conditioned
on the family of events $X_s$ in times $s\in[0,\tau]$
is the same as conditioned only on the event at time $\tau$.
\end{definition}
In simple words it means that the state, $X_t$, of the system at time $t$
 depends on the
past between times $[0,\tau]$ only via the state of
the system at the last time of the past interval, $\tau$.
All necessary information about the past is condensed in the
last moment.

The process depends on its initial value, $X_0$.
Let $\bE_x$ denote the expectation value
assuming that the process started from the point $x\in \bR^d$ or $\bZ^d$, i.e.
$$
   \bE_x \varphi(X_t)= \bE \{\varphi(X_t)\; | \; X_0=x\}.
$$
Here and in the sequel $\varphi$ will denote functions on
the state space, these are also called {\it observables}.
We will not specify exactly the space of observables.

\bigskip

Markov processes are best described by their generators:

\begin{definition}\label{def:gen}
The generator of the Markov process $X_t$ is an operator
acting on the observables $\varphi$ 
and it is given by
$$
    (\cL \varphi)(x):= \frac{d}{d\e}\Big|_{\e=0+0} \bE_x \; \varphi(X_\e).
$$
\end{definition}
This definition is a bit lousy since the function space
on which $\cL$ act is not defined. Typically it is either the space
of continuous functions or the space of
$L^2$-functions on the state space, but for both concept
an extra stucture -- topology or measure -- is needed on the state space.
Moreover, the generator is typically an unbounded operator,
not defined on the whole function space but only on a dense
subset.
For this lecture we will leave these technicalities aside,
but when necessary, we will think of the state space $\bR^d, \bZ^d$
equipped with their natural measures and for the space 
of functions we consider will the $L^2$-space.

\bigskip

The key fact is that the generator tells us everything about the 
Markov process itself. Let us demonstrate it by answering two natural
questions in terms of the generator.

\bigskip

{\bf Question 1:} Starting from $X_0=x$, what is the distribution of
$X_t$?

\medskip

{\it Answer:} Let $\varphi$ be an observable (function on the state
space) and define
\be
    f_t(x):= \bE_x \; \varphi(X_t).
\label{ft}
\ee
We wish to derive an evolution equation for $f_t$:
$$
  \partial_tf_t(x) = \lim_{\e\to 0} \frac{1}{\e}
  \bE_x\Big[ \bE_{X_\e} \varphi(\wt X_t) -\bE_x\;\varphi(X_t)\Big],
$$
since by the Markov property
$$
   \bE_x \; \varphi(X_{t+\e}) = \bE_x \bE_{X_\e} ( \varphi(\wt X_t)), 
$$
where $\wt X_t$ is a new copy of the Markov process started
at the (random) point $X_\e$.
Thus
$$
  \partial_tf_t(x) = \lim_{\e\to 0} \frac{1}{\e}
  \bE_x\Big[ f_t(X_\e)-f_t(x)\Big] = 
\frac{\rd}{\rd\e}\Big|_{\e=0+0} \bE_x \; f_t(X_\e)=(\cL f_t)(x).
$$
Therefore $f_t$, defined in \eqref{ft}, solves the initial
value problem
$$
   \partial_t f_t = \cL f_t\qquad \mbox{with}\quad  f_0 (x) =\varphi(x).
$$
Formally one can write the solution as
$$
   f_t= e^{t\cL}\varphi.
$$
If the observable is a Dirac delta function at a fixed point
$y$ in the state space,
$$
  \varphi(x) = \delta_y(x),
$$
then the solution to the above initial value problem is
called the {\it transition kernel} of the Markov process
and it is denoted by $p_t(x,y)$:
$$
   \partial_t p_t(\cdot, y) = \cL p_t(\cdot , y)
\qquad\mbox{with} \quad p_0(\cdot, y) = \delta_y.
$$
The 
 intuitive meaning of the transition kernel is
$$
  p_t(x,y)\rd y := \mbox{Prob}\{ \mbox{After time $t$ the process is at
$y+\rd y$ if it started at $x$ at time 0}\}.
$$

\bigskip

{\bf Question 2:}  Suppose that the initial value
of the process, $X_0=x$, is distributed
according to a  density function $\psi(x)$ on the state space. What is the probability
that after time $t$ the process is at $y$?

\medskip

{\it Answer:} 
$$
  g_t(y):= \mbox{Prob}(\mbox{$X_t$ at $y$ after time $t$}) = 
\int \psi(x)p_t(x,y)\rd x.
$$
It is an exercise to check (at least formally), that
$g_t$ solves the following initial value problem
$$
    \partial_t g_t = \cL^* g_t \qquad \mbox{with}\quad
    g_0 = \psi,
$$
where $\cL^*$ denotes the adjoint of $\cL$ (with respect
to the standard scalar product of the $L^2$ of the state space).

\subsection{Wiener process and its generator}

The Wiener process is the rigorous mathematical construction of
the Brownian motion. There are various ways to introduce it, we
choose the shortest (but not the most intuitive) definition.

First we need the concept of the  {\it Gaussian process}. We recall that the 
centered Gaussian
random variables have the remarkable property that all
their moments are determined by the covariance matrix
(a random variable is called {\it centered} if its expectation is zero,
$\bE \,X=0$).
If $(X_1, X_2, \ldots, X_{2k})$ is a centered Gaussian vector-valued
random variable, then the higher moments can be computed by
{\bf Wick's formula} from the second moments (or {\it two-point correlations})
\be
   \bE \; X_1X_2\ldots X_{2k} = \sum_{\pi \;\;
\mbox{\footnotesize pairing}} \prod_{(i,j)\in \pi}
  \bE \; X_iX_j,
\label{wick}
\ee
where the summation is over all possible (unordered) pairings of
the indices $\{ 1,2,\ldots 2k\}$.

A stochastic process $X_t\in \bR^d$ is called {\it Gaussian},
if any finite dimensional projection is a Gaussian
vector-valued random variable, i.e. for any $t_1<t_2<\ldots < t_k$,
the vector $(X_{t_1},X_{t_2}, \ldots , X_{t_n})\in \bR^{dn}$ is a Gaussian
random variable.

\begin{definition}
A continuous
Gaussian stochastic process $W_t= (W_t^{(1)},\ldots, W_t^{(d)})
\in \bR^d$, $t\ge 0$, is
called the $d$-dimensional (standard) Wiener process if it satisfies the
following

\begin{itemize}
\item[i)] $W_0=0$
\item[ii)]  $\bE\, W_t =0$
\item[iii)] $\bE \, W_s^{(a)}W_t^{(b)} = \min \{ s,t\}\delta_{ab}$.
\end{itemize}
\end{definition}

From Wick's formula it is clear that all higher moments of $W_t$
are determined.  It is a fact that the above list of
requirements determines the Wiener process uniquely.
We can now extend the central limit theorem to processes.

\begin{theorem}\label{thm:w} Recall the conditions of Theorem~\ref{thm:clt}.
The stochastic process 
$$
    \wt X^\e_T: = \e^{1/2} \sum_{i=1}^{[T\e^{-1}]} v_i
$$
(defined already in \eqref{xtilde}) converges
in distribution (as a stochastic process) to the Wiener process
$$
    \wt X^\e_T\Longrightarrow W_T
$$
as $\e\to0$.
\end{theorem}

The proof of this theorem is not trivial. It is fairly easy
to check that the moments converge, i.e. for any $k\in \bN^d$
multiindex
$$
    \bE \;[\wt X^\e_T]^k \to \bE \; W_T^k
$$
and the same holds even for different times:
$$
   \bE\; \big[ \wt X^\e_{T_1} \big]^{k_1}\big[ \wt X^\e_{T_2}
\big]^{k_2} \ldots \big[ \wt X^\e_{T_m}\big]^{k_m}
   \to \bE\; W_{T_1}^{k_1}W_{T_2}^{k_2}\ldots W_{T_m}^{k_m},
$$
but this is not quite enough. The reason is that a
continuous time process is a collection of
uncountable many random variables and the joint
probability measure on such a big space
is not necessarily determined by finite dimensional
projections. If the process has some additional
regularity property, then yes. The correct notion
is the {\it stochastic equicontinuity} (or Kolmogorov condition)
which provides the necessary compactness (also called tightness in
this context). We will not go into more details 
here, see any standard book on stochastic processes.

\begin{theorem}
The Wiener process on $\bR^d$  is Markovian with generator
$$
  \cL = \frac{1}{2}\Delta.
$$
\end{theorem}

{\it Idea of the proof.} We work in $d=1$ dimension for simplicity.
 The Markovity can be easily checked
from the explicit formula for the correlation function.
The most important ingredient is that the Wiener process
has {\it independent increments:}
\be
  \bE \; (W_t-W_s)W_u =0
\label{stu}
\ee
if $u\leq s \leq t$; i.e. the increment in the time interval
$[s,t]$ is independent of the past. The formula \eqref{stu} is trivial
to check from property (iii).

Now we compute the generator using Definition \ref{def:gen}
$$  \frac{\rd}{\rd\e}\Big|_{\e=0+0} \bE_0\varphi(W_\e) 
  =
 \frac{\rd}{\rd\e}\Big|_{\e=0+0} \bE_0\Bigg( \varphi(0) 
 +\varphi'(0)W_\e+ \frac{1}{2}\varphi''(0) W_\e^2 + \ldots\Bigg)
= \frac{1}{2}\varphi''(0).
$$
Here we used that $\bE_0 W_\e =0$, $\bE_0 W_\e^2=\e$ and
the dots refer to irrelevant terms that are higher order in $\e$.
$\;\;\Box$

Finally we remark that one can easily define a Wiener process with
any nontrivial (non-identity) diffusion coefficient matrix $D$
as long as it is positive definite (just write $D=A^*A$ and
apply the linear transformation $W\to AW$ to the standard Wiener
process).
The generator is then
$$
  \cL = \frac{1}{2}\sum_{i,j=1}^dD_{ij} \partial_i\partial_j.
$$

\subsection{Random jump process
on the sphere $S^{d-1}$ and its generator}\label{sec:jump}

The state space of this process is the unit sphere $S=S^{d-1}$
in $\bR^d$. We are given a function
$$
  \sigma(v,u): S\times S \to \bR_+
$$
which is interpreted as jump rate from the point $v$ to $u$.

The process will be parametrized by a continuous time $t$,
but, unlike the Wiener process, its  trajectories will not be continuous,
rather piecewise constant with some jumps at a discrete set of times.
A good intuitive picture is that the particle
jumping on the unit sphere has a random clock (so-called {\it
exponential clock}) in its pocket. The clock emits a signal
with the standard exponential distribution, i.e. 
$$
  \mbox{Prob} \{ \mbox{there is a signal at $t+\rd t$} \} = e^{-t} \rd t,
$$
and when it ticks, the particle, being at location $v$, jumps
to any other location $u\in S$ according to the distribution
$u\to \sigma(v,u)$ (with the appropriate normalization):
$$
   \mbox{Prob} \{ \mbox{The particle from $v$ jumps to $u+\rd u$}\}
 = \frac{\sigma(v,u) \rd u}{\int \sigma(v,u) \rd u}.
$$
The transition of the process
at an infinitesimal time increment is given by
\be
   v_{t+\e} = \left\{ \begin{array}{ll} u+\rd u & \mbox{with probability}
   \quad \e\sigma(v_t,u)\rd u\\
  v_t & \mbox{with probability}
  \quad 1- \e\int \sigma(v_t,u)\rd u \end{array}\right.
\label{inf}
\ee
up to terms of order $O(\e^2)$ 
as $\e\to 0$.

Thus the generator of the process can be computed
from Definition \ref{def:gen}.
Let
$$
  f_t(v) : = \bE_v \varphi(v_t),
$$
assuming that the process starts from some given point $v_0=v$
and $\varphi$ is an arbitrary observable.
Then
\begin{align}
   \partial_t f_t(v)& =\lim_{\e\to0+0} \frac{1}{\e}\bE_v
   \Big( \bE_{v_\e} \varphi(\wt v_t)-\bE_v \varphi(\wt v_t)\Big)\non \\
   & =\int \rd u \;\sigma (v,u)  \Big[ \bE_u\varphi(\wt v_t)- \bE_v
\varphi(\wt v_t)\Big] \non \\
 & = \int \rd u \;\sigma (v,u)  \big[ f_t(u)-f_t(v)\big], \non
\end{align}
where we used the Markov property in the first line
and the infinitesimal jump rate from \eqref{inf} in the second line.
Note that with probability $1- \e\sigma(v,u) $
we have $v_\e=v$.
Thus the generator of the random jump process
is
\be
  (\cL f)(v) :=\int \rd u \;\sigma (v,u)  \big[ f(u)-f(v)\big].
\label{gen}
\ee  
Note that it has two terms; the first term is
called the {\it gain term} the second one is the {\it loss term}.
The corresponding evolution  equation for the time dependent probability
density of the jump process,
\be
  \partial_t f_t(v) =\int \rd u \;\sigma (v,u)  \big[ f_t(u)-f_t(v)\big]
\label{vboltz}
\ee 
is called {\it linear Boltzmann equation in velocity space}.

The elements of the state space $S$ will be interpreted as
velocities of a moving particle undergoing fictitious
random collisions. Only the velocities are recorded.
A velocity distribution $f(v)$ is called {\it equilibrium distribution}, if
${\cL f} \equiv 0$; it is fairly easy to see that, under
some nondegeneracy condition on $\sigma$, the equilibrium exists uniquely.

\section{Classical mechanics of a single particle}\label{sec:class}
\setcounter{equation}{0}

Classical mechanics of a single particle in $d$-dimensions is
given by a Hamiltonian (energy function) defined
on the phase space $\bR^d\times \bR^d$:
\be
H(v,x):= \frac{1}{2} v^2 +U(x).
\label{clham}
\ee
Here $x$ is the position, $v$ is the momentum coordinate.
For most part of this notes we will not distinguish between
momentum and velocity, since we almost always consider the
standard  kinetic energy $\frac{1}{2}v^2$.

The Hamiltonian equation of motions is the following
set of $2d$ coupled first order differential equations:
\be
      \dot{x}(t)= \partial_v H = v \qquad \dot{v}(t)
 = -\partial_x H = -\nabla U(x).
\label{HAM}
\ee
For example, in case of the free evolution, when the potential
is zero, $U\equiv 0$, we have
\be
 x(t) = x_0 + v_0 t
\label{free}
\ee
i.e. linear motion with a constant speed.

Instead of describing each point particle separately, we
can think of a continuum of (non-interacting) single particles,
described by a phase space density $f(x,v)$.
This function is interpreted to give the number of
particles with given velocity at a given position, more
precisely
$$
  \int_\Delta f(x,v) \rd x \rd v = \mbox{Number of particles at $x$ with
velocity $v$ such that $(x,v)\in \Delta$}.
$$

Another interpretation of the phase space density picture is that
 a single particle with velocity $v_0$ at $x_0$ can
be described by the measure
$$
   f(x,v) = \delta(x-x_0)\delta(v-v_0),\; 
$$
and  it is then natural to consider more general  measures as well.

The system evolves with time, so does its density, i.e.
we consider the time dependent phase space densities, $f_t(x,v)$.
For example in case of a single particle governed by
\eqref{HAM} with initial condition $x(0)=x_0$ and $v(0)=v_0$,
 the evolution of the phase space density is given by
$$
f_t(x,v)=\delta(x-x(t)) \delta(v-v(t)),
$$
where $(x(t),v(t))$ is the Hamiltonian trajectory
computed from \eqref{HAM}.

It is easy to check that if the point particle
trajectories satisfy \eqref{HAM}, then $f_t$ satisfies
the following evolution equation
\be
    (\partial_t + v\cdot \nabla_x) f_t(x,v) = \nabla U(x)\cdot\nabla_v
   f_t(x,v),
\label{li}
\ee
which is called the
{\it Liouville equation}.
The left hand side is called the {\it free streaming term}.
The solution to
\be
 (\partial_t + v\cdot \nabla_x) f_t(x,v) =0
\label{freeev}
\ee
is given by 
the linear transport solution
$$
f_t(x,v) = f_0(x-vt, v),
$$
where $f_0$ is the phase space density at time zero.
This corresponds  to
the free evolution \eqref{free} in the Hamiltonian
particle representation.

\subsection{The linear Boltzmann equation}

The linear Boltzmann equation is a phenomenological
combination of the free flight equation \eqref{freeev}
and the jump process on the sphere of the velocity space
\eqref{vboltz}
\be
 (\partial_t + v\cdot \nabla_x) f_t(x,v)
  =\int \sigma(u,v)\big[f_t(x,u)-f_t(x,v)\big]\rd u.
\label{lb}
\ee
Note that we actually have the adjoint of the 
jump process (observe that $u$ and $v$ 
are interchanged compared with \eqref{vboltz}), so the solution
determines:
$$
f_t(x,v)= \mbox{Prob}\{ \mbox{the process is at $(x,v)$ at time $t$}\},
$$
given the initial probability density $f_0(x,v)$ (Question 2
in Section \ref{sec:markov}).

The requirement that
the velocities are constrained to  the sphere
corresponds to energy conservation. Of course 
there is no real Hamiltonian dynamics behind 
this equation: the jumps are stochastic.

\bigskip

Notice that the free flight 
plus collision process standing  behind
the Boltzmann equation is actually  a random walk in continuous time.
In the standard random walk the particle ``jumps''
in a new random direction after every unit time.
In the Boltzmann process the jumps occur
at random times (``exponential clock''), but
the effect on a long time scale is the same.
In particular, the
long time evolution of the linear Boltzmann equation is diffusion
(Wiener process)
in position space. 

The following theorem formulates this fact more precisely.
We recall that
to translate the velocity process
into position space, one has to integrate the velocity, i.e.
will consider
$$
     x_t =\int_0^t v_s \rd s.
$$

\begin{theorem}
Let $v_t$ be a random velocity jump process
given by the generator \eqref{vboltz}. Then 
$$
                  X_\e(T)=: \e^{1/2}
\int_0^{T/\e} v_t \; \rd t \to   W_T     \qquad \mbox{(in distribution,)}
$$
where $W_T$ is a Wiener process with
 diffusion coeffient being the
velocity autocorrelation 
$$
     D=\int_0^\infty R(t)\rd t, \qquad R(t):= \bE \, v_0 v_t.
$$
Here $\bE$ is with respect to the equilibrium measure
of the jump process on $S^{d-1}$.
\end{theorem}

\section{Quantum mechanics of a single particle}\label{sec:qm}
\setcounter{equation}{0}

\subsection{Wavefunction, Wigner transform}

The state space of a quantum particle in $d$-dimensions
is $L^2(\bR^d)$ or $L^2(\bZ^d)$. Its elements are called $L^2$-wavefunctions and they
are usually 
denoted by $\psi=\psi(x)\in L^2(\bR^d)$ or $L^2(\bZ^d)$.
We assume the normalization, $\|\psi\|_2=1$.
We recall the interpretation of the wave function:
the position space density $|\psi(x)|^2 \rd x$ gives
the probability of finding an electron at a spatial
domain:
$$
   \int_{\Delta} 
|\psi(x)|^2 \rd x = \mbox{Prob} \{ \mbox{the particle's position
 is in $\Delta$}\}.
$$
Similarly, the Fourier transform of $\psi$,
$$
   \wh\psi(v) : = \int_{\bR^d} e^{-iv\cdot x} \psi(x) \rd x,
$$
determines the momentum space density:
$$
   \int_{\Delta} 
|\wh\psi(v)|^2 \rd v = \mbox{Prob} \{ \mbox{the particle's momentum
 is in $\Delta$}\}.
$$
In the lattice case,  i.e. when $\bZ^d$ replaces
$\bR^d$ as the physical space, the Fourier transform is replaced with Fourier series.
In these notes we neglect all $2\pi$'s that otherwise enter the
definition of the Fourier transform.

By the Heisenberg uncertainty principle, one cannot
simultaneously determine the position and the momentum
of a quantum particle, thus the concept of classical phase space
density does not generalize directly to quantum mechanics.
Nevertheless one can define a substitute for it, namely
the {\it Wigner transform}.  For any $L^2$-wavefunction $\psi$
we define the Wigner transform of $\psi$ as
$$
         W_\psi(x, v) := \int \overline\psi \Big( x+ {z\over 2}\Big)
        \psi \Big( x- {z\over 2}\Big) e^{ivz} \rd z,
$$
and we still interpret it as ``quantum phase space density''.

It is easy to check that $W_\psi$ is always real but in general
is not positive (thus it cannot be the density of
a positive measure -- in coincidence with the Heisenberg principle).
However, its marginals  reconstruct the position and momentum
space densities, as the following formulas can be easily checked:
$$
\int W_\psi(x,v) \rd v
=|\psi(x)|^2, \qquad \int W_\psi(x,v) \rd x
=|\wh\psi(v)|^2.
$$
In particular, for  normalized wave functions $\|\psi\|_2=1$, we have
\be
  \iint W_\psi(x,v) \rd v\rd x =1.
\label{wnorm}
\ee

We remark, that for the lattice case some care is needed for
the proper definition of the Wigner transform, since $x\pm \frac{z}{2}$ may
not be a lattice site. The precise definition in this case is
\be
   W_\psi(x, v) := \sum_{y,z\in \bZ^d\atop y+z=2x} e^{iv(y-z)} \bar \psi(y) \psi (z),
\label{Wlattice}
\ee
where $\psi\in \ell^2(\bZ^d)$, i.e. $y,z\in \bZ^d$, but $x\in \big(\bZ/2\big)^d$.
The formulas for marginal of the Wigner transform modify as follows:
$$
   \int W_\psi(x, v) \rd v = 2^d |\psi(x)|^d
$$
if $x\in \bZ^d$ and it is zero if $x\in  \big(\bZ/2\big)^d\setminus \bZ^d$.
We still have
$$
   \int_{ \big(\bZ/2\big)^d} W_\psi(x, v)\rd x = 2^{-d}\sum_{x\in  \big(\bZ/2\big)^d} W_{\psi}(x, v)
  = |\wh\psi(v)|^2
$$
and 
$$
\int_{ \big(\bZ/2\big)^d}\rd x \int \rd v 
 W_\psi(x, v) =\|\psi\|^2.
$$

Often we will use the Wigner transform in the Fourier representation, by
which we will always mean Fourier transform in the first ($x$) variable only, i.e.
with the convention
$$
  \wh f(\xi) = \int e^{-ix\xi}f(x) \rd x,
$$
we define
$$
  \wh W_\psi(\xi, v): = \int e^{-ix\xi} W_\psi(x, v) \rd x.
$$
After a simple calculation, we have
\be
  \wh W_\psi(\xi, v) = \ov{\wh\psi\Big( v-\frac{\xi}{2}\Big)}\wh\psi\Big( v+\frac{\xi}{2}\Big).
\label{wigmom}
\ee
In the discrete case we have $\xi\in R^d/(2\cdot 2\pi \bZ^d)$ and
$$
  \wh W(\xi, v) = \ov{\wh\psi\Big( v-\frac{\xi}{2}\Big)}\wh\psi\Big( v+\frac{\xi}{2}\Big).
$$

More generally, if $J(x,v)$ is a classical phase space observable,
the scalar product
$$
    \langle J, W_\psi\rangle =\int J(x,v)W_\psi(x,v)\rd x \rd v
$$
can be interpreted as the expected value of $J$ in state $\psi$.
Recall that ``honest'' quantum mechanical observables are self-adjoint
operators $\cO$ on $L^2(\bR^d)$ and their expected value
is given by 
$$
\langle \psi,\cO \psi\rangle =\int \overline{ \psi}(x) \; (O\psi)(x) \rd x.
$$
For a large class of observables there is a natural relation between
observables $\cO$ and their phase space representations
(called {\it symbols}) that are functions on the phase space like
$J(x,v)$. For example, if $J$ depends only on $x$ or only on $v$,
then the corresponding operator is just the standard quantization of $J$,
i.e.
$$
    \int J(x) W_\psi(x,v) \rd x \rd v = \langle \psi, J\psi\rangle
$$
where $J$ is a multiplication operator on the right hand side, or
$$
 \int J(v) W_\psi(x,v) \rd x \rd v = \langle \psi, J(-i\nabla)\psi\rangle,
$$
and similar relations hold for the Weyl quantization 
of any symbol $J(x,v)$.

We also remark that the map $\psi \to W_\psi$ is invertible,
i.e. one can fully reconstruct the wave function from
its Wigner transform. On the other hand, not every real function
of two variables $(x,v)$ is the Wigner transform of some
wavefunction.

\subsection{Hamilton operator and the Schr\"odinger equation}

The quantum dynamics is generated by the {\it Hamilton operator}
$$
    H = -\frac{1}{2} \Delta_x + U(x)
$$
acting on $\psi\in L^2(\bR^d)$. The first term is
interpreted as the kinetic energy and it is the
quantization of the classical kinetic energy $\frac{1}{2}v^2$
(compare with \eqref{clham}). The momentum operator is $p=-i\nabla$
and we set the mass equal one, so momentum and velocity coincide.
[To be pedantic, we should use the word momentum everywhere instead
of velocity; since the momentum is the canonical quantity,
then the dispersion relation, $e(p)$, defines the kinetic
energy as a function of the momentum operator, and
its gradient, $v=\nabla e(p)$ is the physical velocity.
Since we will mostly use the dispersion relation $e(p)=\frac{1}{2} p^2$,
the momentum and velocity coincide. However, for the
lattice model the velocity and the momentum will differ.]

The evolution of a state is given by the Schr\"odinger equation
$$
   i\partial_t\psi_t = H\psi_t=
\Big(-\frac{1}{2}\Delta +U\Big)
\psi_t
$$
with a given initial data $\psi_{t=0}=\psi_0$.
Formally, the solution is
$$
\psi_t = e^{-itH}\psi_0.
$$

If $H$ is self-adjoint in $L^2$, then the unitary group $e^{-itH}$ can
be defined by spectral theorem. Almost all Hamilton operators
in mathematical physics are self-adjoint, however the
self-adjointness usually requires some proof. We will neglect
 this issue here, but we only mention
 that self-adjointness is more than the symmetry
of the operator, because $H$ is typically unbounded
when issues about the proper domain of the operator become relevant.

Note the complex $i$ in the Schr\"odinger equation, it plays
an absolutely crucial role. It is responsible for
the wave-like character of quantum mechanics.
The quantum evolution are governed by {\it phase} and
{\it dispersion}. Instead of giving precise explanations, 
look at Fig.~\ref{fig:phase}: the faster the wave oscillates,
the faster it moves. 

\bef\bec
\epsfig{file=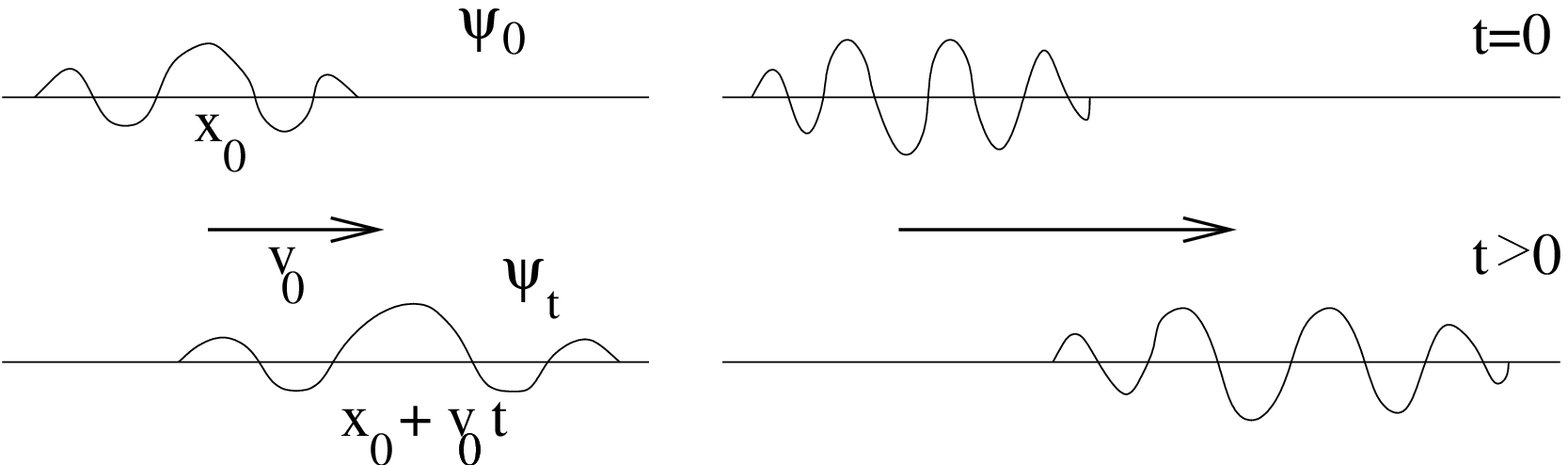, scale=0.7}
\eec
\caption{Evolution of slower (left) and a faster (right) wave packet}
\label{fig:phase}
\eef

We can also justify this picture by some sketchy calculation
(all these can be made rigorous in the so-called semiclassical regime).
We demonstrate that the free 
($U\equiv 0$)
evolution of  
$$
\psi_0(x):= e^{iv_0x}A(x-x_0)
$$
(with some fixed $x_0, v_0$)
after time $t$ is
supported around $x_0+ v_0t$. Here we tacitly assume that $A$ is
not an oscillatory function (e.g. positive) and the only oscillation
is given explicitly in the phase $e^{iv_0x}$, mimicking a plane wave.

We compute the evolution of $\psi_0$:
\begin{align}
  \psi_t(x) & = \int e^{iv(x-y)} e^{-itv^2/2}
 \underbrace{e^{iv_0y} A(y-x_0)}_{=: \psi_0(y)} 
\; \rd y\rd v\non\\
&   = e^{iv_0x_0}\int e^{iv(x-x_0)} e^{-itv^2/2} \wh A(v-v_0) \;\rd v
   \sim \int e^{i\Phi(v)} \wh A(v-v_0) \;\rd v
\non
\end{align}
with a phase factor
$$
     \Phi(v):= v(x-x_0)-\frac{1}{2} tv^2.
$$
We apply the stationary phase argument: the integral
is concentrated around the stationary points of the phase.
Thus
$$
0=\nabla_v \Phi = x-x_0 - tv
$$
 gives $x=x_0+vt\approx x_0+v_0t$
if $\wh A$ is sufficiently localized.

This argument can be made
rigorous if a
scale separation ansatz is used.
We introduce a small parameter $\e$
and assume that
$$
\psi_0(x) = \e^{d/2} e^{iv_0x} A( \; \e(x-x_0) \;),
$$
i.e. the wave has a slowly varying envelope (amplitude)
and fast phase. The prefactor is chosen to keep the
normalization
$\|\psi_0\|=1$ independently of $\e$.
 Such states and their generalizations
of the form
\be
\e^{d/2} e^{iS(\e x)/\e} A( \; \e(x-x_0) \;)
\label{WKB}
\ee
are  called {\it $WKB$-states.} These states
are characterized by a slowly varying amplitude
and wavelength profile and by a fast oscillation.
Similar picture holds if the potential is nonzero, but
slowly varying, i.e. $U(x)$ is replaced with 
$ U(\e x)$.

\subsection{Semiclassics}

The WKB states and the rescaled potential belong to
the {\it semiclassical theory} of quantum mechanics.
A quantum mechanical system is in the 
{\it semiclassical regime} whenever all the data
(potential, magnetic field, initial profile etc.)
live on a much bigger scale than the quantum wavelength.

\bef\bec
\epsfig{file=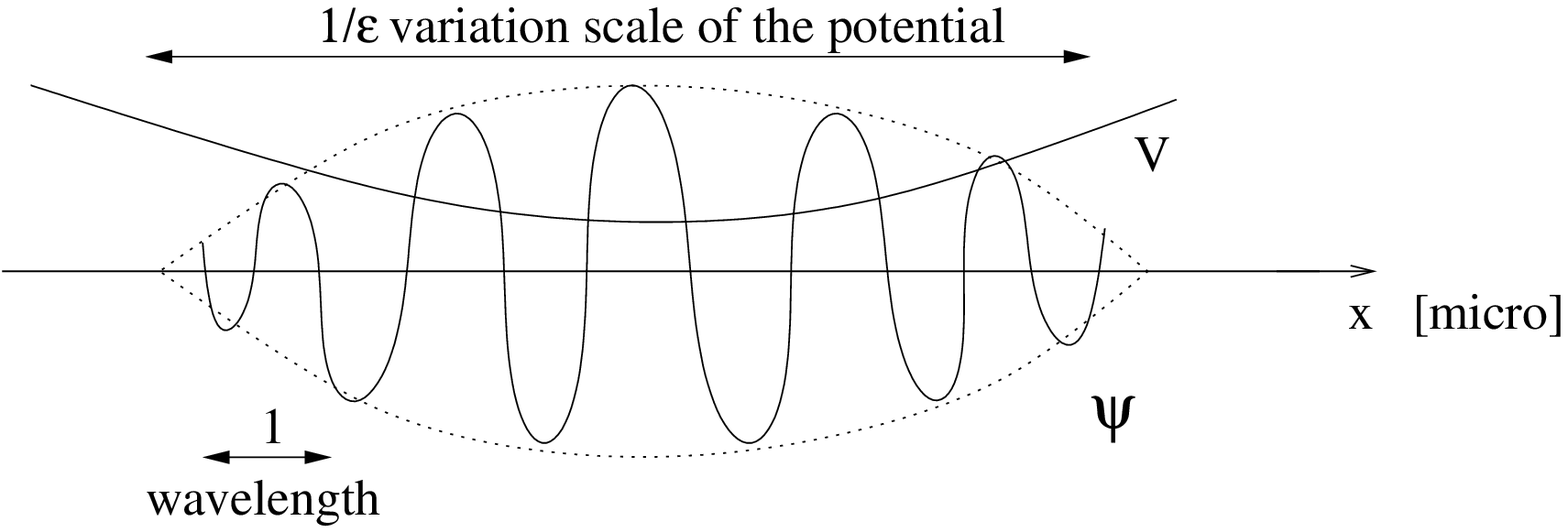, scale=0.7}
\eec
\caption{Semiclassical setup: short wavelength, large scale envelope and potential}
\label{fig:scsec}
\eef

The scale separation parameter is usually explicitly
included in the Schr\"odinger equation, for example:
\be
 i\partial_t \psi_t(x)= \Big[- \frac{1}{2}\Delta_x 
+ U(\e x)\Big]\psi_t(x).
\label{sc1}
\ee
This equation is written in microscopic coordinates.
It becomes more familiar if we rewrite it in
 macroscopic coordinates
$$
(X,T)=(x\e, t\e),
$$
where \eqref{sc1} takes the form
$$
 i\e \partial_T \Psi_T(X)= \Big[- \frac{\e^2}{2}\Delta_X
+ U(X)\Big]\Psi_T(X).
$$
Traditionally $\e$ is denoted $h$ in this equation
and is referred to the (effective) Planck constant, whose
smallness in standard units justifies taking the
limit $h\to 0$ (the so-called {\it semiclassical limit}).
No matter in what coordinates and units, the essential
signature of the semiclassical regime is the
scale separation between the wavelength and any other
lengthscales.

\bigskip

Wigner transform $W_\psi(x,v)$, written originally
in microscopic coordinates, can also be rescaled:
$$
    W^\e_\psi (X,V):= \e^{-d} W_\psi\Big(\frac{X}{\e}, V\Big).
$$
Note that apart from the rescaling of the $X$ variable 
($V$ variable is unscaled!), a normalizing prefactor $\e^{-d}$
is added to keep the integral of $W^\e$ normalized 
(see \eqref{wnorm})
$$
\iint W^\e_\psi(X,V) \rd X \rd V =1.
$$

The following theorem is a simple prototype
of the theorems we wish to derive in more complicated
situations. It shows how the quantum mechanical
Schr\"odinger evolution can be approximated by
a classical equation in a certain limiting regime:

\begin{theorem}
Consider a sequence of initial states $\psi_0^\e$
whose rescaled Wigner transform has a weak limit:
$$
   \lim_{\e\to 0} W_{\psi_0^\e}^\e(X,V) \rightharpoonup W_0(X,V).
$$
For example, the WKB states given in \eqref{WKB} have
a weak limit 
$$
 W_0(X,V) = |A(X)|^2\delta( V-\nabla S(X)).
$$
Let $\psi_t^\e$ denote the solution to the
 semiclassical Schr\"odinger equation
\eqref{sc1} with initial state $\psi_0^\e$.
Then the  weak limit of the rescaled  Wigner transform
of the solution at time $t=T/\e$,
$$
        W_T(X,V): = \lim_{\e\to0}W_{\psi_{T/\e}^\e}^\e ( X, V)
$$
satisfies the Liouville equation:
\be
        (\partial_T + V\cdot \nabla_X)W_T(X,V) =
        \nabla U(X)\cdot \nabla_V W_T(X,V).
\label{lio}\ee
Recall that the Liouville equation is equivalent to
the  classical Hamilton dynamics (see the derivation
of \eqref{li}).
\end{theorem}

In this way, the Liouville equation \eqref{lio} mimics
the semiclassical quantum evolution on large scales
(for a proof, see e.g. \cite{MR}). Notice
that weak limit is taken; this means that the small
scale structure of the Wigner transform $W_{\psi_t^\e}$
is lost. Before the weak limit, the Wigner transform of
$\psi_t^\e$  carries all information about $\psi_t^\e$,
but the weak limit means testing it against functions
that live on the macroscopic scale. This phenomenon
will later be essential to understand why irreversibility
of the Boltzmann equation does not contradict 
the reversibility of Schr\"odinger dynamics.

\section{Random Schr\"odinger operators}\label{sec:rs}
\setcounter{equation}{0}

\subsection{Quantum Lorentz gas or the Anderson model}

Now we introduce our basic model, the quantum Lorentz gas.
Physically, we wish to model electron transport in a disordered
medium. We assume that electrons are non-interacting,
thus we can focus on the evolution of a single electron.
The disordered medium is modelled by a potential
describing random impurities (other models are also possible,
e.g. one could include random perturbations in
the kinetic energy term). This can be viewed as the quantum
analogue of the classical Lorentz gas, where a classical
particle is moving among random obstacles.

\begin{definition} The model of a single particle
 described by the Schr\"odinger
equation on $\bR^d$ or on $\bZ^d$,
$$
i \partial_t \psi_t(x)=  H \psi_t(x), \qquad
 H  = -\Delta_x +  \lambda V_\om(x),
$$
is called the {\bf quantum Lorentz gas},
if $V_\om(x)$ is a random potential at location $x$.
\end{definition}

{\it Notation:} The subscript $\om$  indicates that $V(x)$ is a random variable
that also depends on an element  $\om$ in the probability space.

\medskip

More generally,  we can consider
\be
    H= H_0 + \lambda V,
\label{HH}
\ee
where $H_0$ is a deterministic Hamiltonian,
typically $H_0= -\Delta$.
We remark that a periodic background potential can
be added, i.e.
$$ 
   H= -\Delta_x +U_{per}(x)+  \lambda V_\om(x)
$$
can be investigated with similar methods. We recall that
the operator with a periodic potential and no randomness,
$$
   H_0=-\Delta_x +U_{per}(x),
$$
can be more or less explicitly diagonalized by using
the theory of Bloch waves. The transport properties
are similar to those of the free Laplacian.
This is called the {\it periodic quantum Lorentz gas} and it is
the quantum analogue of the Sinai billiard.
However, while the dynamics of the Sinai billiard has a very complicated
structure, its quantum version is quite straightforward
(the theory of Bloch waves is fairly simple).

We also remark that for the physical content of the model,
it is practically unimportant whether we work on $\bR^d$
or on its lattice approximation $\bZ^d$.
The reason is that we are investigating long time,
large distance phenomenon; the short scale structure
of the space does not matter. However, technically $\bZ^d$
is much harder (a bit unusual, since typically $\bR^d$ is
harder as one has to deal with the ultraviolet regime).
If one works on $\bZ^d$, then the Laplace operator is
interpreted as the discrete Laplace operator on $\bZ^d$, i.e.
\be
    (\Delta f)(x): = 2d \; f(x)-\sum_{|e|=1}f(x+e).
\label{discrlapl}
\ee
In Fourier space this corresponds to the 
dispersion relation 
\be
   e(p) = \sum_{j=1}^d (1- \cos p^{(j)})
\label{disp}
\ee
on the torus $p\in [-\pi,\pi]^d$.

The random potential can be fairly arbitrary, the only
important requirement is that it cannot have a long-distance
correlation in space. For convenience, we will assume
i.i.d. random
potential for the lattice case with a standard
normalization:
$$
\{ V(x)\; : \; x\in \bZ^d\} \quad \mbox{i.i.d} \qquad
\bE\, V(x)=0, \;\; \bE\, V^2(x) =1.
$$
The coupling constant $\lambda$ can then be used to adjust
the strength of the randomness. 
We can also write this random potential as
\be
   V(x) = \sum_{\alpha\in \bZ^d} v_\a \delta(x-\a),
\label{lat}
\ee
where $\{ v_\a\; :\; \a\in \bZ^d\}$ is a collection
of i.i.d. random variables and $\delta$ is the usual
lattice delta function.

For continuous models, the  random
potential can, for example, be given as follows;
$$
 V(x) = \sum_{\a \in \bZ^d} v_\a B(x-\a),
$$
where $B$ is a nice (smooth, compactly supported)
single site potential profile and $\{ v_\a\; :\; \a\in \bZ^d\}$
is a collection of i.i.d. random variables.
It is also possible to let the randomness perturb the location of 
the obstacles instead their strength, i.e.
$$
 V(x) = \sum_{\a \in \bZ^d} B(x-y_\a(\om)),
$$
where, for example, $y_\a(\om)$ is a random point in the
unit cell around $\a\in \bZ^d$, or even more canonically, the collection
$\{ y_\a(\om)\}_\a$ is
just a Poisson point process in $\bR^d$.
 The combination of
these two models is also possible and meaningful, actually
in our work \cite{ESY3} we consider 
\be
      V(x) =\int B(x-y)\rd \mu_\om (y)
\label{VB}
\ee
where $\mu_\om$ is a Poisson point process
with homogeneous unit density and with i.i.d.
coupling constants (weights), i.e.
\be
   \mu_\om = \sum_\a v_\a(\om) \delta_{y_\a(\om)},
\label{poidef}
\ee
where $\{ y_\a(\om)\; : \; \a =1,2,\ldots\}$
is a realization of the  Poission point process
and $\{ v_\a\}$ are independent (from each other and
from $y_\a$ as well) real valued random variables.

\bigskip

The lattice model $-\Delta + \lambda V$
with i.i.d. random potentials \eqref{lat}
is  called {\it Anderson model}. It  was 
invented by Anderson \cite{A} who was the first to
realize that electrons move quite differently
in disordered media than in free space or in a periodic background.
The main phenomenon Anderson discovered was
the {\it Anderson localization}, 
asserting that at sufficiently strong disorder
(or in $d=1,2$ at any nonzero disorder) 
the electron transport stops. We will explain 
this in more details later, here we just
remark that Anderson was awarded the Nobel Prize in 1977
for this work.

However, in  $d\ge 3$ dimensions
electron transport is expected despite the disorder if the disorder is weak.
This is actually what we experience in real life;
the eletric wire is conducting although it does
not have a perfect periodic lattice structure.
However, the nature of the electric transport changes:
we expect that on large scales it can be described by a diffusive equation
(like the heat equation) instead of a wave-type equation
(like the Schr\"odinger equation).

Note that the free 
 Schr\"odinger equation is ballistic (due to wave coherence).
This means that if we define the {\it mean square displacement}
by the expectation value of the observable $x^2$ at state $\psi_t$,
\be
\langle \; x^2\; \rangle_t:= \int  \rd x |\psi_t(x)|^2  x^2 
\qquad \Big( = \int \rd x \big| e^{-it\Delta}\psi_0(x)\big|^2 x^2 \Big),
\label{ms}
\ee
then it behaves as
$$
\langle \; x^2\; \rangle_t\sim t^2
$$
for large $t\gg 1$ as it can be easily computed. In contrast,
the mean square displacement for the heat equation 
scales as $t$ and not as $t^2$ (see \eqref{diffscale}).

Thus the long time transport of the free Schr\"odinger equation
is very different from the heat equation. We nevertheless claim,
that in a weakly disordered medium, the long time
Schr\"odinger evolution can be described by a diffusion (heat)
equation. Our results answer to the intriguing question how the diffusive
character emerges from a wave-type equation.

\subsection{Known results about the Anderson model}

We will not give a full account of all known mathematical results
on the Anderson model, we just mention
the main points.  The main issue is to study the
dichotomic nature of the Anderson model, namely
that at low dimension ($d=1,2$) or at high disorder $(\lambda\ge \lambda_0(d)$)
or  near the spectral edges
of the unperturbed operator $H_0$  the
time evolved state remains localized, while at high
dimension $(d\ge 3)$, at low disorder and away from the spectral edges it
is delocalized.

There are several signatures of (de)localization and
each of them can be used for rigorous definition.
 We list three approaches:

\begin{itemize}

\item[i)] {\it Spectral approach.} If $H$ has pure point (PP) spectrum then
the system is in the localized regime, if $H$ has absolutely continuous (AC)
spectrum then it is in the delocalized regime. 
(The singular continuous spectrum, if appears at all, gives
rise to anomalous diffusions). It is to be noted that
even in the pure point regime the spectrum is dense.

\item[ii)] {\it Dynamical approach.} One considers the mean square
displacement \eqref{ms} and the system is in the localized
regime if 
$$
    \sup_{t\ge 0} \langle x^2\rangle_t <\infty
$$
(other observables can also be considered).

\item[iii)] {\it Conductor or Insulator?} This is the approach
closest to physics and mathematically it has not been sufficiently elaborated
(for a physical review, see \cite{LR}).
In this approach one imposes  an external voltage to the system and
computes the currect and the ohmic resistance.

\end{itemize}

These criteria to distinguish between localized and delocalized
regimes are not fully equivalent
(especially in $d=1$ dimensional systems there are many anomalous
phenomena due to the singular continuous spectrum),
 but we do not go into
more details.

\bigskip

The mathematically most studied approach is the spectral method.
The main results are the following:

\begin{itemize}

\item[i)]  In $d=1$ dimensions all
eigenfunctions are localized for all $\lambda\neq 0$.
(Goldsheid, Molchanov and Pastur \cite{GMP})

\item[ii)] In $d\ge 1$ localization occurs for
large $\lambda$ or for energies near the edges of
the spectrum of $H_0$ (see \eqref{HH}).
This was first proven by the groundbreaking work
of Fr\"ohlich and Spencer \cite{FS} on the exponential
decay of the resolvent via the {\it multiscale method}.
(the absence of AC spectrum
 was then proved by Fr\"ohlich, Martinelli,  Spencer and Scoppola \cite{FMSS}
and the exponential localization by Simon and Wolff \cite{SW}).
Later a different method was found by Aizenman and Molchanov
\cite{AM} ({\it fractional power method}). The spectrum is
not only pure point, but the eigenfunctions decay exponentially
({\it strong localization}).

\item[iii)] Many more localization results were obtained
by variants of these methods for various models,
including magnetic field, acoustic waves, localized
states along sample edges etc. We will not attempt to
give a list of references here.

\end{itemize}

Common in all localization results is that the random potential 
dominates, i.e. in one way or another the free evolution $H_0$
is treated as a perturbation.

It is important to understand that the transport in the
free Schr\"odinger equation is due to the coherence of
the travelling wave. Random potential destroys this coherence;
the stronger the randomness is, the surer is the destruction.
This picture is essential to understand why random potentials
may lead to localization even at energies that belong
to the regime of classical transport. For example in
$d=1$ any small randomness stops transport, although
the classical dynamics is still ballistic at energies that
are higher than the maximal peak of the potential. 
In other words, Anderson localization is a truly quantum
phenomenon, it cannot be explained by a heuristics based
upon classically trapped particles.

\subsection{Major open conjectures about the Anderson model}

All the previous results were for the localization regime, where
one of the two main methods (multiscale analysis or fractional
power method) is applicable.  The complementary regimes
remain unproven. Most notably is the following list of
open questions and conjectures:

\begin{itemize}
\item[i)] [{\bf Extended states conjecture}] 
For small $\lambda\leq \lambda_0(d)$ and in dimension $d\ge 3$ the
spectrum is absolutely continuous away from the edges
of $H_0$. In particular, there exists a threshold
value (called {\it mobility edge})
near the edges of the unperturbed spectrum 
that separates the two spectral types.

This conjecture has been proven only in the following three cases:

\begin{itemize}
\item[a)] Bethe lattice (infinite binary tree)
that roughly corresponds to $d=\infty$ (Klein
\cite{Kl}, recently different proofs were
obtained in \cite{ASW} and \cite{FHS}).

\item[b)] Sufficiently decaying random potential,  $\bE \, V_x^2 = |x|^{-\a}$, for
 $\a > 1$ (Bourgain \cite{B}). The decay is sufficiently strong
such that the typical number of collisions with obstacles is finite.
Note that the randomness is not stationary.

\item[c)] In $d=2$ with a constant magnetic field 
the spectrum cannot be everywhere pure point. (Germinet, Klein and Schenker \cite{GKS})

\end{itemize}

\item[ii)] [{\bf Two dimensional localization}]
 In $d=2$ dimensions
 all eigenfunctions are localized for all $\lambda$.
(i.e. the model in $d=2$ dimensions behaves as in $d=1$).

\item[iii)] [{\bf Quantum Brownian motion conjecture}] 
 For small $\lambda$ and $d \ge 3$,
the location of the electron is governed by a heat equation
in a vague sense:
\be
 \; \partial_t |\psi_t(x)|^2 \sim \Delta_x |\psi_t(x)|^2 \;  
\quad \Longrightarrow \quad
\langle \; x^2  \;\rangle_t
\sim t  \qquad t\gg 1.
\label{vague}
\ee
The precise formulation of the first statement
requires a scaling limit. The second statement about
the diffusive mean square displacement is 
mathematically precise, but what really stands behind
it is a diffusive equation that on large scales 
mimics the Schr\"odinger evolution. 
Moreover, the dynamics of the quantum particle
converges to the Brownian motion as a process as well;
this means that the joint distribution of the
quantum densities $|\psi_t(x)|^2$ at different times
$t_1<t_2<\ldots < t_n$
converges to the corresponding finite dimensional
marginals of the Wiener process.

\end{itemize}

Note that the ``Quantum Brownian motion conjecture''
is much stronger than the ``Extended states conjecture'',
since the former more precisely describes how the states
extend. All  these three open conjectures have been
outstanding for many years and we seem to be far
from their complete solutions.

 Fig.~\ref{fig:phasediag}
depicts the expected phase diagram of the  Anderson model
in dimensions $d\ge 3$.
 The picture shows the different spectral types
(on the horizontal energy axis) at various disorder.
The grey regimes indicate what has actually been proven.

\bef\bec
\epsfig{file=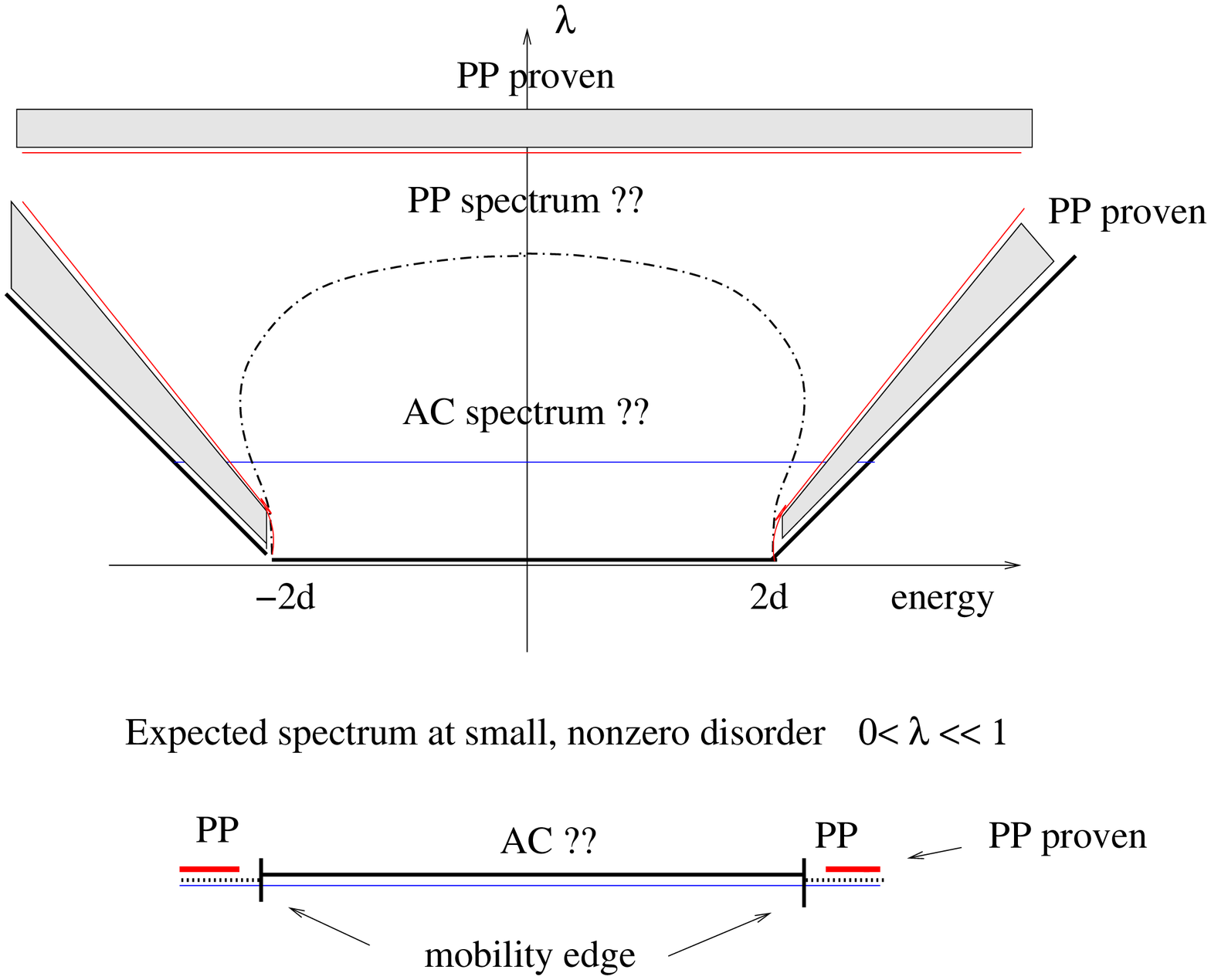, scale=0.55}
\eec
\caption{Phase diagram of the Anderson model in $d=3$}
\label{fig:phasediag}
\eef

Our main result, explained in the next sections, is that
the ``Quantum Brownian motion conjecture'' in the
sense of \eqref{vague} holds {\it in the scaling
limit} up to times $t\sim \lambda^{-2-\kappa}$.
More precisely, we will fix a positive number $\kappa$
and we will consider the family of random Hamiltonians
$$
   H=H_\lambda=-\frac{1}{2} \Delta + \lambda V
$$
parametrized by $\lambda$. We consider their long time evolution
up to times $t\sim \lambda^{-2-\kappa}$ and we take $\lambda\to0$ limit.
We show that, after appropriately rescling the space and time,
the heat equation emerges.
Note that the time scale depends on the coupling parameter.
This result is of course far from either of the conjectures i) and
iii) above, since those conjectures require fixing $\lambda$
(maybe small but fixed) and letting $t\to \infty$. 

We emphasize that the typical number of collisions
is $\lambda^2t$. The reason is that  the quantum rate of collision with
a potential that scales as $O(\lambda)$ is of order $\lambda^2$.
This follows from simple scattering theory:
if 
$$\wt H = -\Delta + \lambda V_0$$
is a Hamiltonian with a single bump potential
(i.e. $V_0$ is  smooth, compactly supported) and $\psi_{in}$
denotes the incoming wave, then after scattering the wave
 splits into two parts;
\be
  e^{-it\wt H}\psi_{in} = \beta e^{it\Delta}\psi_{in} + \psi_{sc}(t) ,
\qquad t\gg 1.
\label{2scat}
\ee
Here $\psi_{sc}(t)$ is the scattered wave (typically spherical
if $V_0$ is spherical)
while the first term describes the transmitted wave that
did not notice the obstacle (apart from its amplitude
is scaled down by a factor $\beta$). Elementary calculation
then shows that the scattered and transmitted waves are (almost)
orthogonal and their amplitudes satisfy 
\be
    \| \psi_{sc}(t)\|^2 = O(\lambda^2), \qquad  \beta^2 = 1- O(\lambda^2).
\label{elem}
\ee
Therefore the incoming wave scatters
with a probability $O(\lambda^2)$. 

 Thus up to time $t$,
the particle encounters $\lambda^2t$ collisions. In our scaling
$$
n:=\mbox{Number of collisions} 
\sim \lambda^2t \sim\lambda^{-\kappa} \to\infty .
$$
The asymptotically infinite number of collisions 
is necessary to detect the diffusion (Brownian motion, heat equation),
similarly as the Brownian motion arises from an increasing
number of independent ``kicks'' (steps in random walk, see
Theorem \ref{thm:w}).

\section{Main result}\label{sec:main}
\setcounter{equation}{0}

In this section we formulate  our main result
precisely, i.e. what we mean by that ``Quantum Brownian motion conjecture''
holds in the scaling limit up to times $t\sim\lambda^{-2-\kappa}$.
Before going into the precise formulation or
the sketch of any proof, we will explain
why it is a difficult problem.

\subsection{Why is this problem difficult?}

The dynamics of a single
quantum particle among random scatterers (obstacles)
is a multiple scattering process. The quantum wave function
bumps into an obstacle along its evolution, and according
to standard scattering theory, it decomposes into
two parts: a wave that goes through the obstacle unnoticing it
and a scattering wave \eqref{2scat}. 
The scattered wave later bumps into another obstacle etc,
giving rise to a complicated multiple scattering picture,
similar to the one on Fig.~\ref{fig:multscat}. Obviously,
the picture becomes more and more complicated as 
the number of collisions, $n$, increases.
Finally, when we perform a measurement, we select a
domain in space and we compute the
wave function at the point. By the superposition principle
of quantum mechanics, at that point {\it all} 
elementary scattering
waves have to be added up; and eventually there exponentially many of them (in
the parameter $\lambda^{-1}$).
They are quantum waves, i.e. complex
valued functions, so they must be added together with their
phases. The cancellations in this sum due to the phases 
are essential even to get the right order of magnitude.
Thus one way or another one must trace all elementary scattering
waves with an enormous precision (the wavelength is order one
in microscopic units but the waves travel at distance $\lambda^{-2}$
between two collisions!)

\bef\bec
\epsfig{file=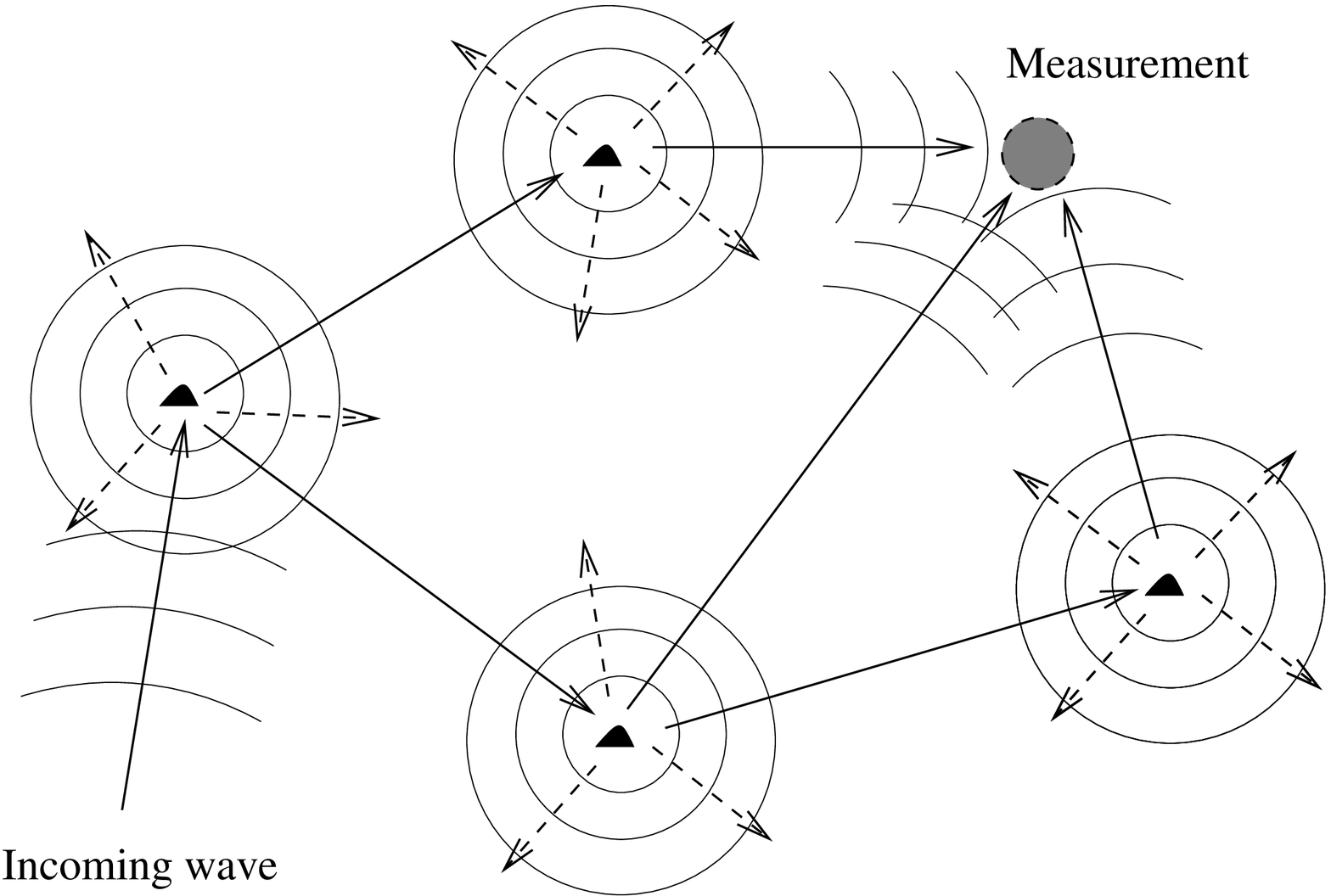, scale=0.65}
\eec
\caption{Schematic picture of multiple scattering}
\label{fig:multscat}
\eef

It is important to emphasize that this system is {\bf not semiclassical}.
Despite the appearance of a small parameter, what matters is
that the typical wavelength remains comparable with
the spatial variation lengthscale of the potential. Thus the
process remains quantum mechanical and cannot be described
by any semiclassical method.

\subsection{Scales}

Since the collision rate is $O(\lambda^2)$,
and the typical velocity of the particle is
unscaled (order 1), the
typical distance between two collisions (the
so-called {\it mean free path})
is $L=O(\lambda^{-2})$ and the time elapsed
 between consecutive collisions is also $O(\lambda^{-2})$.

\subsubsection{Kinetic scale}

In order to see some nontrivial effects of the collisions, we have to run the
dynamics at least up to times of order $\lambda^{-2}$.
Within that time, the typical distance covered is
also of order $\lambda^{-2}$ (always understood in microscopic coordinates).
Thus the simplest scale that yields a nontrivial
result is the so-called {\bf kinetic scaling},
$$
t=\lambda^{-2}T,\qquad x= \lambda^{-2}X, 
\qquad n=\lambda^2t =O(1),
$$
i.e. the space and time are scaled by $\lambda^2$ and
the typical number of collisions $\lambda^2t$ is of order 1.

One convenient way to quantify Fig.~\ref{fig:multscat}
is that one works in a large quadratic box 
whose size is at least comparable with the mean free path
(otherwise the boundary would have a too big effect).
Thus one can consider the box
$$
  \Lambda = [0,L]^d
$$
and put $|\Lambda|=L^d$ obstacles in it, e.g. one at
every lattice site. For definiteness, we work with
the Anderson model, i.e.
$$
   H= -\Delta_x + \lambda\sum_{\a\in \bZ^d}  v_\a\delta(x-\a).
$$
 Each elementary scattering wave
corresponds to a particular (ordered) sequence of random
obstacles.  Since the total number of
obstacles is $L^d$,
after $n$ collisions we have $\sim |\Lambda|^n$
elementary waves to sum up in  Fig.~\ref{fig:multscat}, thus
\be
  \psi_t = \sum_{A} \psi_A,
\label{sum}
\ee
where $A=(\a_1, \a_2, \ldots , \a_n)$ is a sequence of obstacle
locations ($\a_i\in \bZ^d$) and $\psi_A$ describes
the elementary scattering wave that has undergone the collisions
with obstacles at $\a_1, \a_2, \ldots \a_n$ in this order.
The precise definition of $\psi_A$ will come later.

These waves must be summed up together with their phase.
It is easy to see from standard scattering theory that
a spherical wave decays as $\lambda|\mbox{distance}|^{-(d-1)/2}$
(the prefactor $\lambda$ comes from the fact that
the total amplitude of the scattering is scaled down by
a factor of $\lambda$ due to the coupling constant in
front of the potential, see \eqref{elem}).
Since the amplitudes multiply, and since the typical
mean free path is $\lambda^{-2}$, after $n$ collisions
the typical amplitude of $\psi_A$ is
$$
  |\psi_A| \sim \Big[\lambda\Big(\lambda^2\Big)^{\frac{d-1}{2}}\Big]^n
  = \lambda^{dn}.
$$
Thus if we try to sum up these waves in \eqref{sum}
with neglecting their phases, then 
$$
   \sum_A |\psi_A| \sim |\Lambda|^n \lambda^{dn} = \Big(\lambda^{-2d}\Big)^n
\lambda^{dn} = \lambda^{-dn}\to \infty.
$$
However, if we can sum up these waves {\it assuming} that 
their phases are independent, i.e. they could be summed
up as independent random variables, then it is the variance
that is additive (exactly as in the central limit theorem 
in \eqref{eq:quadvar}):
\be
   |\psi_t|^2 = \Big| \sum_A \psi_A\Big|^2 \approx
  \sum_A |\psi_A|^2 \sim  |\Lambda|^n \lambda^{2dn} = O(1).
\label{var}
\ee
Thus it is essential to extract a strong independence
property from the phases. Since phases are determined
by the random obstacles, there is some hope that
at least most of the elementary waves are roughly independent.
This will be our main technical achievement, although it will
be formulated in a different language. 

We remark that in the physics literature, the independence
of phases is often postulated as an Ansatz under the name of
``random phase approximation''. Our result will mathematically
justify this procedure.

\subsubsection{Diffusive scale}

Now we wish to go beyond the kinetic scale and describe a system
with potentially infinitely many collisions. Thus we choose
a longer time scale and we rescale the space appropriately. We obtain
the following {\bf diffusive scaling}:
$$
t = \lambda^{-\kappa}\lambda^{-2}T,
\qquad x=\lambda^{-\kappa/2}\lambda^{-2}T,
\qquad n=\lambda^2t= \lambda^{-\kappa}.
$$
Notice that the time is rescaled by an additional factor
$\lambda^{-\kappa}$ compared with the kinetic scaling,
and space is rescaled by the square root of this
additional factor. This represents the idea that
the model scales diffusively with respect to the units of the kinetic scale.
The total number of collisions is still $n=\lambda^2t$, and now
it tends to infinity.

If we try to translate this scaling into the elementary wave
picture, then first we have to choose a box that is
larger than the largest space scale, i.e. $L\ge \lambda^{-2-\kappa/2}$.
The total number of elementary waves to sum up is
$$
   |\Lambda|^n=
 (\lambda^{-2-\frac{\kappa}{2}})^{dn} \sim \lambda^{-\lambda^{-\kappa}}
$$
i.e. superexponentially large. Even with the assumption that
only the variances have to be summable (see \eqref{var}),
we still get a superexponentially divergent sum:
$$
   |\psi_t|^2  \approx
  \sum_A |\psi_A|^2 \sim  |\Lambda|^n \lambda^{2nd}=
 \Big[\lambda^{-2-\kappa/2}\Big]^{nd} \lambda^{2nd}= 
\lambda^{-nd\kappa/2}= \lambda^{-(const)\lambda^{-\kappa}}
$$
We will have to perform a {\bf renormalization} to prevent this
blow-up.

\subsection{Kinetic scale: (linear) Boltzmann equation}

The main result  on the kinetic scale is the following
theorem:
\begin{theorem} \label{thm:bol} [{\bf 
Boltzmann equation in the kinetic scaling limit}] Let the dimension
be at least $d\ge 2$.
Consider the random Schr\"odinger evolution on $\bR^d$ or $\bZ^d$
$$
  i\partial_t\psi_t = H\psi_t, \qquad H=H_\lambda = -\frac{1}{2}\Delta
+\lambda V(x),
$$
where the potential is spatially uncorrelated
(see more precise conditions below).
Consider the kinetic rescaling
$$
  t=\lambda^{-2}\cT, \qquad x=\lambda^{-2}\cX,
$$
with setting $\e:=\lambda^2$ to be the scale separation
parameter  in space.

Then the weak limit of the expectation of the rescaled Wigner transform of
the solution $\psi_t$ exists,
$$
\bE\, W^\e_{\psi_{\cT/\e}} (\cX, \cV) \rightharpoonup F_\cT(\cX, \cV)
$$
and $F_T$ satisfies the linear Boltzmann  equation,
$$
        \Big(\partial_\cT  + \nabla e(V)\cdot \nabla_\cX\Big) F_\cT( \cX,V)
        =\int dU \sigma (U, V)  \Big[ F_\cT(\cX, U)
         -  F_\cT( \cX,V) \Big]\; .
$$
Here $e(V)$ is the dispersion relation of the free kinetic
energy operator given by $e(V)=\frac{1}{2}V^2$ for the continuous model
and by \eqref{disp} for the discrete model. The collision kernel $\sigma(U,V)$
is explicitly computable and is model dependent (see below).
\end{theorem}

The velocities $U,V$ are interpreted as incoming and outgoing
velocities, respectively, see Fig.~\ref{fig:inout}.
The collision kernel always contains
an {\it onshell condition}, i.e. a
term $\delta(e(U)-e(V))$ guaranteeing energy conservation
(see the examples below).
Thus the right hand side of the linear Boltzmann equation
is exactly the generator of the jump process
on a fixed energy shell (in case of $e(V)=\frac{1}{2}V^2$ it
is on the sphere) as defined in Section \ref{sec:jump}.

\bef\bec
\epsfig{file=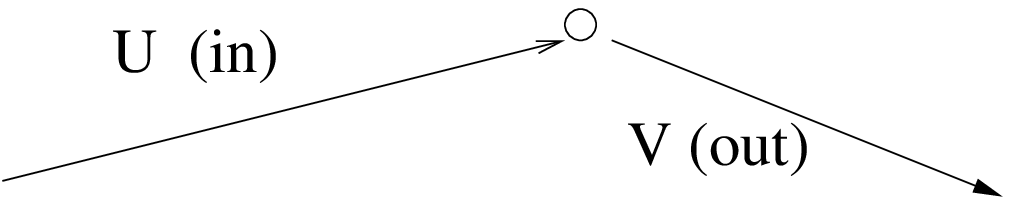, scale=0.85}
\eec
\caption{Incoming and outgoing momentum in a collision}
\label{fig:inout}
\eef

\bigskip

This theorem has been proven in various related models.

\begin{itemize}

\item{}  [{\bf Continuous model on $\bR^d$}] The random potential
is chosen to be a homogeneous Gaussian field, this means that
the random variable $V(x)$ is Gaussian and it is stationary with respect
to the space translations. We assume that
  $\bE \; V(x) =0$, then the distribution of $V(x)$ is
uniquely determined by the two-point correlation function
$$
  R(x-y):= \bE  V(x)V(y),
$$
and we assume that $R$ is a Schwartz function.
The dispersion relation of the free Laplacian on $\bR^d$ is
$$
   e(p):=\frac{1}{2} p^2,
$$
and the Boltzmann collision kernel is computed from
the microscopic data as
\be
\sigma (U, V)= \delta(e(U)-e(V)) |\wh R(U-V)|.
\label{sigm}
\ee
This result was first obtained by Spohn \cite{Sp1}
for short macroscopic time, $T\leq T_0$, and 
later extended to arbitrary times in  a joint work
with Yau \cite{EY} with a different method.
We assume that the dimension $d\ge 2$, with a special
non-degeneracy condition on the initial data in case of $d=2$
forbidding concentration of mass at zero momentum. 
The $d=1$ case is special; due to strong interference
effects no Markovian limiting dynamics is expected.

\item{}  [{\bf Discrete model + Non Gaussian}]
The method of \cite{EY} was extended by Chen in
two different directions \cite{Ch}. He considered the
discrete model, i.e. the original Anderson model and
he allowed non-Gaussian randomness as well. In the case of
Laplacian on $\bZ^d$
the dispersion relation is more complicated \eqref{disp},
but the collision kernel is simpler;
\be
 \sigma (U, V)= \delta(e(U)-e(V)),
\label{unif}
\ee
i.e. it is simply the uniform  measure on the energy surface
given by the energy of the incoming velocity.

\item{} 
[{\bf Phonon model}] In this model the random potential
is replaced by  a heat bath of non-interacting
bosonic particles that interact only with
the single electron. {\it Formally}, this
model leads to an electron in
a time dependent random potential, but the system is
still Hamiltonian. This model is the quantum analogue
of Einstein's picture. The precise formulation
of the result is a bit more complicated,
since the phonon bath has to be mathematically introduced,
so we will not do it here. The interested reader
can consult with \cite{E}. More recently, De Roeck and Fr\"ohlich
\cite{DF} have proved  diffusion
for any times  in $d\ge 4$ dimensions
with an additional strong spin coupling that enhances
the loss of memory.

\item{}[{\bf Cubic Schr\"odinger equation with random initial data}]
A nonlinear version of the Boltzmann equation was proven 
in \cite{LS2} where the Schr\"odinger equation contained a weak cubic
nonlinearity and the initial data was drawn from near thermal
equilibrium.

\item{}
[{\bf Wave propagation in random medium}]
It is well known that in a system of harmonically coupled
oscillators the wave propagate ballistically. If
the masses of the oscillators deviate randomly 
from a constant value, the wave propagation can change.
The effect is similar to the disturbances in
the free Schr\"odinger propagation;
the ballistic transport is a coherence effect that can
be destroyed by small perturbations. This model,
following the technique of \cite{EY}, has been
worked out in \cite{LS}.

\item{} [{\bf Low density models}] There is another way
to reduce the effects of possible recollisions apart
from introducing a small coupling constant: one can
consider a low density of obstacles. 
Let $V_0$ be a radially symmetric Schwartz function  
with a sufficiently small weighted Sobolev norm:
$$
   \| \langle x \rangle^{N(d)}\langle \nabla \rangle^{N(d)}V_0\|_\infty
  \qquad \mbox{is sufficiently small for some $N(d)$ sufficiently large.}
$$
The Hamiltonian of the model is
given by
$$
 H=   -\frac{1}{2} \Delta_x + \sum_{\a=1}^M V_0(x-x_\a)
       \qquad \mbox{in a box}\; [-L, L]^d, \quad L\gg \e^{-1}
$$
acting on $\bR^d$, $d\ge 3$, where
$\{ x_\a\}_{\a=1, \ldots M}$ denotes a collection of random i.i.d. points.
 with density 
$$
 \varrho : = {M\over L^d}\to 0.
$$
The kinetic scaling corresponds to choosing 
$$
x\sim \e^{-1}, \;\;  t\sim \e^{-1}, \qquad \e=\varrho,
$$
and then letting $\e\to 0$.  In this case 
convergence to the linear Boltzmann equation 
in the spirit of Theorem \ref{thm:bol} was
proven in \cite{EY1}, \cite{EE} (although these papers
consider the $d=3$ case, the extension to higher dimensions
is straightforward). The dispersion
relation is $e(p)=\sfrac{1}{2}p^2$ and the collision kernel is
$$
\sigma(U, V)= |T(U, V)|^2 \delta(U^2-V^2)
$$
where
 $T(U, V)$
is the quantum scattering cross section for $-\sfrac{1}{2}\Delta + V_0$.

\end{itemize}

\bef\bec
\epsfig{file=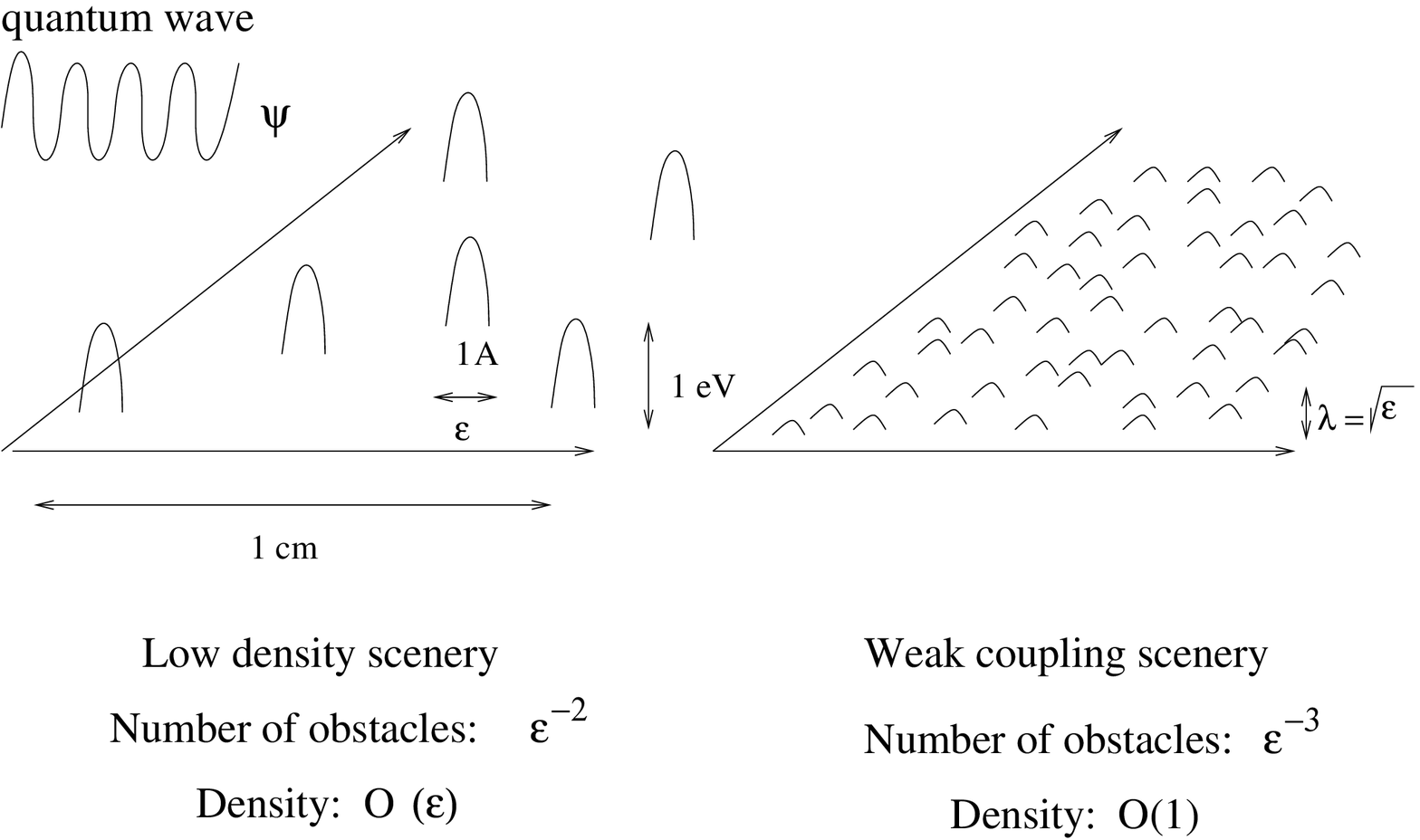, scale=0.65}
\eec
\caption{Low density and weak coupling sceneries}
\label{fig:scenery}
\eef

It is amusing the compare the weak-coupling and the low-density sceneries;
they indeed describe different physical models
(Fig.~\ref{fig:scenery}). Although in both cases
the result is the linear Boltzmann equation, the
microscopic properties of the model influence the collision kernel.
In particular, in the low density model the full single-bump
scattering information is needed (while the function $\wh R$ 
\eqref{sigm} in
the weak coupling model can be viewed as the Born approximation
of the full scattering cross section).
The scaling is chosen such that the typical number of collisions
remains finite in the scaling limit.
Note that in both models the wavelength is comparable with
the spatial variation of the potential; both live on the microscopic
scale. Thus neither model is semiclassical.

We make one more interesting remark. When compared with
the corresponding classical Hamiltonian,
the low density model yields the (linear) Boltzmann equation
both in classical and quantum mechanics (although
the collision kernels are different). 
The weak coupling model yields Boltzmann equation
when starting from quantum mechanics, however it
yields a random walk on the energy shell (for the limiting velocity
process $v_t$) when
starting from classical mechanics \cite{KP} and \cite{KR}.
 For further 
references, see \cite{EY}.

\bigskip

Fig.~\ref{fig:twoscale1} depicts schematically the
microscopic and macroscopic scales in Theorem \ref{thm:bol}.
The right side of the picture is under a microscope which
is ``sees'' scales comparable with the wavelength (Angstrom scale).
Only one obstacle  and only one outgoing wave
are pictured. In reality the outgoing wave goes in
all directions. On this scale the Schr\"odinger equation holds:
$$
 i\partial_t\psi_t(x)  = \Big[
        -\Delta_x + \lambda V(x)\Big]\psi_t(x).
$$
On the left side we zoomed our ``microscope'' out. Only a few
obstacles are pictured that are touched by a
single elementary wave function. All other trajectories
are also possible. The final claim is that
on this scale the dynamics can be described by the Boltzmann equation
$$
        \Big(\partial_\cT  + \nabla e(V) \cdot\nabla_\cX\Big) F_\cT(\cX, V)
        = \int \rd U \sigma(U, V)\Big[ F_\cT(\cX, U)-F_\cT(\cX, V)\Big].
$$
Actually the  obstacles  (black dots) on the left picture are
only fictitious: recall that there are no physical obstacles behind
the Boltzmann equation. It represents
a process, where physical obstacles are replaced by
a stochastic dynamics: the  particle
has an exponential clock and it randomly decides 
to change its trajectory i.e. to behave as if there
were an obstacle (recall Section \ref{sec:jump}).
It is crucial to make this small distinction
because it is exactly the fictitious obstacles
are the signatures that all recollisions have been
eliminated  hence this guarantees the Markov property
for the limit dynamics.

\bef\bec
\epsfig{file=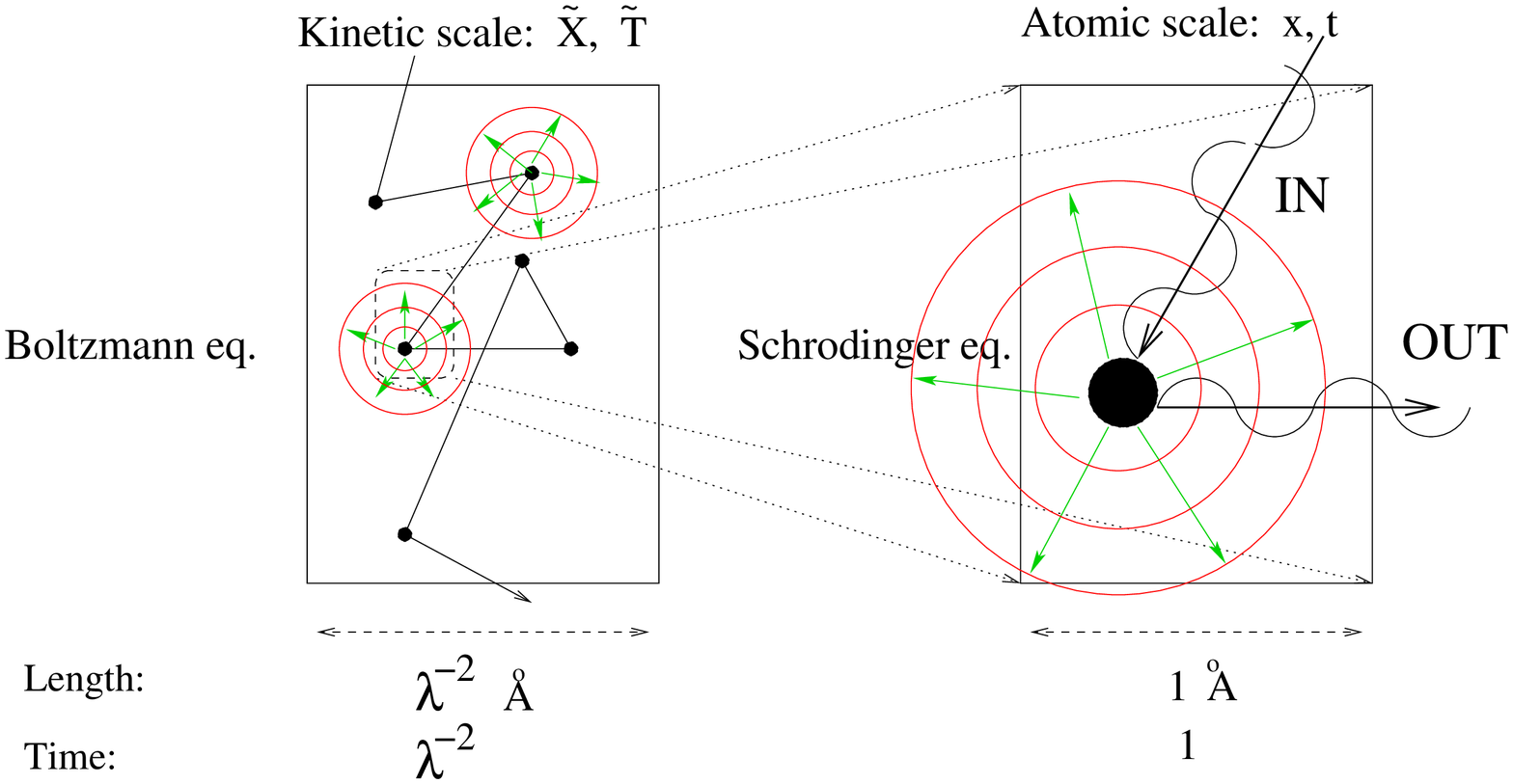, scale=0.70}
\eec
\caption{Macroscopic and microscopic scales}
\label{fig:twoscale1}
\eef

\bef\bec
\epsfig{file=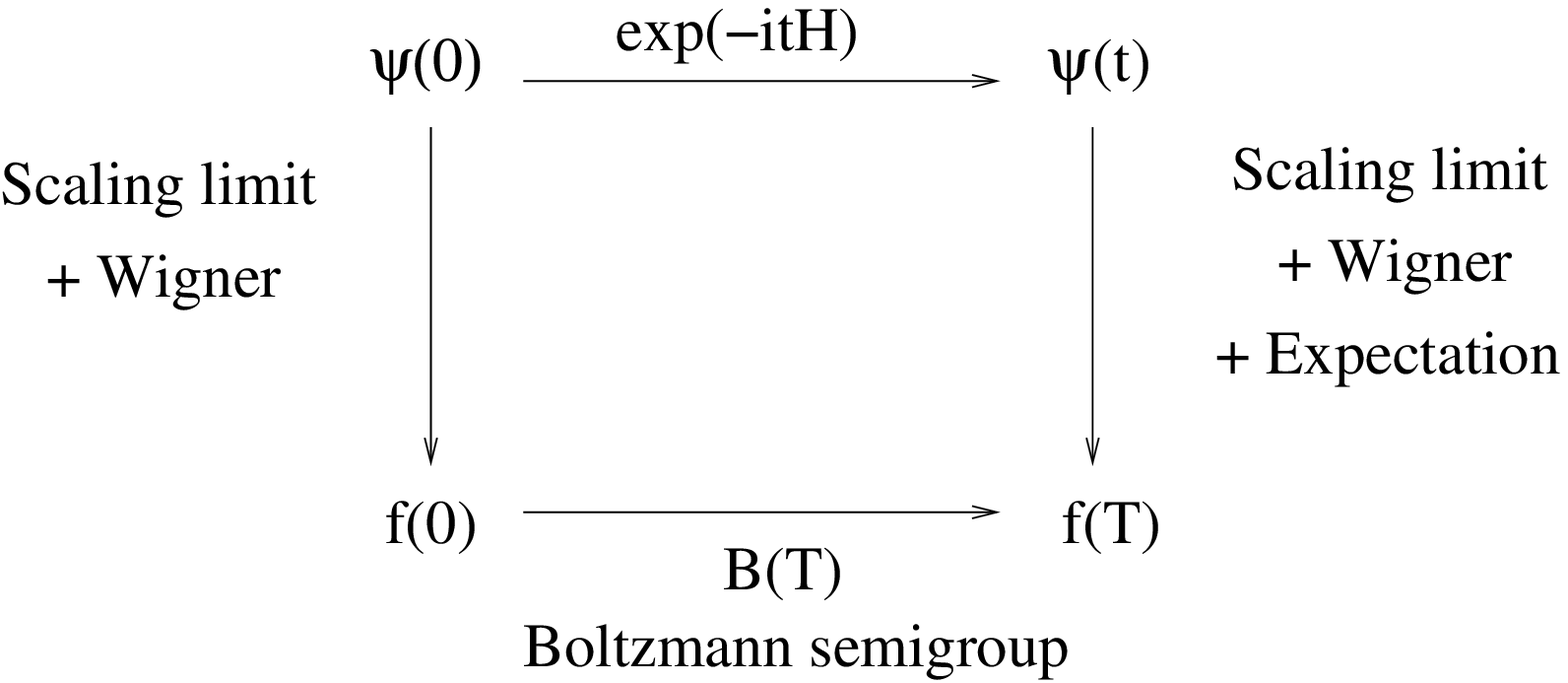, scale=0.70}
\eec
\caption{Boltzmann semigroup models Schr\"odinger evolution}
\label{fig:comm}
\eef
\bigskip

The main message of the result is that
the long time $(t= T\e^{-1})$ Schr\"odinger evolution can
modelled by a finite time ($T$) Boltzmann evolution
on the macroscopic scale.
Of course the detailed short scale information is lost,
which explains  irreversibility.
The effective limiting equation is classical, but quantum features are retained
in  the collision kernel.

Note that this approximation is not just conceptually interesting
but it also gives rise to an enormous computational simplification.
Solving the Schr\"odinger equation directly for a long time in an
complicated environment is a computationally impossible task.
In contrast, the Boltzmann equation needs to be solved
on $O(1)$ times and there is no complicated environment.
 This is expressed schematically
in the ``commutative diagram'' Fig.~\ref{fig:comm}.
The top horizontal line is the true (Schr\"odinger) evolution. 
The theorem says that one can ``translate'' the quantum initial
data by Wigner transform and scaling limit into a phase
space density $f(0)= f_0(X,V)$ and one can use the
much simpler (and shorter time) Boltzmann evolution
to $f_0$ to obtain the Wigner transform (after expectation
and scaling limit) of the true evolution $\psi_t$.

\subsection{Diffusive scale: Heat equation}

We consider the same model with a weakly coupled random
potential as before. For definiteness, we work on the lattice $\bZ^d$,
$d\ge 3$,
the Hamiltonian is given by
$$
    H = -\Delta + \lambda V,
$$
where the discrete Laplacian is given by \eqref{discrlapl} 
with dispersion relation \eqref{disp} and 
where the random potential is
$$
  V(x) = \sum_{\a\in \bZ^d}
    v_\a \delta (x-\a).
$$
The random variables $v_\a$ are i.i.d. with moments $m_k = \bE \, v_\a^k$
satisfying
$$
  m_1=m_3=m_5=0,\quad m_{2d}<\infty.
$$
Let
$$
  \psi_t : = e^{-itH}\psi_0,
$$
and we rescale the Wigner transform as before:
$$
   W_{\psi_t}\to W_{\psi_t}^\e =  \e^{-d} W_{\psi_t}(X/\e, v).
$$
We remark that in case the discrete
lattice  $\bZ^d$ the definition of the Wigner function $W_\psi(x,v)$
is given by \eqref{Wlattice}.
Our main result is the following
\begin{theorem} \label{thm:diff} {\bf [Quantum diffusion in the discrete
 case]} 
For any  dimension $d\ge 3$  
there exists  $\kappa_0(d)>0$
[in $d=3$ one can choose $\kappa_0(3) =\frac{1}{10000}$],
such that for any $\kappa\le \kappa_0(d)$ 
and any $\psi_0\in L^2(\bZ^d)$ the following holds.
In the diffusive scaling
$$
t  = \lambda^{-\kappa} \, \lambda^{-2} T, \qquad
x=  \lambda^{-\kappa/2} \, \lambda^{-2} X, \qquad
 \e = \lambda^{-\kappa/2-2},
$$
 we have  that
\be
 \int_{\{e(v)=e\}} \bE  \; 
 W^\e_{\psi_t} (  X, v) \;  \rd v
\rightharpoonup f_T(X, e)  \qquad \mbox{weakly as $\lambda\to0$},
\label{shell}
\ee
and for almost all $e>0$  the limiting function satisfies the heat equation 
\be
\partial_T f_T( X, e) = \nabla_X\cdot D(e)
\nabla_X f_T( X, e)
\label{heate}
\ee
with a diffusion matrix given by
\be
D_{ij}(e)= \Big\langle \; \nabla e(v) \otimes \nabla e(v) 
\Big\rangle_e , \qquad 
 \langle f(v)\rangle_e = \mbox{Average of $f$ on} \; \{ v\; : \; e(v)=e \}
\label{diffm}
\ee
and with initial state 
$$
    f_0(X, e):=\delta(X)\int \delta(e(v)-e) |\wh\psi_0(v)|^2 \rd v.
$$
The weak convergence in \eqref{weaklim} means
that for any Schwartz function $J(x,v)$ on $\bR^d\times \bR^d$ we have
$$
   \lim_{\e\to0} \int_{\big( \frac{\e}{2}\bZ\big)^d} \rd X \int \rd v 
   J(X, v)  \bE  \;  W^\e_{\psi(\lambda^{-\kappa-2}T)} (  X, v) 
  = \int_{\bR^d}\rd X\int \rd v J(X, v) f_T(X, e(v))
$$
uniformly in $T$ on $[0,T_0]$, where $T_0$ is an
arbitrary fixed number.
\end{theorem}
This result is proven in \cite{ESY4}. The related continuous
model is treated in \cite{ESY2, ESY3}. A concise summary of the
ideas can
be found in the expository paper \cite{ESY1}.
Note again that weak limit means that 
only macroscopic 
observables can be controlled.
Moreover, the theorem does not keep track of the velocity any more,
since on the largest diffusive scale the  velocity of the particle
is not well defined (similarly as the Brownian motion has
no derivative). Thus we have to integrate out the velocity
variable on a  fixed energy shell (the energy remains 
macroscopic observable). This justifies the additional
velocity integration 
in \eqref{shell}.

The diffusion matrix is the velocity autocorrelation matrix
(computed in equilibrium)
obtained from the Boltzmann equation
\be
   D_{Boltz}(e):= \int_0^\infty \cE_e \;\big[ \nabla e(v(0))\otimes \nabla e(v(t))\big],
 \rd t
\label{dedef}
\ee
similarly to Theorem \ref{thm:vv}. Here $\cE_e$ denotes the
expectation value of the Markov process $v(t)$ described
by the linear Boltzmann equation as its generator if the initial
velocity $v(0)$ is distributed according to the equilibrium measure
of the jump process with generator \eqref{gen} on the 
energy surface $\{ e(v) = e\}$ with the Boltzmann collision kernel
\eqref{unif}. 
Since this kernel is the uniform measure on the energy surface,
the Boltzmann  velocity process has no memory and thus
it is possible to compute the velocity autocorrelation 
just by averaging with respect to the uniform  measure
on  $\{ e(v) = e\}$:
\be
   D_{Boltz}(e)= \int_0^\infty \cE_e \, \big[ \nabla e(v(0))\otimes \nabla e(v(t))
\big] \rd t
   = \Big\langle \; \nabla e(v) \otimes \nabla e(v) 
\Big\rangle_e .
\label{gkdirect}
\ee
In particular, Theorem \ref{thm:diff} states that
the diffusion matrix $D(e)$ obtained from the quantum evolution
coincides with the velocity autocorrelation matrix
of the linear Boltzmann equation.

For completeness, we state the result also for the continuum model.

\begin{theorem} \label{thm:diffcont} {\bf [Quantum diffusion in the
continuum
 case]} 
Let the dimension be at least $d\ge 3$ and $\psi_0\in L^2(\bR^d)$.
Consider the diffusive scaling
$$
t  = \lambda^{-\kappa} \, \lambda^{-2} T, \qquad
x=  \lambda^{-\kappa/2} \, \lambda^{-2} X, \qquad
 \e = \lambda^{-\kappa/2-2}.
$$
Let
$$
   \psi_t = e^{itH}\psi_0, \qquad H= -\frac{1}{2}\Delta+\lambda V,
$$
where the random potential is given by \eqref{VB} with a
single site profile $B$ that is a spherically symmetric Schwartz function
with 0 in the support of its Fourier transform, $0\in \mbox{supp}\; \wh B$.
For $d\ge3$ there exists $\kappa_0(d)>0$ (in $d=3$ one can choose
$\kappa_0(3) = 1/370$) such that for any 
$\kappa < \kappa_0(d)$  we have 
\be\label{weaklim}
 \int_{\{e(v)=e\}} \bE  \; 
 W^\e_{\psi_t} (  X, v) \;  \rd v
\rightharpoonup f_T(X, e)  \qquad \mbox{(weakly as $\lambda\to0$)},
\ee
and for almost all energies $e>0$
the limiting function satisfies the heat equation
$$
\partial_T f_T( X, e) = D_e\Delta_X f_T( X, e)
$$
 with initial state 
$$
    f_0(X, e):=\delta(X)\int \delta(e(v)-e) |\wh\psi_0(v)|^2 \rd v.
$$
The diffusion coefficient is given by
\be
   D_e:= \frac{1}{d}\int_0^\infty \cE_e \;\big[  v(0)\cdot v(t)\big] \;\rd t,
\label{gk}
\ee
where $\cE_e$ is the expectation value for the random
jump process on the  energy surface $\{ e(v)=e\}$ with generator
$$
   \sigma(u,v) = |\wh B(u-v)|^2\delta(e(u)-e(v)).
$$
\end{theorem}

The condition $0\in \mbox{supp} \; \wh B$ is not essential
for the proof, but the statement of the theorem needs to
be modified if the support is separated away from zero.
In this case, the low momentum component of the initial
wave function moves ballistically since the diameter
of the energy surface is smaller than the minimal range of $\wh B$.
The rest of the wave function still moves diffusively.
\bigskip

\bef\bec
\epsfig{file=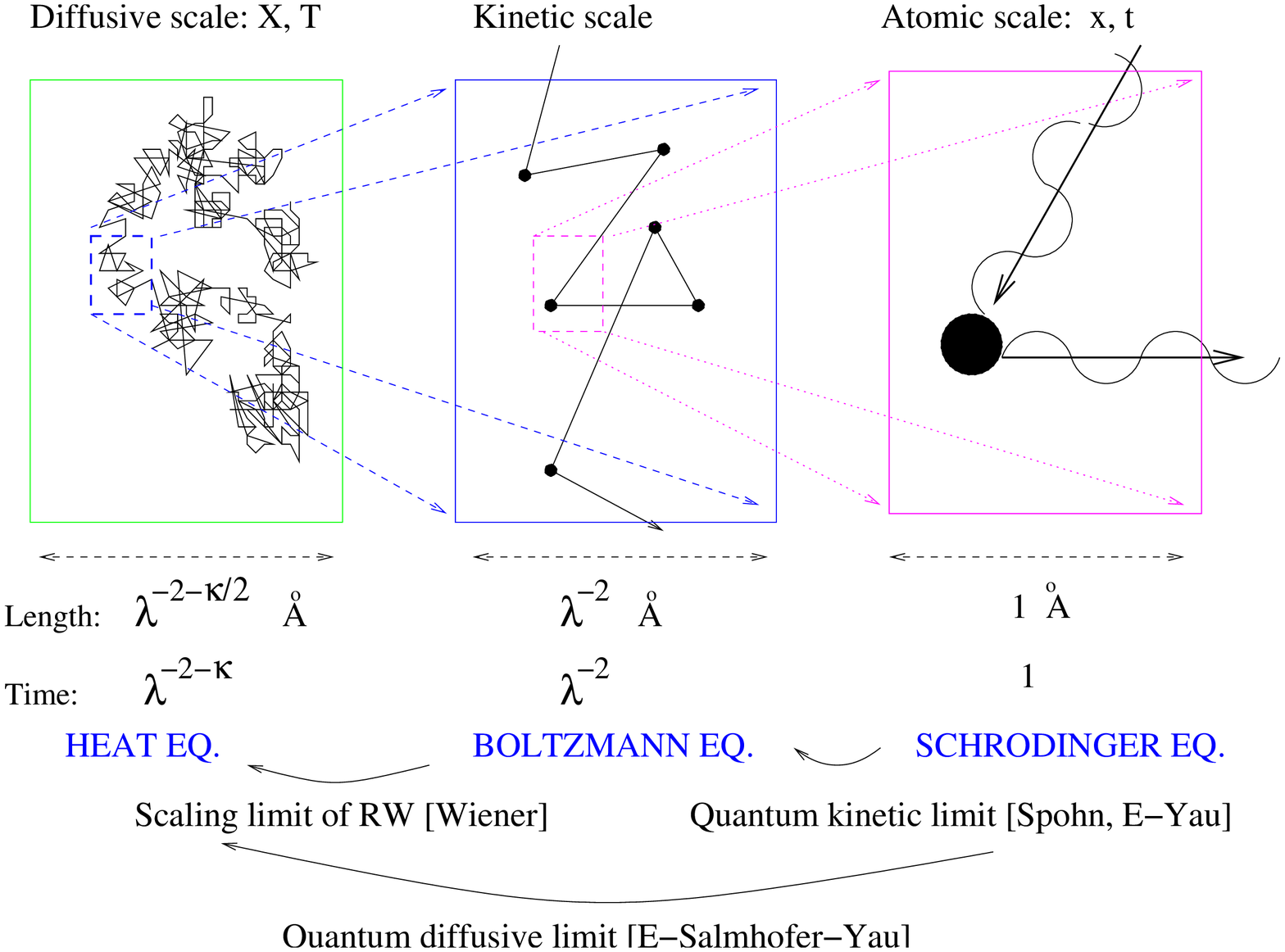, scale=0.70}
\eec
\caption{Three scales: diffusive, kinetic and atomic}
\label{fig:threescale1}
\eef

Fig.~\ref{fig:threescale1} shows schematically the three scales
we discussed. Notice that going from the Boltzmann scale 
to the diffusive scale is essentially the same as
going from random walk to Brownian motion in Wiener's construction
(Theorems \ref{thm:vv} and \ref{thm:w}).
However it is misleading to try to prove Theorem
\ref{thm:diff} by a two-step limiting argument, first
using the kinetic limit (Theorem \ref{thm:bol}) then combining
it with Wiener's argument. There is only one limiting parameter
in the problem, one cannot take first the kinetic limit,
then the additional $\lambda^{-\kappa}\to \infty$ limit
modelling the long kinetic time scale. The correct
procedure is to consider the Schr\"odinger evolution
and run it up to the diffusive time scale. In this way,
the intermediate Boltzmann scale can give only a hint
what to expect but it cannot be used for proof.

In fact, this is not only a mathematical subtlety. Correlations
that are small on the kinetic scale, and were neglected
in the first limit, can become relevant on longer time scales.
Theorem \ref{thm:diff} shows that it is not the case;
at least up to times $t\sim \lambda^{-2-\kappa}$ the
correlations are not strong enough to destroy
the diffusive picture coming from the long
time limit of the Boltzmann equation (in particular
the diffusion coefficient can be correctly computed
from the Boltzmann equation). However, this
should not be taken for granted, and in fact
it is not expected to hold in $d=2$ dimensions
for exponentially long times, $t\sim \exp(\lambda^{-1})$.
Theorem \ref{thm:bol} on the kinetic limit is valid
in $d=2$ as well and Wiener's argument is dimension
independent. On the other hand if the diffusive
evolution were true up to any time scales, then
in particular the state would be delocalized,
in contrast to the conjecture that 
the random Schr\"odinger operator in $d=2$ dimensions
is always localized.

\section{Feynman graphs (without repetition)}\label{sec:fey}
\setcounter{equation}{0}

\subsection{Derivation of Feynman graphs}

Feynman graphs are extremely powerful graphical representations
of perturbation expansions. They have been primarily used
in many-body theories (they were invented for QED), but
they are very convenient for our problem as well to organize 
perturbation expansion.
They have been used by physicist to study random Schr\"odinger
operators, a good overview from the physics point of view is
\cite{VW}.

We will derive the Feynman graphs we need for this presentation.
Recall that
\be
     H=-\Delta +\lambda V,  \qquad
V=  \sum_{\alpha \in \bZ^d} V_\alpha,  \quad \bE V_\alpha = 0,
\label{model}
\ee
where $V_\a$ is a single-bump potential around $\a\in\bZ^d$, for
example
$$
   V_\a(x) =v_\a\delta(x-\a).
$$
We consider the potential as a small perturbation of $-\Delta$
and we expand the unitary evolution by the identity (Duhamel formula)
\be
\psi_t  = e^{-it H }\psi_0 = e^{it \Delta }\psi_0 -i\lambda\int_0^t
e^{-i(t-s) H  } V  e^{is \Delta }\psi_0 \;  \rd s .
\label{duhh}
\ee
This identity can be seen by differentiating $U(t):= e^{-itH}e^{itH_0}$
where $H= H_0+\lambda V$, $H_0=-\Delta$:
$$
  \frac{\rd U}{\rd t} = e^{-itH}(-iH + iH_0) e^{itH_0} = 
  e^{-itH}(-i\lambda V) e^{itH_0},
$$
thus we can integrate back from $0$ to $t$
$$
 U(t) = I + \int_0^t \rd s \; e^{-i(t-s)H}(-i\lambda V) e^{i(t-s)H_0}
$$
and multiply by $e^{-itH_0}$ from the right.

We can iterate the expansion \eqref{duhh} by repeating the same identity
for $e^{-i(t-s)H}$. For example, the next step is
$$
e^{-it H }\psi_0 = e^{it \Delta }\psi_0 -i \lambda\int_0^t
e^{i(t-s) \Delta  } V  e^{is \Delta }\psi_0 \;  \rd s 
+\lambda^2
\int_0^t\rd s_1\int_0^{s_1} \rd s_2 \; e^{-i(t-s_1-s_2)H}
Ve^{is_2\Delta}Ve^{is_1\Delta}\psi_0
$$
etc. We see that at each step we obtain a new {\it fully expanded}
term (second term), characterized by the absence of the unitary evolution
of $H$; only free evolutions $e^{-is\Delta}$ appear in these terms. There is always
 one last term that contains the full evolution and
that will be estimated trivially after sufficient number of 
expansions.
More precisely, we can expand
\be
   \psi_t = e^{-itH}\psi_0 = \sum_{n=0}^{N-1} \psi^{(n)}(t) + \Psi_N(t),
\label{finiteduh}
\ee
where
$$
  \psi^{(n)}(t): = (-i\lambda)^n \int_{\bR_+^{n+1}} \rd s_0 \rd s_1 \ldots \rd s_n 
  \; \delta\Big( t- \sum_{j=0}^n s_j\Big)
  e^{is_0\Delta} V e^{is_1\Delta} V\ldots V e^{is_n\Delta}\psi_0
$$
and
\be
 \Psi_N(t): = (-i\lambda) \int_0^t \rd s \; e^{-i(t-s)H} V \psi^{(N-1)} (s).
\label{def:PSI}
\ee
Recalling that each $V$ is a  big summation \eqref{model}, we arrive at
the following {\it Duhamel formula}
\be
\quad \psi_t =\sum_{n=0}^{N-1}
\sum_{A}  \psi_{A} + \mbox{full evolution term}, 
\label{fullduh}
\ee
where the summation is over collision histories:
$$
A:= (\a_1, \a_2, \cdots, \a_n),  \qquad \a_j\in \bZ^d,
$$
and each elementary wave function is given by 
\be
\psi_{A}: = \psi_A(t)= (-i\lambda)^n \int \rd\mu_{n,t} (\us)\;
e^{is_0  \Delta } 
\; V_{\alpha_1}   \cdots V_{\alpha_n} e^{is_{n} \Delta}  \;
\psi_0 .
\label{psii}
\ee
Here, for simplicity, we introduced the notation
\be
   \int \rd\mu_{n,t} (\us) := \int_{\bR^{n+1}_+}
\delta\Big(t- \sum_{j=0}^n s_j\Big) \rd s_0 \rd s_1 \ldots \rd s_n
\label{timeorder}
\ee
for the integration over the simplex $s_0+s_1+ \ldots + s_n=t$, $s_j \ge 0$.

If the cardinality of $A$ is $n=|A|$, then we say that
the elementary wave function is of order $n$ and their sum 
for a fixed $n$ is
denoted by
$$
   \psi^{(n)}(t) : =\sum_{A\;: \; |A|=n} \psi_A(t).
$$
Note that the full evolution term $\Psi_N$ has a similar structure, but
the leftmost unitary evolution $e^{is_0\Delta}$ is replaced
with $e^{-is_0H}$. 

We remark that not every elementary wave function 
has to be expanded up to the same order.   The expansion can be adjusted at every step,
i.e. depending on the past collision history, encoded in the sequence
$A$, we can decice if an elementary wave function $\psi_A$ should be
expanded further or we stop the expansion. This will be essential
to organize our expansion without overexpanding: once the collision 
history $A$ indicates that $\psi_A$ is already small (e.g. it contains
recollisions), we will not expand it further.

Each $\psi_A$ can be symbolically represented as shown
on Fig.~\ref{fig:psi}, where lines represent the free
propagators and bullets represent the potential terms (collisions).
Note the times between collisions is not recorded in the graphical
picture since they are  integrated out.

\bef\bec
\epsfig{file=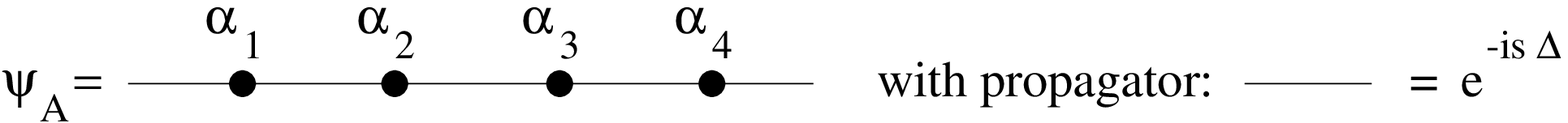, scale=0.80}
\eec
\caption{Representation of a wave function with collision history $A$}
\label{fig:psi}
\eef

We need to compute expectation values of quadratic functionals
of $\psi$ like the Wigner transform. For simplicity we will
show how the $L^2$-norm can be treated, the Wigner transform
is very similar (see next Section).

Suppose for the moment that there is no repetition in $A$,
i.e. all $\a_j$'s are distinct.
To compute $\bE \|\psi\|^2 = \bE \; \ov{\psi}\, \psi$, we need 
to compute
$$
\sum_{n,n'}\sum_{A\; : |A|=n} \sum_{B: |B|=n'}
\bE  \; \langle \psi_B \; ,  { \psi}_A  \rangle.
$$
 This is zero unless 
$B$ is a permutation of $A$, because individual potentials
have zero expectation. Higher order moments of the same potential $V_\a$ are excluded 
since we assumed that $A$ and $B$ have no repetition.
Thus the only nonzero contributions come from second moments, i.e. from
pairings of each element of $A$ with an element of $B$,
in particular, $|A|=|B|$ is forced, i.e. $n=n'$.

In conclusion, with the tacit assumption that
there are no repetitions, we can express $\bE\|\psi\|^2$
as a summation over the order $n$ and
 over permutations between the sets $A$ and $B$
with $|A|=|B|=n$:
\be
\sum_{n,n'}\sum_{A\; : |A|=n}^* \sum_{B: |B|=n'}^*
\bE  \; \langle \psi_B \; ,  { \psi}_A  \rangle 
=\sum_{n}\sum_{\pi\in S_n}\sum_{A\; : |A|=n}^* 
\bE  \; \langle \psi_{\pi(A)} \; ,  { \psi}_A  \rangle
 =: \sum_{n}\sum_{\pi\in S_n} \mbox{Val}(\pi),
\label{nn'}
\ee
where $S_n$ is set of permutations on $n$ elements
and upper star denotes summation over non-repetitive sequences.
$\mbox{Val}(\pi)$ is defined by the last formula
by summing up all sequences $A$; this summation 
 will be computed explicitly
and it will give rise to delta functions among momenta.

Drawing the two horizontal lines that represent $\psi_A$ and $\psi_B$
parallel and connecting the paired bullets, we obtain 
Fig.~\ref{fig:feyn1}, where the first summation is over the
order $n$, the second summation is over all permutations
$\pi\in S_n$. 
a graph, called the {\it Feynman graph of $\pi$}.
We will give an explicit formula
for the value of each Feynman graph, here denoted by $\mbox{Val}(\pi)$.

\bef\bec
\epsfig{file=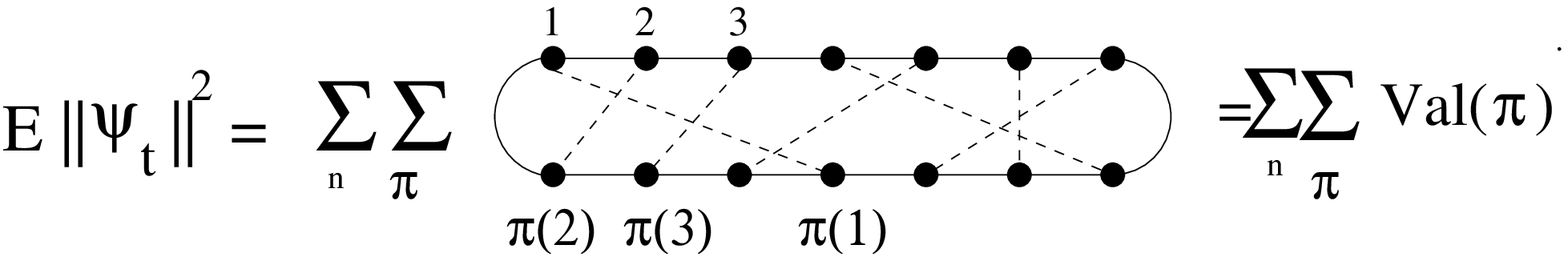, scale=0.65}
\eec
\caption{Representation of the $L^2$-norm as  sum of Feynman graphs}
\label{fig:feyn1}
\eef

\subsection{$L^2$-norm vs. Wigner transform}\label{sec:l2}

Eventually we need to express 
$$
   \bE \langle J, W_\psi\rangle : = \int J(x,v) \bE\, W_\psi(x,v) \rd x \rd v
   = \int \wh J(\xi,v) \bE\, \wh W_\psi(\xi,v) \rd \xi \rd v
 = \bE \langle \wh J, \wh W_\psi\rangle
$$
(recall that hat means Fourier transform only in the $x$ variable).
However, we have 
\begin{lemma}\label{lm:cont} For any Schwarz function $J$,
the quadratic functional $\psi \mapsto \bE \langle \wh J, \wh W_\psi\rangle$ is
continuous in the $L^2$-norm.
\end{lemma}
{\it Proof.} Let $\psi, \phi\in L^2$ and set
$$
   \Omega: = \bE \Big(  \langle \wh J, \wh W_\psi\rangle- \langle \wh J, \wh W_\phi\rangle \Big)
  = \bE \int \rd\xi \rd v \wh J(\xi, v) \Big[ \wh W_\psi (\xi, v) -
\wh W_\phi (\xi, v)\Big].
$$
Let $v_\pm = v\pm \frac{\xi}{2}$ and write
$$
   \Big[ \cdot \Big] = \ov {\wh \psi (v_-)}\big( \wh\psi (v_+) - \wh\phi(v_+)\big)
   + \big( \ov{ \wh \psi(v_-)} -  \ov{ \wh \phi(v_-)}\big)\wh \phi(v_+).
$$
Thus
\begin{align}
   |\Omega| & \le \int \rd \xi \rd v |\wh J(\xi, v)| \; \bE \big| \big[\cdot \big]\big| 
   \le \int \rd\xi \sup_v|\wh J(\xi, v)| \int \rd v \; \bE \big| \big[\cdot \big]\big| \no \\
   &\le  \int \rd\xi \sup_v|\wh J(\xi, v)| \int \rd v \; \bE 
  \Big[  |\wh \psi (v_-)| \cdot | \wh\psi (v_+) - \wh\phi(v_+)| +
   | \wh \psi(v_-) -  \ov{ \wh \phi(v_-)}| \cdot |\wh \phi(v_+)|\Big]. \no
\end{align}
By Schwarz inequality,
\begin{align}
   \int \rd v \bE \Big(  |\wh \psi (v_-)| \cdot | \wh\psi (v_+) - \wh\phi(v_+)|  \Big) 
 & \le \Big( \int \rd v \bE |\wh\psi(v_-)|^2\Big)^{1/2} \Big( \int \rd v \bE
   |\wh\psi(v_+)-\wh\phi(v_+)|^2\Big)^{1/2}\no \\
  & =  \Big( \int \rd v \bE |\wh\psi(v)|^2\Big)^{1/2} \Big( \int \rd v \bE
   |\wh\psi(v)-\wh\phi(v)|^2\Big)^{1/2}\no\\
 & = \Big( \bE \|\wh\psi\|^2\Big)^{1/2} \Big( \bE \|\wh\psi -\wh\phi\|^2\Big)^{1/2}.
\end{align}
Thus
$$
  |\Omega|\le \Big( \int \rd\xi \sup_v |\wh J(\xi, v)|\Big) \Big( \sqrt{\bE \|\psi\|^2}
   +  \sqrt{\bE \|\phi\|^2}\Big)\Big(  \bE \|\wh\psi -\wh\phi\|^2\Big)^{1/2},
$$
which completes the proof of the Lemma. \qed

\medskip

This lemma, in particular, guarantees that it is sufficient to consider nice
initial data, i.e. we can assume that $\psi_0$ is a Schwarz function.
More importantly, this lemma guarantees that it is
sufficient to control the Feynman diagrammatic expansion of
the $L^2$-norm of $\psi$; the same expansion for the Wigner transform
will be automatically controlled. Thus all estimates of the error terms can be done
on the level of the $L^2$-norm; the more complicated arguments 
appearing in the Wigner transform are necessary only for the
explicit computation  of the main term.

\subsection{Stopping the expansion}

We will never work with infinite expansions, see \eqref{finiteduh},
but then we will have to control the last, fully expanded term. 
This will be done by using the unitarity of the full evolution $e^{-i(t-s)H}$
in the definition of $\Psi_N$, see \eqref{def:PSI}.
More precisely, we have:
\begin{lemma}[Unitarity bound]\label{lm:unit} Assume that $\|\psi_0\|=1$.
According to \eqref{finiteduh}, we write $\psi(t) = \Phi_N(t) + \Psi_N(t)$ 
with $\Phi_N  =\sum_{n=0}^{N-1} \psi^{(n)}$ containing the fully expanded
terms. Then
$$
  \Big| \bE \Big[ \langle J, W_{\psi(t)}\rangle - \langle J, W_{\Phi_N(t)}\rangle \Big]
 \Big| \le \Big( 1 + \sqrt{\bE \| \Phi_N\|^2}\Big) t 
   \Big[ \sup_{0\le s\le t} \bE \| \lambda V\psi^{(N-1)}(s)\|^2\Big]^{1/2}.
$$ 
\end{lemma}
In other words, the difference between the true wave function $\psi(t)$
and its $N$-th order approximation $\Phi_N$ {\it can be expressed in terms
of fully expanded quantities}, but we have to pay  an additional price $t$.
This additional $t$ factor will be compensated by a sufficiently long
expansion, i.e. by choosing $N$ large enough so that $\psi^{(N-1)}(s)$ is small.
In practice we will combine this procedure by 
stopping the expansion for each elementary wave function $\psi_A$
separately,  depending on the collision history. Once $\psi_A$ is
sufficiently small to compensate for the $t$ factor lost in 
the unitarity estimate, we can stop its expansion.

\medskip

{\it Proof.} We apply Lemma \ref{lm:cont} with $\psi = \psi(t)$ and $\phi= \Phi_N$,
so that $\psi-\phi = \Psi_N$. Then we have, for any $N$, that
$$
    \Big| \bE \Big[ \langle J, W_{\psi(t)}\rangle - \langle J, W_{\Phi_N(t)}\rangle \Big]
  \le \Big( 1 + \sqrt{\bE \| \Phi_N\|^2}\Big) \Big( \bE \|\Psi_N\|^2\Big)^{1/2},
$$
since $\|\psi_t \|= \|\psi_0\|=1$. Then  using the unitarity, i.e. that
$$
  \Big\| e^{-i(t-s)H} \lambda V \psi^{(N-1)}(s)\Big\| = \Big\| \lambda V \psi^{(N-1)}(s)\Big\|,
$$
we can estimate
\begin{align}
\bE \|\Psi_N\|^2 & = \bE \Big\| \int_0^t \rd s e^{-i(t-s)H} \lambda V \psi^{(N-1)}(s)\Big\|^2 \no\\ 
& \le \bE  \Big(\int_0^t \rd s \big\|\lambda V \psi^{(N-1)}(s)\big\| \Big)^2\no\\
&\le t \int_0^t \rd s \; \bE \big\|\lambda V \psi^{(N-1)}(s)\big\|^2,
\end{align}
which proves Lemma \ref{lm:unit}. \qed

\subsection{Outline of the proof of the Boltzmann equation with Feynman graphs}

The main observation is that among the $n!$ possible permutations
of order $n$,
only one, the identity permutation contributes to the limiting
Boltzmann equation. Actually it has to be like that,
since this is the only graph whose collision history
can be interpreted classically, since here $A=B$ as sets,
so the two wavefunctions $\psi_A$ and $\psi_B$ visit
the same obstacles in the same order. If there were a
discrepancy between the order in which the obstacles in
$A$ and $B$ are visited, then no classical collision history could be
assigned to that pairing.

The Feynman graph of the identity permutation is called 
{\it the ladder graph}, see Fig.~\ref{fig:feyn2}.
Its value can be computed explicitly
and it resembles to the $n$-th term in the Taylor
series of the exponential function.

\bef\bec
\epsfig{file=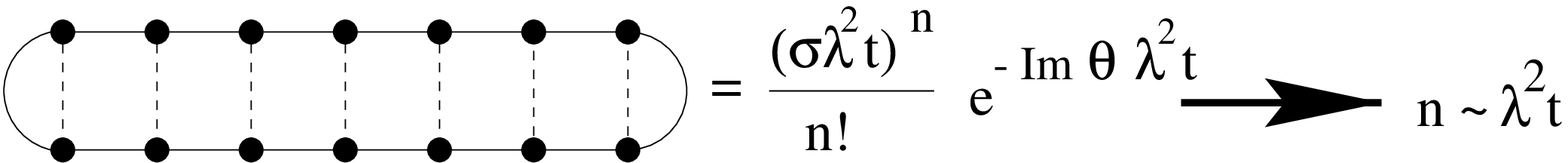, scale=0.70}
\eec
\caption{Contribution of the ladder graph}
\label{fig:feyn2}
\eef

The ladder with  $n$ pairings corresponds 
to a classical process with $n$ collisions. If the collision
rate is $\sigma\lambda^2$, then the classical probability of $n$
collisions is given by the (unnormalized)
Poisson distribution formula
$$
   \mbox{Prob ($n$ collisions)} = \frac{ (\sigma \lambda^2 t)^n}{n!}.
$$
This is clearly summable
for the kinetic scale $t\sim \lambda^{-2}$, and the typical number
of collisions is the value $n$ giving the largest term, i.e.
$n\sim \lambda^2 t$.
The exponential damping factor
 $e^{-\mbox{Im}(\theta) \lambda^2 t}$ in  Fig.~\ref{fig:feyn2}
comes from renormalization (see later in Section \ref{sec:renorm}
where the constants $\sigma$ and $\theta$ will also be defined).

We note that from the Duhamel formula leading to the Feynman graphs,
it is not at all clear that the value of the
 ladder graphs is of order one (in the 
limiting parameter $\lambda$). According to the Duhamel representation, they 
are highly oscillatory integrals of the form (written in momentum
space)
\be
    \mbox{Val($id_n$)} \sim \lambda^{2n}\Big( \int_0^t e^{is_1p_1^2 } \int_0^{s_1}
    e^{is_2p_2^2 }  \ldots \int_0^{s_n} \Big) \Big( 
\int_0^t e^{-is_1'p_1^2 } \int_0^{s_1'}
    e^{-is_2'p_2^2 }  \ldots \int_0^{s_n'}
 \Big) .
\label{timetime}
\ee
 Here
$id_n$ is the identity permutation in $S_n$ which generates the ladder.
If one estimates these integrals naively, neglecting all oscillations, one obtains
$$
 \mbox{Val($id_n$)} \le 
    \frac{\lambda^{2n}t^{2n}}{(n!)^2},
$$
which is completely wrong, since it does not respect the power counting that
$\lambda^2t$ should be the relevant parameter.

\bigskip

Assuming that the ladder can be computed precisely, with all oscillations
taken into account, one still has
to deal with the remaining $n!-1$ non-ladder graphs. 
Their explicit formulas will reveal that their values are
small, but there are many of them, so it is not
clear at all if they are still summable to something negligible.

It is instructive to compare the value of the non-ladder graphs
with the ladder graph because that is of order one
(see Section \ref{sec:ex} for a detailed computation).
 Fig.~\ref{fig:feyn3}
contains schematic estimates of simple crossing graphs that
can be checked by direct computation (once the explicit
formula for $\mbox{Val}(\pi)$ is established in the next section).
It indicates that the value of a non-ladder graph is
a positive power of $\lambda$ that is related with
the combinatorial complexity of the permutation (e.g. 
the number of ``crosses'').

\bef\bec
\epsfig{file=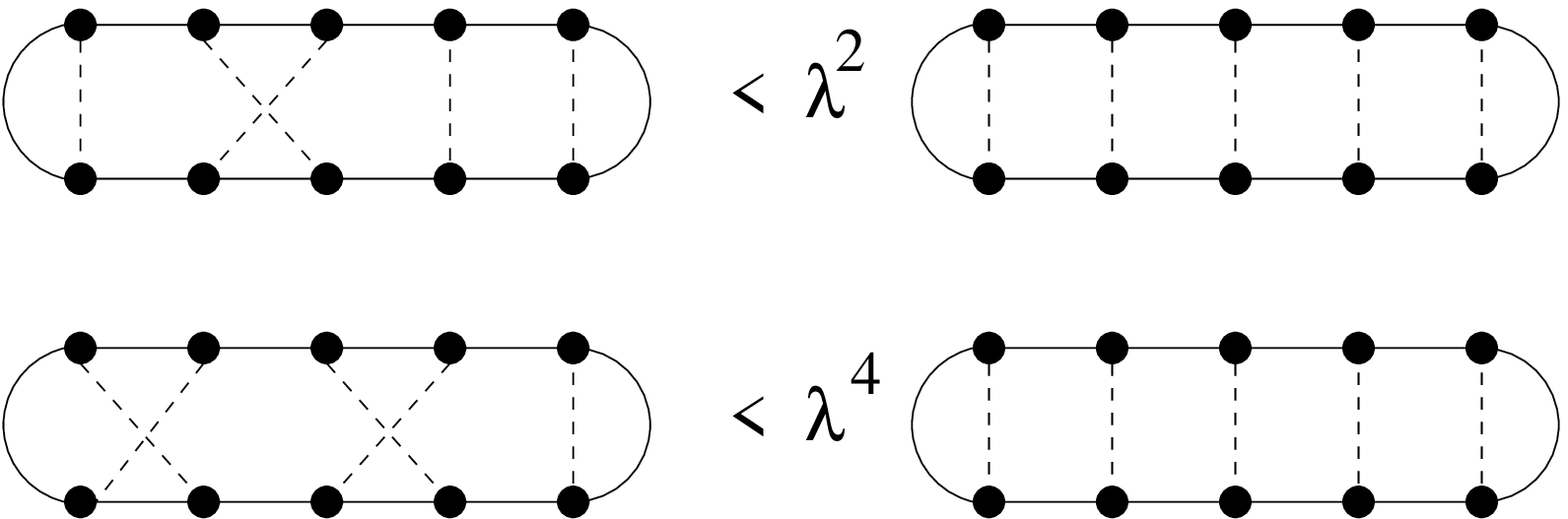, scale=0.70}
\eec
\caption{Comparing the value of various crossing graphs}
\label{fig:feyn3}
\eef

Accepting for the moment that all non-ladder graphs are at least
by an order $\lambda^2$ smaller than the ladder graphs, we obtain the following 
majorizing estimate for the summation of the terms 
in the Duhamel expansion:
$$
 \sum_n \sum_\pi |\mbox{Val}(\pi)|\leq
 \sum_n  \frac{ (\sigma \lambda^2 t)^n}{n!} \Big( \underbrace{1}_{A = B} + 
 \underbrace{ (n!-1) \mbox{(small)} }_{A\neq B}\Big).
$$
This series clearly 
converges for short kinetic time ($\lambda^2 t = T\leq T_0$)
but not for longer times, since the two $n!$'s cancel each other
and the remaining series is geometric. This was the basic
idea of Spohn's proof \cite{Sp1} and it clearly shows the
limitation of the method to short macroscopic times.

The source of the $1/n!$ is actually the time-ordered
multiple time integration present in \eqref{timeorder}; the total volume of the
simplex
\be
   \{ (s_1, s_2, \ldots, s_n)\; : \; 0\le s_1\le\ldots \le s_n\le t\}
\label{timeorder11}
\ee
is $t^n/n!$. However, \eqref{timetime} contains two such multiple time integrations,
but it turns out that only one of them is effective.
The source of the competing $n!$ is the
pairing that comes from taking the expectation, e.g. 
in case of Gaussian randomness this appears explicitly
in the Wick theorem.
The reason for this mechanism is the fact that in
quantum mechanics the physical quantities are always
quadratic in $\psi$, thus we have to pair two collision histories
in all possible ways even if only one of them has a
classical meaning. This additional $n!$ is very typical in all
perturbative quantum mechanical argument.

This was only the treatment of the fully expanded terms,
the  error term in \eqref{fullduh} needs
a separate estimate. As explained in details in Lemma \ref{lm:unit},
here we simply use the unitarity
of the full evolution
\be
    \Big\| \int_0^t e^{i(t-s)H}
\underbrace{\int V e^{is_1\Delta} V e^{is_2\Delta}\ldots
    \rd \mu_{n,s}(\us)   }_{\psi_s^\#}  \Big\| \leq \underbrace{t}_{price}
    \sup_s \| \psi_s^\#\|.
\label{unit}
\ee
In this estimate we lose a factor of $t$, but we end up
with a fully expanded term $\psi_s^\#$ which can also be
expanded into Feynman graphs.

\bigskip

How to go beyond Spohn's argument that is valid only up
to short macroscopic time?
 Notice that ``most'' graphs have many ``crosses'', 
 hence they are expected to be much smaller due to phase decoherence
than the naive ``one-cross'' estimate indicates.
One can thus distinguish graphs with one, two and
more crossings. From the graphs with more crossings,
additional $\lambda^2$-factors can be obtained.
On the other hand,
the combinatorics of the graphs with a few crossings
is much smaller than $n!$.

On kinetic scale the following estimate suffices
(here $N$ denotes the threshold so that the Duhamel
expansion is stopped after $N$ collisions):
\begin{align}
\bE\|\psi_t\|^2 \leq &
  \sum_{n=0}^{N-1}  \frac{ (\sigma \lambda^2 t)^n}{n!}
 \Big( \underbrace{1}_{\mbox{ladder}} + 
   \underbrace{n \lambda^2}_{\mbox{one cross}} + 
   \underbrace{ n! \lambda^4}_{\mbox{rest}}\Big)
\non\\
&   +  \underbrace{t}_{unit. price} 
\frac{ (\sigma \lambda^2 t)^N}{N!} \Big( \underbrace{1}_{\mbox{ladder}} + 
   \underbrace{N \lambda^2}_{\mbox{one cross}} + 
   \underbrace{ N^2 \lambda^4}_{\mbox{two cross}} 
+\underbrace{ N! \lambda^6}_{\mbox{rest}}\Big). \non
\end{align}
Finally, we can optimize 
$$
N=N(\lambda) \sim (\log\lambda) /(\log\log \lambda)
$$
to get the convergence. This estimate
gives the kinetic (Boltzmann) limit for all fixed $T$ 
 ($t= T\lambda^{-2}$). This concludes the outline of 
the proof of Theorem \ref{thm:bol}.

For the proof of Theorem \ref{thm:diff}, i.e. going
to diffusive time scales, one needs 
to classify essentially {\bf all} diagrams; it will not be enough
to separate a few crosses and treat all other graphs
identically.

\section{Key ideas of the proof of the diffusion  (Theorems~\ref{thm:diff}
and \ref{thm:diffcont})}
\setcounter{equation}{0}

We start with an apologetical remark. For pedagogical reasons, we will 
present the proof of a mixed model that actually does not exists.
The reason is that the dispersion relation is simpler
in the continuum case, but the random potential is simpler
in the discrete case. The simplicity of the dispersion
relation is related to the fact that in continuum the level sets 
of the dispersion relation, $\{ v\; : \; e(v)=e\}$,
are convex (spheres), thus several estimates are more effective.
The simplicity of the random potential in the discrete case is obvious: no additional
form-factor $B$ and no additional Poisson point process needs to be 
treated. Moreover, the ultraviolet problem
is automatically absent, while in the continuous case the additional
large momentum decay needs to be gained from the decay of
$\wh B$.  Therefore, we will work in the discrete
model (i.e. no form factors $B$, no Poisson points and no large momentum problem),
and  we will denote the dispersion relation in general by $e(p)$.
All formulas hold for both dispersion relations, except some
estimates  that will be commented on.
The necessary modifications from this ``fake'' model
to the two real models are technically sometimes laborous,
but not fundamental. We will comment on them at the
end of the proof.

\subsection{Stopping rules}\label{sec:stop}

The Duhamel formula has the advantage that it can be stopped
at different number of terms depending on the collision
history of each elementary wave function. Thus the threshold $N$ does not have to be
chosen uniformly for each term in advance; looking
at each term
$$ 
 (-i\lambda)^n \int
e^{-is_0 H } 
\; V_{\alpha_1}   \cdots V_{\alpha_n} e^{is_{n} \Delta}  \;
\psi_0  \; \rd\mu_{n,t}(\us)
$$
separately,
we can decide if we wish to expand $e^{-is_0H}$ further, 
or we rather use the unitary estimate \eqref{unit}.
The decision may depend on the sequence $(\a_1, \ldots, \a_n)$.

Set $K:=(\lambda^2 t)\lambda^{-\delta}$ (with some $\delta>0$) as an absolute
upper threshold for the number of collisions in
an expanded term (notice that
$K  $ is much larger
typical number of collisions). We stop the expansion if
we {\it either} have reached $K$ expanded potential terms
{\it or} we see a repeated $\a$-label. This procedure has the
advantage that repetition terms do not pile up.
This stopping rule leads us to the following version
of the Duhamel formula:
\be
  \psi_t = \sum_{n=0}^{K-1} \psi_{n,t}^{nr} +\int_0^t e^{-i(t-s)H}
\underbrace{\Big( \psi^{nr}_{*K,s} + \sum_{n=0}^{K} 
\psi^{rep}_{*n,s}\Big)}_{=:\psi^{err}_s} \rd s
\label{stoprule}
\ee
\be
  \mbox{with}\qquad
\psi_{*n,t}^{nr} := \sum_{A\; :\; \mbox{\footnotesize nonrep.}}\psi_{*A,t},
\qquad \psi_{*n,s}^{rep} := \sum_{|A|=n\atop \mbox{\footnotesize first rep.
at $\alpha_n$}}
\psi_{*A,s}
\label{sta}
\ee
Here $\psi^{nr}$ contains collision histories with no repetition,
while the repetition terms $\psi^{rep}$ contain exactly
one repetition at the last obstacle, since once
such a repetition is found, the expansion is stopped.
In particular,
 the second sum above runs over $A=(\a_1, \ldots \a_n)$ where
$\a_n=\a_j$ for some $j<n$ and this is the first repetition.
Actually the precise definition of the elementary
wave functions in \eqref{sta} is somewhat different
from \eqref{psii} (this fact is indicated by the stars),
 because the expansion starts with a potential
term while $\psi_A$ in \eqref{psii} starts with a free evolution.

\bigskip

The backbone of our argument is the following three theorems:

\begin{theorem}\label{thm:negl}
{\bf [Error terms are negligible]} We have the following estimate
$$
  \sup_{s\le t} \bE \| \psi_s^{err}\| = o(t^{-2}) .
$$
In particular, by the unitarity estimate this trivially implies
that
\be
\bE \Big\| \int_0^t e^{-i(t-s)H} 
\psi_s^{err} \rd s\Big\|^2=o(1).
\label{errsmall}
\ee
\end{theorem}
\begin{theorem}\label{thm:only} {\bf [Only the ladder contributes]}
For $n\leq K= (\lambda^2t)\lambda^{-\delta} =O( \lambda^{-\kappa-\delta})$
we have
\begin{align}
   \bE \| \psi_{n,t}^{nr} \|^2 & = \mbox{Val(id}_n) + o(1)
\label{onlyladder}
\\
   \bE \, W_{\psi_{n,t}^{nr}} & = \mbox{Val}_{\mbox{\footnotesize Wig}}
(id_n) + o(1).
\label{onlyladderwigner}
\end{align}
\end{theorem}
Here $ \mbox{Val}_{\mbox{\footnotesize Wig}}$ denotes the value
of the Feynman graphs that obtained by expanding the
Wigner transform instead of the $L^2$-norm of $\psi_t$. 
Note that \eqref{onlyladderwigner} follows from \eqref{onlyladder}
by (the proof of) Lemma~\ref{lm:cont}.

\begin{theorem} \label{thm:mainterm}
{\bf[Wigner transform of the main term]}
The total contribution of the ladder terms up to $K$ collisions,
$$
   \sum_{n=0}^K \mbox{Val}_{\mbox{\footnotesize Wig}}(id_n),
$$
satisfies the heat equation.
\end{theorem}

In the following sections we
 will focus on the non-repetition terms with $n\le K$, i.e.
on Theorem \ref{thm:only}. We will only discuss the proof for the $L^2$-norm
 \eqref{onlyladder}, the proof of \eqref{onlyladderwigner}
is a trivial modification.
 Theorem \ref{thm:negl}
is a fairly long case by case study, but it essentially
relies on the same ideas that are behind Theorem \ref{thm:only}.
Some of these cases are sketched in Section \ref{sec:rep}.
Finally, Theorem \ref{thm:mainterm} is a non-trivial
but fairly explicit calculation that we sketch
in Section~\ref{sec:mainterm}.

\subsection{Feynman diagrams in the momentum-space formalism}\label{sec:momspace}

Now we express the value of a Feynman diagram explicitly
in momentum representation.
We recall the formula \eqref{psii} and
Fig.~\ref{fig:psi} for the representation
of the  collision history of an elementary wave function.
We will express it in momentum space.
 This means that we assign a running momentum,
$p_0, p_1, \ldots , p_n$ to each edge of the graph in Fig.~\ref{fig:psi}.
Note that we work on the lattice, so all momenta are on the torus, $p\in {\bf T}^d
=[-\pi, \pi]^d$;
in the continuum model, the momenta would run over all $\bR^d$.
The propagators thus become multiplications with $e^{-is_j e(p_j)}$
and the potentials become convolutions:
\begin{align}
   \wh \psi_{A, t} (p) &  = 
 \int_{({\bf T}^d)^n} \prod_{j=1}^n \rd p_j  \int \rd \mu_{n,t}(\us) \, 
  e^{-is_0e(p)}\wh V_{\a_1}(p-p_1)   e^{-is_1e(p_1)} \wh V_{\a_2}(p_1-p_2) \ldots \wh\psi_0(p_n)
\non \\
   &   = \underbrace{e^{\eta t}}_{\eta:=1/t} \int_{( {\bf T}^d)^n} \prod_{j=1}^n \rd p_j 
    \int_{-\infty}^{\infty}\rd\a\;  e^{-i\a t} \prod_{j=1}^n 
\frac{1}{\a - e(p_j) + i\eta}
    \wh V_{\alpha_j}(p_{j-1}-p_j) \wh\psi_0(p_n). \label{psiform}
\end{align}
In the second step we used the identity
\begin{align}
   \int_{\bR^{n+1}_+} \delta\Big( t-\sum_{j=0}^n s_j\Big) \rd s_0 \rd s_1\ldots \rd s_n\; &
   e^{-is_0e(p_0)} e^{-is_1e(p_1)} \ldots e^{-is_ne(p_n)} \non\\
& =e^{i\eta t}\int_\bR \rd\a \,e^{-i\a t}\prod_{j=1}^n \frac{1}{\a - e(p_j) +i\eta}
\end{align}
that holds for any $\eta>0$ (we will choose $\eta=1/t$ in the
applications to neutralize the exponential prefactor).  
This identity can be obtained from writing
$$
 \delta\Big( t-\sum_{j=0}^n s_j\Big) =\int e^{-i\a (t-\sum_{j=0}^n s_j)}\rd \a
$$
and integrating out the times. [Note that the letter $\a$ 
is used here and in the sequel as the conjugate variable to time, while previously
$\a_j$ denoted obstacles. We hope this does not lead to confusion,
unfortunately this convention was adopted in our papers as well.]

The $L^2$-norm is computed by doubling the formula \eqref{psiform}. We will
use the letters $p_1, \ldots p_n$ for the momenta in $\psi_A$
and primed momenta for $\psi_B$:
$$
  \bE \| \psi_t^{nr} \|^2 = \bE \Big\| \sum_{A\;:\;
  \mbox{\footnotesize nonrep}} \psi_A\Big\|^2 = \sum_{A,B \; \mbox{\footnotesize nonrep} }\bE 
  \langle \psi_A, \psi_B\rangle  =\sum_n \sum_{\pi\in\Pi_n} \mbox{Val}(\pi),
$$
\begin{align}
  \mbox{Val}(\pi): = & \; e^{2\eta t}
  \int_{-\infty}^{\infty}\rd\a \rd\beta\;  e^{i(\a-\beta) t} 
  \prod_{j=0}^n \frac{1}{\a - e(p_j) - i\eta}  \; 
\frac{1} {\beta - e(p_j') + i\eta}  \label{valdef}
 \\ &  \times \lambda^{2n} \prod_{j=1}^n 
\delta\Big( (p_{j-1}-p_j) - (p_{\pi(j)-1}' - p_{\pi(j)}')\Big)
  | \wh\psi_0(p_n)|^2 \delta(p_n-p_n')\prod_{j=0}^n \rd p_j \rd p_j' .
\non
\end{align}
After the pairing, here we computed explicitly the
 pair expectations:
\be
   \bE \sum_{\a\in \bZ^d}\overline{\wh V_\a(p)} \wh V_\a(q)
 = \sum_{\a\in \bZ^d} e^{i\alpha (p-q)}=\delta(p-q),
\label{VVV}
\ee
and we also used that only terms with
 $B=\pi(A)$ are nonzero. This formula holds for the simplest
random potential \eqref{lat} in the discrete model;
for the continuum model \eqref{VB} we have
\be
   \bE \sum_\a\overline{\wh V_\a(p)} \wh V_\a(q)
 = |\wh B(p)|^2\delta(p-q)
\label{VVVcont}
\ee
According to the definition of $V_\om$ in the continuum model \eqref{poidef}, 
$\a$ labels
the realizations of the Poisson point process in the last formula \eqref{VVVcont}.

The presence of the additional delta function, $\delta(p_n-p_n')$,  in \eqref{valdef}  is
due to the fact  that
we compute the $L^2$-norm. We remark that the analogous
formula in the expansion for Wigner transform differs only in this factor;
the arguments of the two wave functions in
the momentum representation of the Wigner transform are
shifted by a fixed value $\xi$ \eqref{wigmom}, thus the corresponding
delta function will be $\delta(p_n-p_n'-\xi)$.

A careful reader may notice that the non-repetition condition
on $A$  imposes a restriction on the summation 
in \eqref{VVV}; in fact the summation for $\a_j$ is not over the whole $\bZ^d$,
but only for those elements
that are distinct from the other $\a_i$'s. These terms we
will add to complete the sum \eqref{VVV} and then we will
estimate their effect separately. They correspond to higher
order cumulants and their contribution is negligible, but
technically they cause serious complications; essentially a
complete cumulant expansion needs to be organized.
We will shortly comment on them 
in Section~\ref{sec:replump}; we will neglect this issue for the moment.

This calculation gives  that
$$
    \bE \| \psi_t^{nr} \|^2 =\sum_n\sum_{\pi\in S_n} \mbox{Val}(\pi)
$$
as it is pictured in Fig.~\ref{fig:feyn1}. More concisely,
\be
 \mbox{Val}(\pi) = \lambda^{2n} e^{2\eta t} \int  \rd\bp \rd \bp' \rd\a \rd\beta
\; e^{it(\alpha-\beta)} \prod_{j=0}^n
 \frac{1}{\a - {e}(p_j) -i\eta}
 \; \frac{1}{\beta - e(p_j') +i\eta} \; \Delta_\pi(\bp, \bp')\; | \wh\psi_0(p_n)|^2  ,
\label{valpi}
\ee
where  $\bp$ and $\bp'$ stand for the collection of integration momenta
and
$$
\Delta_\pi(\bp,\bp'):= \delta(p_n-p_n')\prod_{j=1}^n 
\delta\Big( (p_{j-1}-p_j) - (p_{\pi(j)-1}' - p_{\pi(j)}')\Big)
$$
contains the product of all delta functions.
These delta functions can be obtained from the graph: they
express the Kirchoff law at each pair of vertices,
 i.e. the signed sum of the four momenta attached to any
 paired vertices must be zero. It is one of the main advantages
of the Feynman graph representation that the
complicated structure of momentum delta functions
can be easily read off from the graph.

We remark that in \eqref{valpi} we omitted the integration domains;
the momentum integrations runs through the momentum space, i.e. each $p_j$
and $p_j'$ is integrated over $\bR^d$ or ${\bf T}^d$, depending
whether we consider the continuum or the lattice model. The $\rd\a$ and $\rd\beta$
integrations always run through the reals. We will adopt this short
notation in the future as well:
$$ 
\int  \rd\bp \rd \bp' \rd\a \rd\beta = \int_{({\bf T}^d)^{n+1}}\prod_{j=0}^n \rd p_j
\int_{({\bf T}^d)^{n+1}}\prod_{j=0}^n \rd p_j' \int_\bR \rd \a \int_\bR \rd\beta.
$$
We will also often use momentum integrations without indicating their
domains, which is always $\bR^d$ or ${\bf T}^d$.

\subsection{Lower order examples}\label{sec:lowerorder}

In this section we compute explicitly a few low order diagrams
for the Wigner transform. Recalling that $\Phi_N = \sum_{n=0}^{N-1}\psi^{(n)}$
represents the fully expanded terms and recalling
the Wigner transform in momentum space \eqref{wigmom},
 we can write the
(rescaled) Wigner transform of $\Phi_N$ as follows:
$$
    \wh W_{\Phi_N}^\e(\xi, v) = \sum_{n'=0}^{N-1}  \sum_{ n=0}^{N-1} 
\ov { \wh \psi^{(n')}_t \big( v- \frac{\e}{2}\xi\big)}
 \wh \psi^{( n)}_t \big( v+\frac{\e}{2}\xi\big) =: \sum_{n,n' =0}^{N-1}
  \wh W^\e_{n, n', t}(\xi, v).
$$
Set $k_n = v +\frac{\e}{2}\xi$ and $k'_{n'} :=  v -\frac{\e}{2}\xi$,
then we can write
$$
   \wh \psi_t (k_n) = (-i\lambda)^n \int\rd\mu_{n,t}(\us) e^{-is_n e(k_n)} 
  \int \prod_{j=0}^{n-1} \Big( \rd k_j \; e^{-is_j e(k_j)}\Big) \prod_{j=1}^n 
\wh V(k_j-k_{j-1})\wh \psi_0(k),
$$
which can be doubled to obtain $ \wh W^\e_{n, n', t}$.
Notice that we shifted all integration variables, i.e. we use
$k_j = v_j +\frac{\e}{2}\xi$. 

\medskip

\underline{Case 1}: $n=n'=0$.  In this case there is no need for taking expectation and we have
$$
  \wh W^\e_{n, n', t}(\xi, v) = e^{it( e(k'_0)- e(k_0))} \wh \psi_0(k_0) \ov {\wh \psi_0(k'_0)}
  =  e^{it( e(k'_0)- e(k_0))} \wh W^\e_{0}(\xi,v)
$$
where $ W^\e_{0}$ is the Wigner transform of the initial wave function $\psi_0$.
Suppose, for simplicity, that the initial wave function is unscaled, i.e.
it is supported near the origin in position space, uniformly in $\lambda$.
Then it is easy to show that
its rescaled Wigner transform converges to the Dirac delta measure in position space:
$$
   \lim_{\e\to 0} W^\e_{\psi_0}(X, v)\rd X \rd v = \delta(X)|\wh\psi_0(v)|^2 \rd X \rd v
$$
since for any fixed $\xi$,
\be
   \wh \psi_0(k_0) \ov {\wh \psi_0(k'_0)} \to |\wh\psi_0(v)|^2
\label{pointw}
\ee
as $\e\to0$.  In the sequel we will use  
the continuous dispersion relation $e(k)=  \frac{1}{2}k^2$ for simplicity
because it produces explicit formulas. In the general case, similar
formulas are obtained by Taylor expanding $e(k)$ in $\e$ up
to the first order term and higher order terms are negligible but
need to be estimated.

 Since
$$
   e(k'_0)- e(k_0) =  \frac{1}{2}\Big( v-\frac{\e}{2}\xi\Big)^2 - 
 \frac{1}{2}\Big( v+\frac{\e}{2}\xi\Big)^2 = - \e v\cdot\xi,
$$
we get
that in the kinetic limit, when $\lambda^2 t=T$ is fixed and $\e=\lambda^2$
$$
  \lim_{\e\to 0} \int \wh J(\xi, v)  \wh W^\e_{n, n', t}(\xi, v)\rd \xi \rd v
  =  \int \wh J(\xi, v)  |\wh \psi_0(v)|^2 e^{-iTv\cdot \xi}\rd \xi \rd v 
  = \int J(Tv, v)  |\wh \psi_0(v)|^2  \rd v,
$$
by using dominated convergence and \eqref{pointw}. The last formula is the weak formulation
of the evolution of initial phase space measure $\delta(X)|\wh \psi_0(v)|^2\rd X\rd v$
under the free motion along straight lines.

\medskip
\underline{Case 2}: $n=0$, $n'=1$ or $n=1$, $n'=0$. Since there is one single
potential in the expansion, these terms are zero after taking the expectation:
$$
  \bE \, \wh W^\e_{1, 0, t}(\xi, v) = \bE \,  \wh W^\e_{0,1, t}(\xi, v)=0.
$$

\medskip
\underline{Case 3}: $n=n'=1$. We have
\begin{align}
  \wh W^\e_{1,1, t}(\xi, v) = &(-i\lambda) \int_0^t \rd s \; e^{-i(t-s) e(k_1)}\int \rd k_0
  \wh V(k_1-k_0) e^{-is e(k_0)} \wh \psi_0(k_0)\no\\
  &\times  (i\lambda) \int_0^t \rd s' \; e^{i(t-s') e(k'_1)}\int \rd k'_0
  \ov {\wh V(k'_1-k'_0)} e^{is' e(k'_0)} \ov{\wh \psi_0(k'_0)}\label{timedirect} .
\end{align}
The expectation value acts only on the potentials, and we have
(using now the correct continuum potential)
$$
 \bE \wh V(k_1-k_0) \ov {\wh V(k'_1-k'_0)}
 = |\wh B(k_1-k_0)|^2 \delta\big( k'_1- k'_0 - (k_1-k_0)\big).
$$
After collecting the exponents, we obtain
\begin{align}
  \bE \wh W^\e_{1,1, t}(\xi, v) 
  = \lambda^2\int & \rd k_0 \rd k'_0 |\wh B(k_1-k_0)|^2 \delta\big( k'_1- k'_0 - (k_1-k_0)\big) \no\\
   & \times
 e^{it\big(e(k'_1)- e(k_1)\big)} Q_t\big( e(k_1)- e(k_0)\big) \ov Q_t \big( e(k'_1)- e(k'_0)\big)
  \wh \psi_0(k_0) \ov{\wh \psi_0(k'_0)},\no 
\end{align}
where we introduced the function
$$
  Q_t(a): = \frac{e^{ita}-1}{ia}.
$$
Consider first the case $\xi=0$, i.e. when the Wigner transform
becomes just the momentum space density. Then $v=k_1=k'_1$ and
thus $k_0=k'_0$ by the delta function, so $e(k_1)- e(k_0) =  e(k'_1)- e(k'_0)$ and we get
$$   
\bE \wh W^\e_{1,1, t}(0, v) = \lambda^2\int \rd k_0  |\wh B(k_1-k_0)|^2 | \wh \psi_0(k_0)|^2
  \big|Q_t\big( e(k_1)- e(k_0)\big)\big|^2.
$$
Clearly
$$
 t^{-1} |Q_t(A)|^2 \to 2\pi  \delta(A)
$$
as $t\to \infty$ (since $\int \big(\frac{\sin x}{x}\big)^2\rd x =\pi$), so we obtain
that in the limit $t\to\infty$, $\lambda^2 t=T$ fixed,
\be
\bE \wh W^\e_{1,1, t}(0, v) \to T \int \rd k_0  |\wh B(k_1-k_0)|^2 
 2\pi \delta\big(  e(k_1)- e(k_0)\big) | \wh \psi_0(k_0)|^2
 = T \int \rd k_0 \sigma(k_1, k_0) | \wh \psi_0(k_0)|^2 
\label{w11}
\ee
(recall $k_1=v$ in this case),
where the collision kernel is given by
$$
  \sigma(k_1, k_0): =  |\wh B(k_1-k_0)|^2 
 2\pi \delta\big(  e(k_1)- e(k_0)\big).
$$

A similar but somewhat more involved calculation gives the result for a general $\xi$,
after testing it against a smooth function $J$. Then it is easier to use 
\eqref{timedirect} directly, and we get, after taking expectation and changing variables,
$$
  \bE \wh W^\e_{1,1, t}(\xi, v)  = \lambda^2 \int_0^t\rd s\int_0^t\rd s' \int \rd v_0
 \; e^{i\Phi/2}  |\wh B(v-v_0)|^2
    \wh W^\e_{0}(\xi, v_0)
$$
with a total phase factor 
$$
  \Phi:= -(t-s)\Big( v+ \frac{\e}{2}\xi\Big)^2  - s\Big( v_0+ \frac{\e}{2}\xi\Big)^2
 + (t-s')\Big( v- \frac{\e}{2}\xi\Big)^2+ s' \Big( v_0- \frac{\e}{2}\xi\Big)^2
$$
$$
   = (s-s')(v^2-v_0^2) - 2\e\Big[ \big(t-\frac{s+s'}{2}\big)v -\frac{s+s'}{2}v_0\Big]\cdot \xi .
$$
After changing variables: $b = s-s'$ and $T_0 = \e\,  \frac{s+s'}{2}$, we notice that the first
summand gives
\be
   \int \rd b \; e^{ib(v^2-v_0^2)} =2\pi \delta(v^2-v_0^2),
\label{bs}
\ee
and from the second one we have
$$
  \bE \wh W^\e_{1,1, t}(\xi, v)  =  \int_0^T\rd T_0 \int \rd v_0
 \; e^{-i\big[ (T-T_0) v - T_0 v_0\big]\cdot \xi} \sigma(v, v_0)
    \wh W^\e_{0}(\xi, v_0).
$$
These steps, especially \eqref{bs},
 can be made rigorous only if tested against a smooth function $J$, so we have
$$
  \bE \langle J, W_{1,1,t}\rangle \to \int_0^T\rd T_0 \int \rd v_0 J\Big( (T-T_0) v - T_0 v_0, v\Big)
  \sigma(v, v_0)
    |\wh \psi_0(v_0)|^2.
$$
The dynamics described in the first variable of $J$ is a free motion with
velocity $v_0$ up to time $T_0$, then a fictitious collision happens 
that changes the velocity from $v_0$ to $v$ (and this process
is given by the rate $\sigma(v, v_0)$) and then free evolution
continues during the remaining time $T-T_0$ with velocity $v$.
This is exactly the one ``gain'' collision term in the Boltzmann equation.

\medskip
\underline{Case 4}: $n=0$, $n'=2$ or $n=2$, $n'=0$.

The calculation is similar to the previous case, we just record the
result:
$$
  \bE \wh W_{2,0,t}(\xi, v) = -\lambda^2 t \wh \psi_0(k_1) \ov{\wh \psi_0(k'_1)}
 \; e^{-it\big( e(k_1)- e(k'_1)\big)}
\int \rd q
   |\wh B(k_1-q)|^2 
 t R\big( t(e(k_1)-e(q))\big),
$$
where
$$
  R(u) = \frac{e^{iu}- iu-1}{u^2}.
$$
Simple calculation shows that $tR(tu) \to \pi\delta(u)$ as $t\to\infty$.
Thus we have
$$
   \bE \wh W_{2,0,t}(\xi, v) \to - \frac{T}{2} |\wh\psi_0(v)|^2 e^{-iTv\cdot \xi} \sigma(v),
$$
where we defined
$$
\sigma(v):=  \int \sigma(v, q) \rd q.
$$
Similar result holds for $\wh W_{0,2,t}$ and combining it with \eqref{w11},  we obtain the
conservation of the $L^2$-norm up to second order in $\lambda$, since with $\xi=0$ we have
$$
   \int \rd v\Big[ \bE \wh W_{1,1,t}(0, v)+  \bE \wh W_{2,0,t}(0, v) +  
\bE \wh W_{0,2,t}(0, v) \Big]
 =0.
$$
For general $\xi$ and after testing against a smooth function, we obtain
$$
 \bE \langle J, W_{2,0,t}\rangle \to -\frac{T}{2} \int \rd v  J(Tv, v) \sigma(v) 
|\wh \psi_0(v)|^2 
$$
which corresponds to the loss term in the Boltzmann equation.

\medskip

Apart from revealing how the Boltzmann equation emerges
from the quantum expansion, the above calculation carries
another important observation. Notice that terms with $n\ne n'$
did give rise to non-negligible contributions
despite the earlier statement \eqref{nn'} that $n=n'$ is
forced by the non-repetition rule (which is correct)
and that repetitive collision sequences are negligible
(which is, apparently, not correct). The next section will explain this.

\subsection{Self-energy renormalization}\label{sec:renorm}

The previous simple explicit calculations  showed that not all repetition
terms are neglible, in fact 
immediate recollisions are of order one and they -- but only they --
 have to be treated differently. We say that a collision sequence
$(\a_1, \a_2, \ldots)$ has an {\it immediate recollision}
if $\a_i=\a_{i+1}$ for some $i$.  This modification must also be 
implemented in the stopping rule, such that immediate recollisions
do not qualify for recollisions for the purpose of
collecting reasons to stop the expansion (see Section~\ref{sec:stop}).

\bef\bec
\epsfig{file=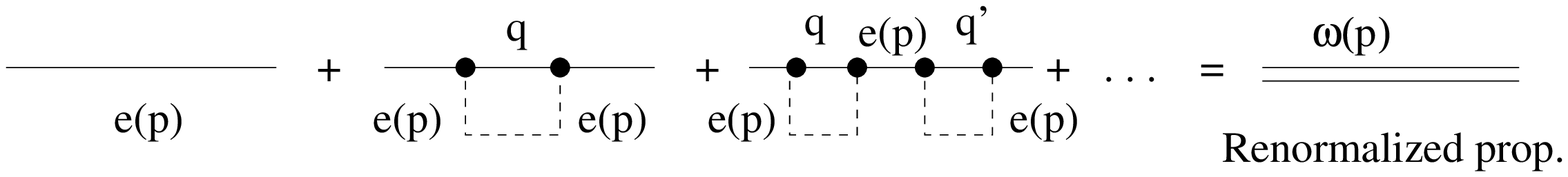, scale=0.70}
\eec
\caption{Schematic picture of the renormalization}
\label{fig:renor1}
\eef

Fortunately immediate repetitions appear very locally in
the Feynman graph and they can be resummed according
to the schematic picture on Fig.~\ref{fig:renor1}.
The net effect is that the free propagator $e(p)=\frac{1}{2}\, p^2$
needs to be changed to another propagator $\om(p)$
that differs from $e(p)$ by an $O(\lambda^2)$ amount.
More precisely, using the time independent formalism,
second line of \eqref{psiform}, we notice that the first
term in Fig.~\ref{fig:renor1} is represented by 
\be
  \frac{1}{\a- e(p) - i\eta}
\label{propp}
\ee
(which is also often called propagator). The second term carries
a loop integration (momentum $q$), and its contribution is
$$
  \frac{1}{(\a- e(p) - i\eta)^2} \; \int \frac{\lambda^2 |\wh B(p-q)|^2 }{ \a- e(q) - i\eta} \, \rd q.
$$
This formula is written for the continuum model and
the integration domain is the whole momentum space $\bR^d$. For the lattice model $\wh B$ is
absent and the integration is over ${\bf T}^d$.

The third term carries two independent loop integration (momenta $q$ and
$q'$), and its contribution is
$$
  \frac{1}{(\a- e(p) - i\eta)^3} \;
 \int \frac{\lambda^2 |\wh B(p-q)|^2 \rd q}{ \a- e(q) - i\eta}
 \int \frac{\lambda^2 |\wh B(p-q')|^2 \rd q'}{ \a- e(q') - i\eta}.
$$
Setting
$$
  \Theta_\eta(p,\a): = \frac{|\wh B(p-q)|^2 \rd q}{ \a- e(q) - i\eta}
$$
and noticing that due to the almost singularities of the 
$(\a- e(p) - i\eta)^{-k}$ prefactors
 the main contribution comes
from $\a \sim e(p)$, we can set
\be
  \theta(p): =\lim_{\eta\to 0+}\Theta\big(p, e(p)\big).
\label{thep}
\ee
(Similar, in fact easier formulas hold for the lattice model).
Therefore, modulo negligible errors, the sum of the 
graphs in Fig.~\ref{fig:renor1} give rise to the following
geometric series:
$$
  \frac{1}{\a- e(p) - i\eta} 
 +  \frac{\lambda^2\theta(p)}{(\a- e(p) - i\eta)^2}
 +  \frac{\big(\lambda^2\theta(p)\big)^2}{(\a- e(p) - i\eta)^3} + \ldots
 = \frac{1}{\a- \big( e(p) + \lambda^2\theta(p)\big) - i\eta} 
$$
This justifies to define the {\it renormalized propagator}
$$
  \om(p): = e(p) + \lambda^2\theta(p)
$$
and the above calculation indicates that all immediate recollisions
can be taken into account by simply replacing $e(p)$ with $\om(p)$.
This is in fact can be rigorously proved up to the leading order we are interested.

An alternative way to see the renormalization is
to reorganize how the original Hamiltonian
is split into main and perturbation terms:
$$
   H = \underbrace{e(p) + \lambda^2 \theta(p)}_{\omega(p)} + \lambda V -
   \lambda^2 \theta(p).
$$
The precise definition of the correction term $\theta(p)$ is determined
by the following self-consistent equation:
$$
    \theta(p) :=  \int \!\! \frac{\rd q}{\om(p)-\om(q) + i0},
$$
the formula \eqref{thep} is in fact only the solution
to this equation up to order $\lambda^2$, but for
our purposes such precision is sufficient.
The  imaginary part of $\theta$ can also be computed as
$$
    \sigma(p) := \eta \int \frac{\rd q}{|\om(p) - \om(q) + i\eta|^2}
 \to \mbox{Im} \theta(p)  \qquad \eta\to 0
$$
and notice that it is not zero. In particular, the
renormalization  {\it regularizes the propagator}:
while the trivial supremum bound on the original propagator is
$$
 \sup_p \Big| \frac{1}{\a- e(p) - i\eta}\Big|  \le \eta^{-1}
$$
the similar bound on the renormalized propagator is much better:
\be
  \Big| \frac{1}{\a- \om(p) - i\eta}\Big|  \le \frac{1}{\lambda^2+\eta}.
\label{linf}
\ee
Strictly speaking, this bound does not hold if $p\approx 0$
since $\mbox{Im} \,\theta(p)$ vanishes at the origin, but 
such regime in the momentum space has a small volume, since
it is given by an approximate point singularity, while
the (almost) singularity manifold of \eqref{propp} 
is large, it has codimension one.

The bound \eqref{linf} will play a crucial role in our estimates.
Recall that due to the exponential prefactor $e^{2\eta t }$
in \eqref{valpi}, eventually we will have to choose $\eta \sim 1/t$.
Thus in the diffusive scaling, when $t\gg \lambda^{-2}$, the bound
\eqref{linf} is a substantial improvement.

The precise formulas are not particularly interesting; the main point
is that after renormalization: only the ladder has classical contribution
and gives the limiting equation. If one uses the renormalized propagators,
then one can assume that no immediate repetitions occur in the
expansion (in practice they do occur, but they are algebraically 
cancelled out by the renormalization of the propagator).
Only after this renormalization will the value of the ladder
graph given in Fig.~\ref{fig:feyn2}  be correct 
with the exponential damping factor. From now
on we will assume that the renormalization is
performed, the propagator is $\om(p)$  and there is no immediate recollision.
In particular, the statements of the key Theorems~\ref{thm:negl} and \ref{thm:only}
are understood {\it after this renormalization}.

\subsection{Control of the crossing terms}

The key to the proof of Theorem \ref{thm:only} is a good
classification of all Feynman diagrams based upon the complexity
of the permutation $\pi: A\to B$. This complexity
is  expressed by a degree $d(\pi)$ that is defined, temporarily, as follows.
We point out that this is a little simplified definition,
the final definition is a bit more involved and
is given later in Definition~\ref{def:slope}.

\begin{definition}
Given a permutation $\pi\in S_n$ on $\{1, 2, \ldots, n\}$,
 an index $i$ is called {\bf ladder index}
if $|\pi(i)-\pi(i-1)|=1$ or  $|\pi(i)-\pi(i+1)|=1$.
 The {\bf degree of the permutation $\pi$} is defined as
\be
 d(\pi)  = \#\{
 \mbox{non-ladder indices}\}.
\label{def:dtemp}
\ee
\end{definition}

\bef\bec
\epsfig{file=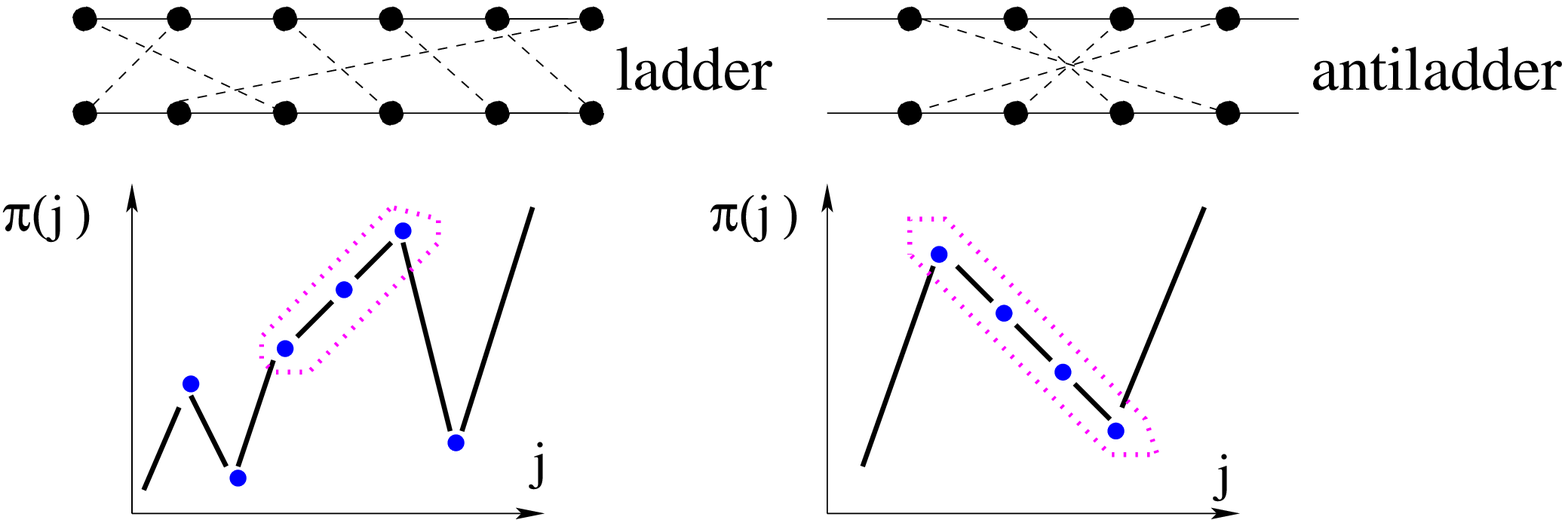, scale=0.80}
\eec
\caption{Ladder and antiladder}
\label{fig:lad11}
\eef
Two examples are shown on Fig.~\ref{fig:lad11}.  The top pictures
show the pairing in the Feynman diagrams, the bottom pictures
show the graph of the permutation  as a function
$$
\pi:\{1,2,\ldots , n\}\to\{1,2,\ldots , n\},
$$
 where the set $\{1,2,\ldots , n\}$ is naturally embedded into
the reals and we conveniently joined the discrete
points of the graph of $\pi$. The dotted region encircles the
ladder indices: notice that a long antiladder also has many
ladder indices. This indeed shows that it is not really
the number of total crosses in the Feynman graph that is responsible for 
the smallness of the value of the Feynman graph,
rather the disordered structure and that is more precisely expressed
by the non-ladder indices.

The main philosophy is that
most permutations have a high degree (combinatorial complexity).
The key estimates are the following two lemmas.
The first lemma estimates the number of permutations
with a given degree; the proof is an elementary combinatorial counting
and will not be presented here.

\begin{lemma}
The number of permutations with a given degree is estimated as
$$
    \# \{ \pi\; : \; d(\pi)=d\} \le (Cn)^d.
$$
\end{lemma}

The second lemma is indeed the hard part of the proof;
it claims that the value of the Feynman graph 
decreases polynomially (in $\lambda$) as the degree
increases. Thus we can
{\it gain a $\lambda$ factor  per each non-ladder vertex}
of the Feynman graph.

\begin{lemma}\label{lm:hard} There exists  some positive $\kappa$,
depending on the dimension $d$, such that
\be
\mbox{Val}(\pi) \leq (C\lambda)^{\kappa d(\pi)}. 
\label{star}
\ee
\end{lemma}

Combining these two Lemmas, we can easily control
the series $\sum_\pi \mbox{Val}(\pi)$:
$$
   \sum_{\pi\in S_n}
 \mbox{Val}(\pi) =\sum_{d=0}^\infty \sum_{\pi: d(\pi)=d} \mbox{Val}(\pi)
    =\sum_d C^d n^d\lambda^{\kappa d}  <\infty 
$$
if $n \leq  K\sim \lambda^{-\kappa}$ (assume $\delta=0$ for simplicity).
 Since $n\sim\lambda^2 t$, get convergence for
 $t\leq c\lambda^{-2-\kappa}$, i.e. the
$\kappa$ from Lemma \ref{lm:hard} determines the
time scale for which our proof is valid.

\bigskip
We remark that, although for technical reasons we can prove
\eqref{star} only for very small $\kappa$, it 
 should be valid up to $\kappa=2$ but not beyond.
To see this, recall that the best possible estimate
for a single Feynman graph is $O(\lambda^{2n})$ and their
total number is $n!$. Since
$$
   \lambda^{2n} n!\approx (\lambda^2n)^n\approx (\lambda^4t)^n,
$$
the summation over all Feynman graphs will diverge
if $t\ge \lambda^{-4}$. 
This means that this method is limited up to times $t\ll \lambda^{-4}$.
Going beyond $t\sim\lambda^{-4}$
requires a second resummation procedure, namely the resummation
of the so-called {\it four-legged} subdiagrams.
In other words, one cannot afford to estimate each Feynman
graph individually, a certain cancellation mechanism among
them has to be found.
Similar resummations have been done in many-body problems
in euclidean (imaginary time) theories but not in real time.
In the current problem it is not clear even on the intuitive
level which diagrams cancel each other.

\subsection{An example}\label{sec:ex}

In this very concrete example we indicate why the cross is smaller than
the ladder by a factor $\lambda^2$, i.e. we justify the first estimate in 
Fig.~\ref{fig:feyn3}. We recall that the value of a Feynman graph of order $n$ is
\be
 \mbox{Val}(\pi) =\lambda^{2n}e^{2\eta t}\int \!\! \rd\bp \rd \bp' \rd\a \rd\beta\; e^{it(\alpha-\beta)}
 \prod_{j=0}^n
 \frac{1}{\a - \ov\om(p_j) -i\eta}
 \; \frac{1}{\beta - \om(p_j') +i\eta} \; \Delta_\pi(\bp, \bp')\; ,
\label{example}
\ee
where the delta function comes from momentum 
conservation. Notice that the integrand is a function
that are almost singular on the level sets of
the dispersion relation, i.e. on the manifolds
$$
\a = \mbox{Re}\, \om (p_j) = e(p_j) + \lambda^2 \mbox{Re} \,\theta(p_j),
\qquad \beta = \mbox{Re}\, \om (p_j') = e(p_j') + \lambda^2 \mbox{Re} \, \theta(p_j').
$$
For the typical values of $\a$ and $\beta$, these are manifolds of codimension $d$
in the $2(n+1)d$ dimensional space of momenta. The singularities
are regularized by the imaginary parts of the
denominators, $\mbox{Im} \, \om(p_j)+\eta$, which are typically 
positive quantities of order $O(\lambda^2+\eta)$. 
The main contribution to the integral comes from regimes of
integration near these ``almost singularity'' manifolds.
We remark that in the continuum model the momentum space 
extends to infinity, so in principle one has to control the integrals in
the large momentum (ultraviolet) regime as well, but this is
only a technicality.

The main complication in evaluating and estimating the integral \eqref{example}
comes from the
 delta functions in $\Delta_\pi(\bp, \bp')$ since they
 may enhance singularity overlaps and may increase the value
of the integral. Without these delta functions, each momentum
integration $\rd p_j$ and $\rd p_j'$ could be performed
independently and the total value of the Feynman graph
would be very small, of order $\lambda^{2n} |\log \lambda|^{2(n+1)}$.
It is exactly the overlap of these (almost) singularities, forced
by the delta functions, that is responsible for the
correct size of the integral.

\medskip

To illustrate this effect, 
we consider the simplest example for a cross and compare it with
the direct pairing (ladder). The notations are found 
on Fig.~\ref{fig:lad2}. For simplicity, we assume that
$\lambda^2=\eta$, i.e. we are in the kinetic regime
and the extra regularization coming from the
renormalization, $\mbox{Im} \, \theta$, is not important.

\bef\bec
\epsfig{file=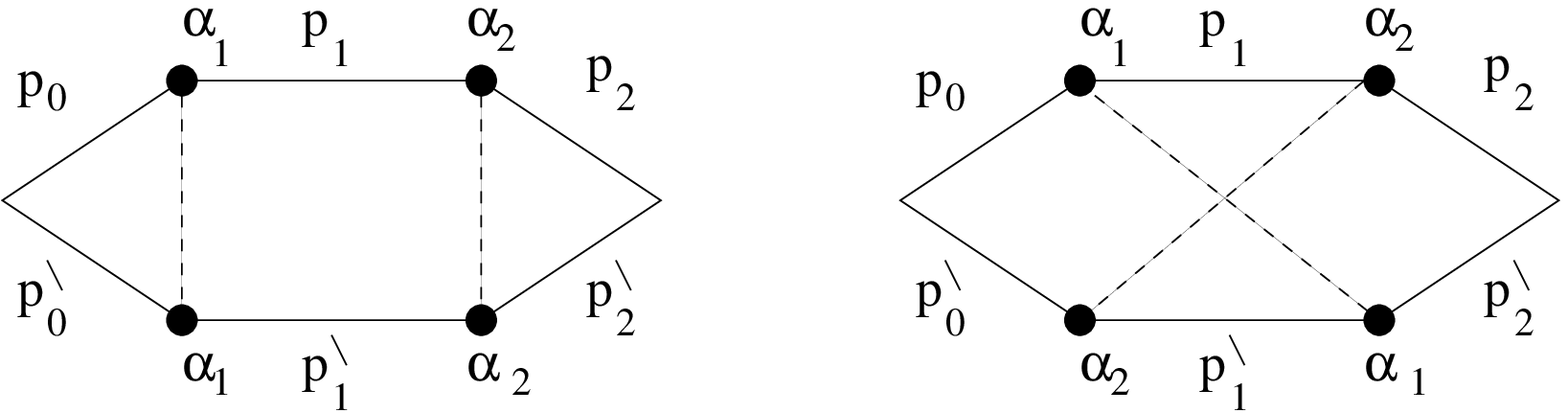, scale=0.80}
\eec
\caption{Ladder and cross}
\label{fig:lad2}
\eef

In case of the ladder, the delta functions force all
paired momenta to be exactly the same (assuming that
$p_0=p_0'$ since we compute the $L^2$-norm of $\psi_t$), i.e.
$$
 \mbox{Ladder}\quad\Longrightarrow
\quad p_j=p_j', \quad \forall j=0,1,2.
$$
In contrast, the
crossing delta functions yield the following relations:
$$
p_0'= p_0,\quad  p_1'= p_0-p_1+p_2, \quad p_2'=p_2.
$$

Now we can compute explicitly:
\begin{align}
 \mbox{Val(ladder)}=
\lambda^4\, e^{2\eta t}\int & \rd\a \rd\beta \; e^{i(\a-\beta)t} \int 
\frac{1}{\a - \om(p_0) -i\eta} 
\; \frac{1}{\a - \om(p_1) -i\eta} \; \frac{1}{\a -\om( p_2) -i\eta}
\non\\
&  \times  \frac{1}{\beta - \om(p_0) +i\eta} \; 
\frac{1}{\beta - \om(p_1) +i\eta} \; 
\frac{1}{\beta - \om(p_2) +i\eta}  \rd p_0\rd p_1\rd p_2.
\label{la}
\end{align}
Choosing $\eta = 1/t$, 
with a simple analysis one can verify that
the main contribution comes from the regime
where  $|\a -\beta|\lesssim t^{-1}$, thus effectively
all singularities overlap (with a precision $\eta$).
We have
\be
     \int \frac{\rd p}{|\alpha -\om(p) + i\eta|^2} 
\sim \frac{1}{|\mbox{Im}\, \om|} \sim \lambda^{-2}.
\label{2}
\ee
The intermediate relations are not completely correct as they
stand because $\omega$ depends on the momentum
and its imaginary part actually vanishes at $p=0$.
However, the volume of this region is small. Moreover,
in the continuum case the ultraviolet regime needs
attention as well, but in that case the decaying form factor $|\wh B|^2$
is also present.
A more careful calculation shows that
the final relation in \eqref{2} is nevertheless correct.

Thus, effectively, we have
\be
    \mbox{Val(Ladder)} \sim \lambda^4 (\lambda^{-2})^2 \sim O(1)
\label{laddd}
\ee
modulo logarithmic corrections.
Here \eqref{2} has been used twice and 
the last pair of denomitors integrated out by $\rd\a\rd\beta$
collecting a logarithmic term:
\be
  \int \frac{\rd \a}{|\a - \om(p_0) -i\eta|}= O(|\log \eta|)
\label{log}
\ee
(recall that $\eta=1/t\sim \lambda^2$). This estimate again is not completely
correct as it stands, because the integral in \eqref{log} is logarithmically
divergent at infinity, but the ultraviolet regime is
always trivial in this problem. Technically, for example,
here one can save a little more $\alpha$ decay from 
the other denominators in \eqref{la}.
Both inequalities \eqref{2} and \eqref{log} hold both
in the discrete and continuum case. In  Appendix \ref{ineq}
we listed them and some more complicated related estimates
more precisely that will also be  used.

As a rule of thumb, we should keep in mind that the main
contribution in all our integrals come from the regimes
where:
\begin{itemize}

\item[(i)]  The two dual variables to the time are close
with a precision $1/t$, i.e.
$$
    |\alpha -\beta|\leq 1/t;
$$
\item[(ii)] The momenta are of order one and away from zero
(i.e. there is no ultraviolet or infrared issue in this problem);

\item[(iii)] The variables $\alpha, \beta$ are also of order
one, i.e. there is no divergence at infinity for their integrations.

\end{itemize}

By an {\it alternative argument} one can
simply 
estimate all (but one) $\beta$-denominators in \eqref{la} by 
an $L^\infty$-bound \eqref{linf}, i.e. by 
$O(\lambda^{-2})$,
then integrate out $\rd\beta$ and then
 all  $\alpha$-denominators, using \eqref{log} and the similar $L^1$-bound
\be
  \int \frac{\rd p}{|\a - \om(p) -i\eta|} = O(|\log \eta|)
\label{l1}
\ee
for the momentum integrals.
The result is
$$
   \mbox{Val(Ladder)} \lesssim \lambda^4  (\lambda^{-2})^2 
(\log \lambda)^4 =  (\log \lambda)^4
$$
which coincides with the previous calculation \eqref{laddd}
modulo logarithmic terms. In fact, one can verify 
that these are not only upper estimates for the ladder, but
in fact the value of the $n$-th order ladder is
\be
    \mbox{Val(Ladder)} \sim O_n(1)
\label{laddval}
\ee
as $\lambda\to0$, 
modulo $|\log \lambda|$ corrections.

\medskip

The similar calculation for the crossed diagram yields:
\begin{align}
\mbox{Val(cross)}=
 \lambda^4 \int & \rd\a \rd\beta \; e^{i(\a-\beta)t}
 \int \frac{1}{\a - \om(p_0) -i\eta} 
\; \frac{1}{\a - \om(p_1) -i\eta} \; \frac{1}{\a -\om( p_2) -i\eta}
\non\\
&  \times  \frac{1}{\beta - \om(p_0) +i\eta} \; 
\frac{1}{\beta - \om(p_0-p_1+p_2) +i\eta} \; 
\frac{1}{\beta - \om(p_2) +i\eta}  \rd p_0\rd p_1\rd p_2.
\end{align}
Assuming again that the main contribution is from the regime where
$\a\sim \beta$ (with precision $1/t$), we notice that
 spherical singularities of the two middle denominators
 overlap only at a
point singularity
\be
     \int \rd p_1 \frac{1}{|\alpha -\om(p_1) - i\eta|}\;
 \frac{1}{|\alpha -\om(p_0-p_1+p_2) + i\eta|} 
     \sim \frac{1}{|p_0+p_2| +\lambda^2}
\label{pointsing}
\ee
(see also Fig.~\ref{fig:point} and Lemma~\ref{lemma:main} for a more precise estimate).
In three dimensions the point singularity is harmless (will
disappear by the next integration using \eqref{withp} from Lemma~\ref{lemma:main}),
and we thus obtain
$$
  \mbox{Val(cross)}\leq \lambda^2\mbox{Val(ladder)}
$$
(modulo logarithms).

\bef\bec
\epsfig{file=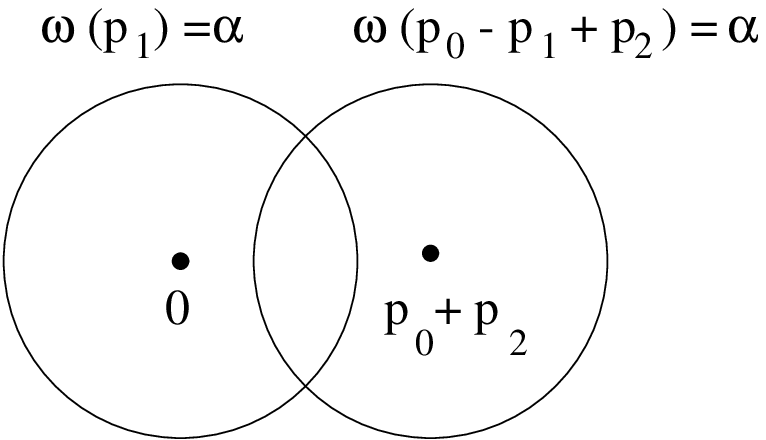, scale=0.80}
\eec
\caption{Overlap of the shifted level sets in the $p_1$-space}
\label{fig:point}
\eef

We emphasize that inequality \eqref{pointsing} in this form holds
only for the continuous dispersion relation (see \eqref{withoutp}
in Appendix \ref{ineq}), i.e. {\it if 
the level sets of the dispersion relation are convex},  since 
it is relied on the fact that a convex set and its shifted
copy overlap transversally, thus a  small neighborhood of these sets
(where $|\a - e(p)|$ is small) have a small intersection.
This is wrong for the level sets of the discrete dispersion  relation
which, for a certain range of $\a$, is not a convex set (see Fig.~\ref{fig:disprel}). 
However, \eqref{pointsing} holds with an additional
factor $\eta^{-3/4}$ on the left hand side (see \eqref{nopont}),
which is a weaker estimate than in the continuous
case but it is still useful because it is stronger than
the trivial estimate $\eta^{-1}$ obtained by by taking the $L^\infty$-norm of
one propagator on the left hand side of  \eqref{pointsing}.

\bef\bec
\epsfig{file=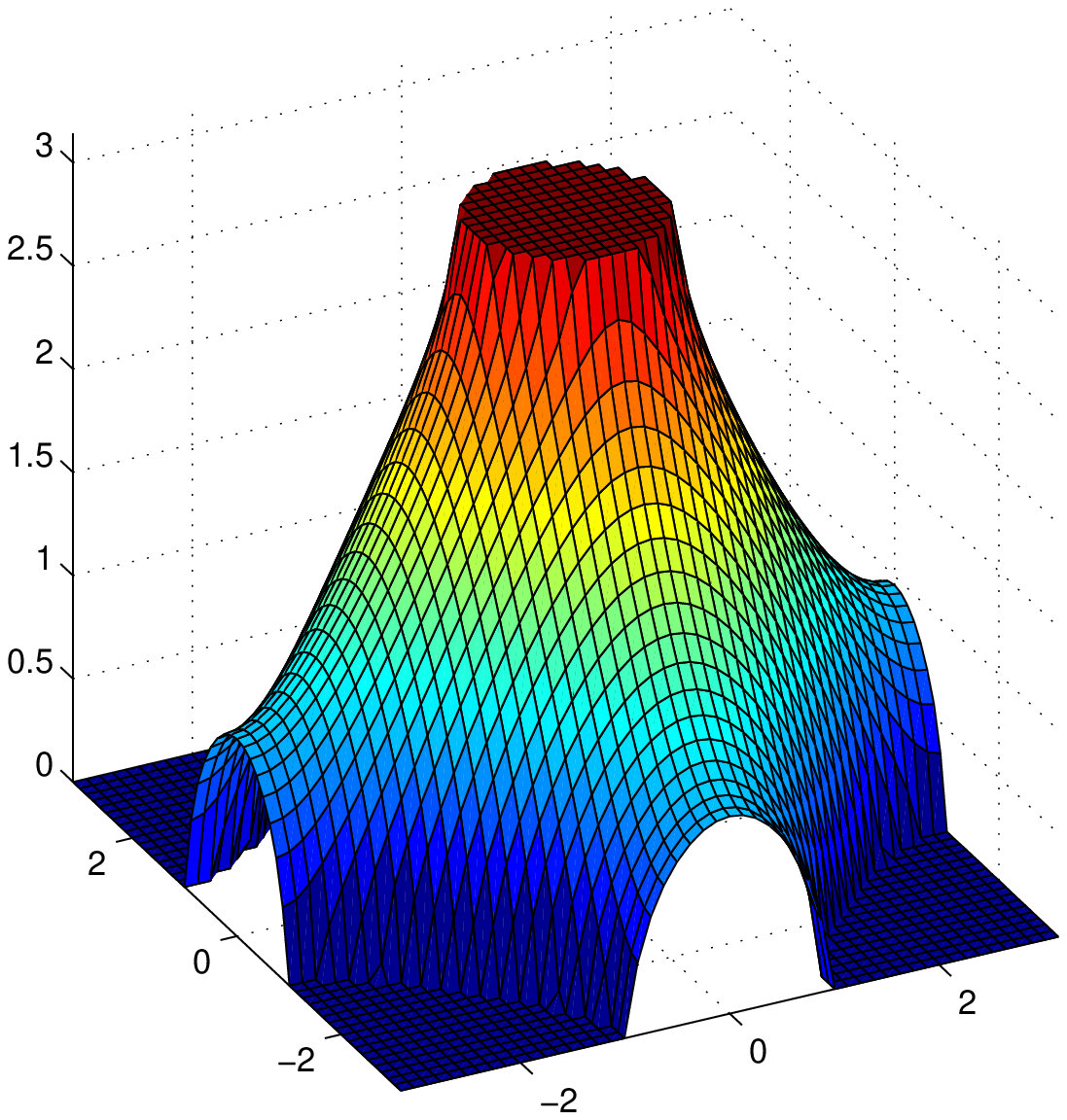, scale=0.70}
\eec
\caption{A level set of the discrete dispersion relation $e(p)$ from \eqref{disp}}
\label{fig:disprel}
\eef

\section{Integration of general Feynman graphs}\label{sec:genint}
\setcounter{equation}{0}

\subsection{Formulas and the definition of the degree}

We introduce a different encoding for the permutation $\pi$  by using 
permutation matrices. To follow the standard convention of 
labelling the rows and columns of a matrix by positive natural
numbers, we will shift all indices in \eqref{valpi} by one, i.e.
from now on we will work with the formula
\be
 \mbox{Val}(\pi) =\lambda^{2n} e^{2\eta t}\int \!\! \rd\bp \rd \bp' \rd\a \rd\beta
\; e^{it(\alpha-\beta)} \prod_{j=1}^{n+1}
 \frac{1}{\a - \ov\om(p_j) -i\eta}
 \; \frac{1}{\beta - \om(p_j') +i\eta} \; \Delta_\pi(\bp, \bp')\; ,
\label{valpi1}
\ee
where 
$$
\Delta_\pi(\bp,\bp'):= \delta(p_{n+1}-p_{n+1}')\prod_{j=1}^{n} 
\delta\Big( (p_{j+1}-p_j) - (p_{\pi(j)+1}' - p_{\pi(j)}')\Big).
$$
We introduce a convenient  notation.
For any  $(n+1)\times (n+1)$ matrix $M$ and for any vector
of momenta $\bp=(p_1, \ldots p_{n+1})$, we let $M\bp$
denote the following $(n+1)$-vector
of momenta
\be
   M\bp : = \Big( \sum_{j=1}^{n+1} M_{1j} p_j, \;  
 \sum_{j=1}^{n+1} M_{2j} p_j, \ldots \Big) \; .
\label{def:Mp}
\ee

The permutation $\pi\in \cS_n$ acting on the indices $\{1, 2, \ldots, n\}$
in \eqref{valpi1} 
will be encoded by an $(n+1)\times(n+1)$ matrix $M(\pi)$ defined as
follows
\be
       M_{ij}(\pi): = \left\{ \begin{array}{cll} 1 & \qquad \mbox{if}
    \quad & \tpi(j-1) < i \leq \tpi(j)\\
       -1 & \qquad \mbox{if} \quad & \tpi(j) <i \leq \tpi(j-1)\\
       0 & \qquad \mbox{otherwise,} \quad & \;  \end{array} \right.
    \label{def:Mmat}
\ee
where, by definition, $\tpi$ is the {\bf extension} of $\pi$ to
 a permutation of $\{ 0, 1, \ldots, n+1\}$
by $\tsi(0):=0$ and $\tsi(n+1):=n+1$. 
In particular $[M\bp]_1= p_1$,
$[M\bp]_{n+1}= p_{n+1}$.
It is easy to check that
\be
     \Delta_\pi(\bp,\bp')=
    \prod_{j=1}^{n+1}\delta\Big( \; p_j' - [M\bp]_j 
    \Big)\; ,
\label{M}
\ee
in other words, each $p'$-momentum can be expressed as a linear
combination of $p$-momenta, the matrix $M$ encodes the corresponding
coefficients, and these are all the relations among the $p$ and $p'$
momenta that are enforced by $\Delta_\pi$. In particular, all $p$-momenta
are independent.  

The  rule to express $p'$-momenta in terms of $p$-momenta is
 transparent in the graphical representation
of the Feynman graph: the momentum $p_j$ appears in 
those $p_i'$-momenta which fall into its "domain of dependence",
i.e. the section between the image of the two endpoints of
$p_j$, and the sign depends on the ordering of these images
(Fig.~\ref{fig:domdep}).  Notice that the roles of $\bp$ and $\bp'$
are symmetric, we could have expressed the $p$-momenta in terms of
$p'$-momenta as well. It follows from this symmetry that
$M(\pi^{-1})= [M(\pi)]^{-1}$ and
$$
       \Delta_\pi(\bp,\bp')=
    \prod_{j=1}^{n+1}\delta\Big( \; p_j - [M^{-1}\bp']_j 
    \Big)
$$
also holds.

A little linear algebra and combinatorics reveals that $M$
is actually
a {\bf totally unimodular} matrix, which means that all
its subdeterminants are 0 or $\pm 1$. This will mean
that the Jacobians of the necessary 
changes of variables are always controlled.

\bef\bec
\epsfig{file=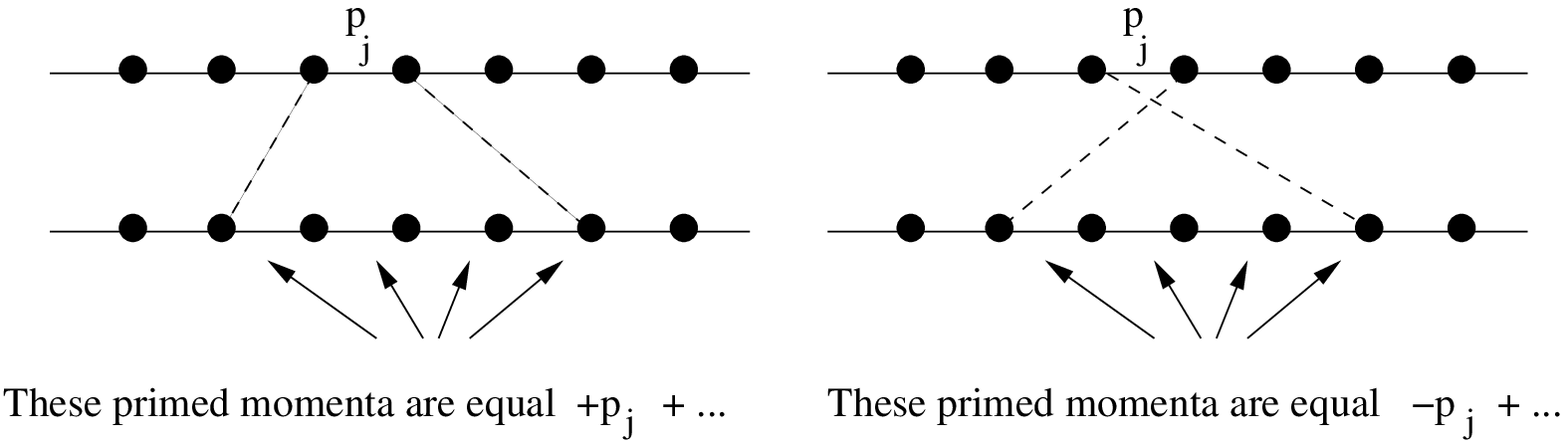, scale=0.8}
\eec
\caption{Domain of dependencies of the momenta}\label{fig:domdep}
\eef

\medskip

The following definition is crucial. It establishes the necessary concepts
to measure the complexity of  a permutation.

\begin{definition}[Valley, peak, slope and ladder]\label{def:slope}
Given a permutation $\pi\in \cS_n$ let $\tsi$ be its extension. A point
 $(j, \pi(j))$, $j\in I_n=\{1,2, \ldots, n\}$,
on the  graph of $\pi$ is called {\bf peak} 
if $ \pi(j) < \min\{ \tpi(j-1), \tpi(j+1)\}$,
it is called {\bf valley} if $\pi(j) > \max\{ \tpi(j-1),\tpi(j+1)\}$.
Furthermore, 
if $\pi(j)-1 \in \{ \tpi(j-1), \tpi(j+1)\}$ and $(j,  \pi(j))$ is not
a valley, then the point $(j, \pi(j))$, $j\in I_n$, is called {\bf ladder}.
Finally, a point  $(j, \pi(j))$, $j\in I_n$, on the graph
of  $\pi$ is called {\bf slope} if it is not  a peak, valley or ladder.

Let $I=\{ 1, 2, \ldots, n+1\}$ denote the set of row indices of $M$.
This set is partitioned
into five disjoint subsets, $I=I_p\cup I_v \cup I_\ell\cup I_s\cup I_{last}$,
 such that $I_{last}:= \{ n+1\}$ is the last index, and
 $i\in I_p, I_v, I_\ell$ or $I_s$ depending
on whether $(\pi^{-1}(i),i)$ is a peak, valley, ladder or slope, respectively.
The cardinalities of these sets are denoted by $p:=|I_p|$, $v:=|I_v|$,
$\ell:= |I_\ell|$ and $s:= |I_s|$. The dependence on $\pi$ is
indicated as $p=p(\pi)$ etc. if necessary.
We define the {\bf degree} of the permutation $\pi$ as
\be
{\rm deg} (\pi): =d(\pi):= n-\ell(\pi)\; .
\label{def:d}
\ee
 \end{definition}

{\it Remarks:} (i) The terminology of peak, valley, slope, ladder
comes from the graph of the permutation $\tpi$ 
drawn in a coordinate system where the axis of the dependent variable,
$\pi(j)$, is oriented downward (see Fig.~\ref{fig:slope6pi}).
It immediately follows from the definition of the extension $\tpi$
that the number of peaks and valleys is the same:
$$
p= v.
$$
By the partitioning of $I$, we also have
$$
   p+v+\ell+s+1 = n+1.
$$

(ii) The nonzero
entries in the matrix $M(\sigma)$
follow the same geometric pattern as the graph: each downward
segment of the graph corresponds to a column with
a few consecutive 1's, upward segments correspond to
columns with $(-1)$'s.  These blocks of nonzero entries
in each column will be called the {\bf tower} of that column.
On Fig.~\ref{fig:slope6pi} we also
pictured the towers of $M(\pi)$ as rectangles.

(iii) Because our choice of orientation of the vertical axis 
follows the convention of labelling rows of a matrix,
a peak is a local minimum of  $j\to \pi(j)$. 
We fix the convention that  the notions ``higher'' or ``lower'' 
for objects related to the vertical axis (e.g. row indices)
always refer to the graphical picture.
In particular the ``bottom'' or the ``lowest element'' of a tower is 
located in the row with the highest index.

Also, a point on the graph of the function $j\to \pi(j)$ is traditionally
denoted by $(j,\pi(j))$, where the first coordinate
$j$ runs  on the horizontal axis,
while in the labelling  of the  $(i,j)$--matrix element $M_{ij}$
of a matrix $M$ the first coordinate $i$ labels rows, i.e.
it runs vertically. 
To avoid confusion, we will always specify whether
a double index $(i,j)$ refers to a point on the graph of $\pi$
or a matrix element.

(iv) We note that for the special  case of the identity permutation
$\pi = id=id_n$ we have $I_p=I_s=I_v=\emptyset$,
and $I_\ell=\{1, 2, \ldots , n\}$. In particular, 
${\rm deg}(id)=0$
and ${\rm deg}(\pi)\ge2$ for any other permutation $\pi\neq id$.

(v) Note that Definition~\ref{def:d} slightly differs from the
preliminary definition of the degree given in \eqref{def:dtemp}.
Not every index participating in a ladder
are defined to be a ladder index; the top index of
the ladder is excluded (to avoid overcounting)
and the bottom index of a ladder is also excluded
if it is a valley, since the valleys will play a special role.
The precise definition of the degree given in Definition~\ref{def:d}
is not canonical, other variants are also possible;
the essential point is that long ladders should reduce the
degree.

\bigskip

\bef\bec
\epsfig{file=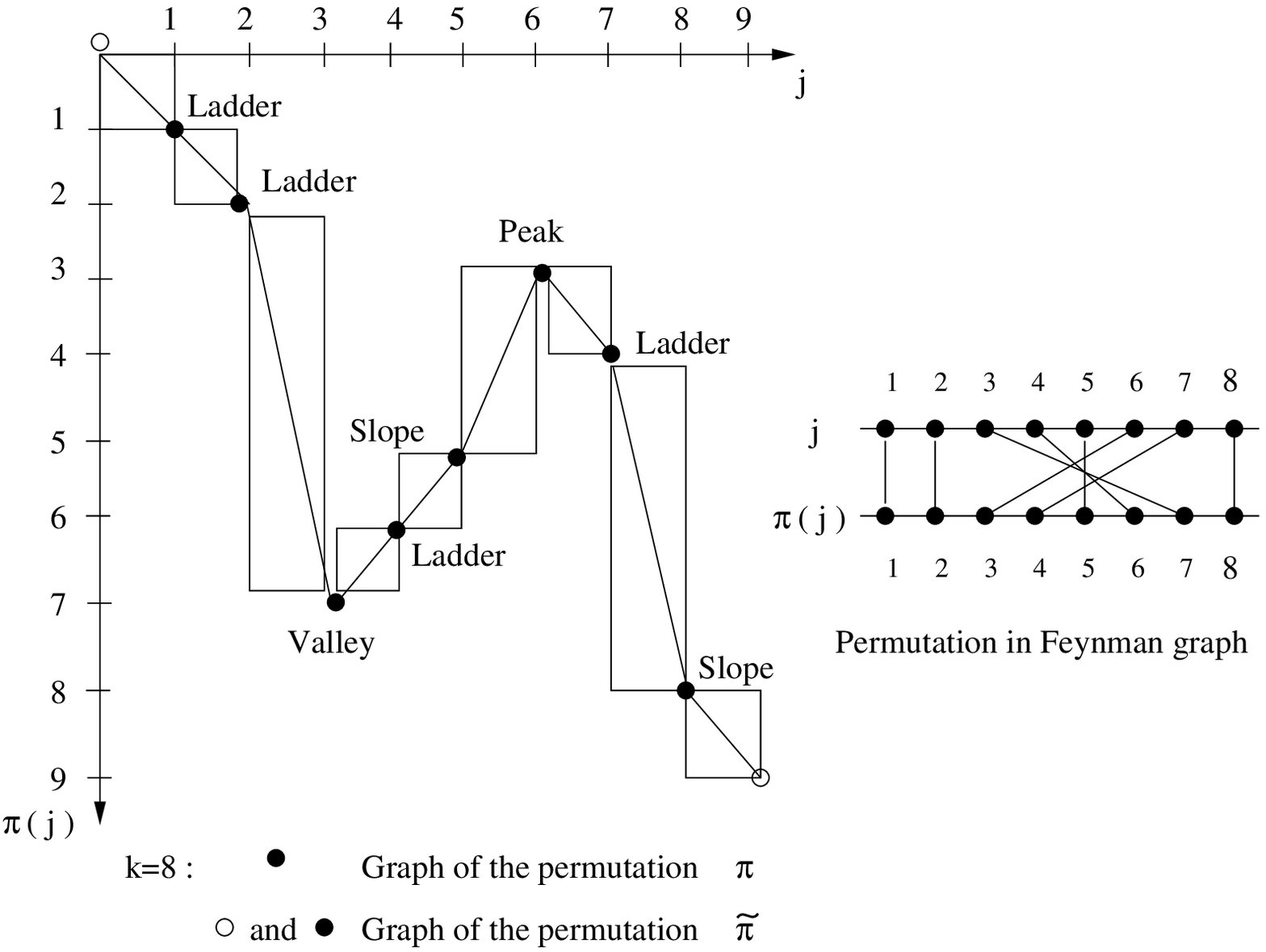, scale=0.9}
\eec
\caption{Graph of a permutation with the towers}\label{fig:slope6pi}
\eef

An example is shown on Fig.~\ref{fig:slope6pi} with $n=8$.
The matrix corresponding to the permutation on this figure is the following
(zero entries are left empty)
\be
    M(\pi):= \begin{pmatrix}
  1 & &  &  & &  &  & &\cr
&   1 & &  &  & &  &  & \cr
& & 1&  &  & &  &  & \cr
&& 1&  &  &-1 & 1 &  & \cr
& & 1&  &  &-1 &  & 1 & \cr
& & 1&  & -1 & &  & 1 & \cr
& & 1& -1 &  & &  & 1 & \cr
& & &  &  & &  & 1 & \cr
&  & &  &  & &  &  & 1 
\end{pmatrix}
\begin{array}{cl}
1 & \ell \cr
2 & \ell \cr
3 & p \cr
4 & \ell \cr
5 & s \cr
6 & \ell \cr
7 & v \cr
8 & s \cr
9 &  (last)
\end{array}  
\label{Mex}
\ee
The numbers on the right indicate the column indices and the
letters show whether it is peak/slope/valley/ladder or last. 
In this case $I_p= \{ 3 \}$, $I_v=\{ 7\}$, $I_s = \{5,8\}$.
$I_\ell=\{ 1,2, 4,6\}$, $I_{last}=\{ 9\}$
and ${\rm deg}(\s)=4$.

\subsection{Failed attempts to integrate out Feynman diagrams}

To estimate the value of a Feynman graph,
based upon \eqref{valpi1} and \eqref{M},
our task is to
successively integrate out all $p_j$'s in
the multiple integral of the form
$$
    Q(M): = \lambda^{2n}
    \int \rd\alpha \rd\beta \int
    \rd \bp
    \prod_{i=1}^{n+1}
    \frac{1}{|\alpha- \om(p_i)-i\eta|}\;\frac{1}{\big|\beta-\om\big(
    \sum_{j=1}^{n+1} M_{ij}p_j \big)+i\eta\big|}
$$
As usual, the unspecified domains of integrations for the $\a$ and $\beta$ variables is $\bR$, 
and for the $\rd\bp$ momentum variables is
$({\bf T}^d)^{n+1}$.
As one $p_j$ is integrated out, the matrix $M$ needs to be
updated and effective algorithm is needed to keep track of these
changes.

Why is it hard to compute or estimate such a quite explicit
integral? The problem is that there is  nowhere to start:
each integration variable $p_j$ may appear in many denominators: apparently
there is no ``easy'' integration variable at the beginning.

\bigskip

As a {\bf first attempt}, we can try to decouple the interdependent
denominators by Schwarz or H\"older
inequalities. It turns out that they cannot be used
effectively, since by their application we will lose the non-overlapping effects imposed by
the crossing. If one tries to decouple these integrals trivially,
one obtains the ladder and gains nothing.
In contrast to a complicated crossing diagram,
the ladder is easily computable 
because $p_i=p_i'$ means that the  integrals decouple:
\begin{align}
      Q(M)\leq \lambda^{2n}
    \int \rd\alpha \rd\beta \int &
    \rd \bp  \frac{1}{|\alpha- \om(p_{n+1})-i\eta|}\;
    \frac{1}{\big|\beta-\om(p_{n+1})+i\eta\big|} 
\label{firstattempt}\\
& \times
   \Bigg[
    \prod_{i=1}^{n}
    \frac{1}{|\alpha- \om(p_i)-i\eta|^2} + \frac{1}{\big|\beta-\om\big(
    \sum_{j=1}^{n} \wt M_{ij}p_j \big)+i\eta\big|^2}\Bigg], \non
\end{align}
where $\wt M$ denotes the $n\times n$ upper minor of $M$ and 
we used that $M$ and $\wt M$ differ only with an entry 1
in the diagonal, so in particular $\sum_{j=1}^{n+1} M_{n+1,j}p_j=p_{n+1}$
and $\mbox{det} (M) = \mbox{det} (\wt M)$.
Changing variables in the second term ($p_i'=  \sum_{j=1}^{n} \wt M_{ij}p_j$)
and using that the Jacobian is one ($\mbox{det}(M)=\pm 1$), we
see that the first and second terms in the parenthesis are exactly the same.
Thus
$$ 
     Q(M)\leq 2\lambda^{2n}
    \int \rd\alpha \rd\beta \int
    \rd \bp  \frac{1}{|\alpha- \om(p_{n+1})-i\eta|}\;
    \frac{1}{\big|\beta-\om(p_{n+1})+i\eta\big|} 
    \prod_{i=1}^{n}
    \frac{1}{|\alpha- \om(p_i)-i\eta|^2} ,
$$
and we can succesively integrate out the momenta $p_1, p_2, \ldots p_n$
and then finally $\alpha, \beta$. Using the inequalities \eqref{2} and 
\eqref{log}, we obtain (with the choice $\eta=1/t$)
$$
    |Val(\pi)|\leq e^{2\eta t}
  Q(M) \leq C \lambda^{2n} \lambda^{-2n} (|\log \eta|^2) = C (|\log \eta|^2)
$$
i.e. essentially the same estimate as the value  of the ladder \eqref{laddval}.

\bigskip

As a {\bf second attempt}, we can trivially estimate all (but one)
$\beta$-denominators,
by their $L^\infty$ norm
\be
   \Big|\frac{1}{\beta - \om(\ldots) +i\eta}\Big|
  \leq \frac{1}{\lambda^2 |\mbox{Im} \om|}\sim \lambda^{-2} 
\label{linfty}
\ee
(see \eqref{linf} and the remark afterward on its limitations),
and then integrate out all $p_j$ one by one
by using the $L^1$-bound  \eqref{l1}
$$
\int \frac{\rd p}{|\a-\om(p)+i\eta|}=O(|\log\eta|).
$$
This gives
$$ 
  |Val(\pi)|\leq O(|\log\eta|^{n+2}),
$$
i.e. it is essentially order 1 with logarithmic corrections.
Note that this second method gives a worse exponent for the
logarithm than the first one, nevertheless this method will be the good
starting point.

\medskip

The second attempt is a standard integration procedure in diagrammatic perturbation theory.
The basic situation is that  given a large graph with momenta assigned to the edges,
each edge carries a propagator, i.e. a function depending in this momenta, and
 momenta are subject to momentum conservation (Kirchoff law)
at each vertex. The task is to integrate out all momenta.
The idea is to use $L^\infty$-bounds for the propagator on certain edges in the graph
to decouple the rest and use independent $L^1$-bounds 
for the remaining integrations.
Typically one constructs a spanning tree in the graph,
then each additional edge creates a loop.
The Kirchoff delta functions amount to expressing
 all tree momenta in terms of loop
momenta in a trivial way. Then one uses
$L^\infty$-bound on ``tree''-propagators to free
up all Kirchoff delta functions, integrate out
the ``tree''-variables freely and then use
 $L^1$-bound on the ``loop''-propagators that are
now independent. 

This procedure can be used to obtain a rough bound on
values of very general Feynman diagrams subject to Kirchoff laws.
We will formalize this procedure 
in Appendix \ref{app:B}.  Since typically $L^1$ and $L^\infty$ bounds
on the propagators scale with a different power of the key
parameter of the problem (in our case $\lambda$), we will
call this  method {\it the power counting estimate}.

\bigskip

In our case simple graph theoretical counting shows that
$$
\mbox{Number of tree momenta = number of loop momenta = $n+1$}.
$$
In fact, after identifying the paired vertices in the  Feynman
diagram (after closing the two long horizontal lines at the
 two ends), we obtain a new graph where  one can easily see that 
the edges in the lower horizontal line, i.e. the edges
 corresponding $p_j'$-momenta, form a spanning tree and
the edges of  all $p_j$-variables form loops. 
Thus the delta functions
in \eqref{M} represent the way how the tree momenta are expressed
in terms of the loop momenta.

Since each  $L^\infty$-bound on ``tree''-propagators costs $\lambda^{-2}$
by \eqref{linfty}, the total estimate would be of order
$$
\lambda^{2n}\lambda^{-2(n+1)}
$$
with actually logarithmic factors. But due to the additional $\a, \beta$
integrations, one $L^\infty$-bound can be saved (modulo logarithm) by using \eqref{log}. So the
total estimate is, in fact,
$$
\lambda^{2n}\lambda^{-2n} = O(1)
$$
modulo logarithmic factors, i.e. the same size as
for the ladder diagrams. The conclusion is that even in the second attempt,
with the systematic power counting estimate, we did not gain
from the crossing structure of the permutation either.

\subsection{The new algorithm to integrate out Feynman diagrams}

In our new algorithm, we
use the  $L^\infty$-bound on the propagators for a carefully selected
subset of the ``tree''-variables: the ones that lie ``above the peak''.
The selection is shown in Fig.~\ref{fig:val1} for a concrete example. Notice that the segments in
the horizontal axis correspond to $p_1, p_2, \ldots, p_{n+1}$, i.e. the  loop momenta,
and the segments in the vertical axis correspond to the tree momenta.

We will first explain the main idea, then we will work out a concrete
example to illustrate the algorithm.
After drawing the graph of the permutation, 
a structure of valleys and peaks emerges.
By carefully examining the relation between this
graph and the matrix $M$, one notices that
if one estimates only those  ``tree''-propagators that lie above a
peak by the trivial $L^\infty$-bound, then all the
remaining propagators can be successively integrated out
by selecting the integration variables $p_j$ in
an appropriate order (dictated by the graph). Each integration
involves no more than two propagators at one time, hence it can be
estimated by elementary calculus (using estimates collected
in Appendix~\ref{ineq}). 

In fact, this is correct if there are no ladder indices;
but momenta corresponding to ladder indices can be integrated
out locally (see later). 
As it turns out, ladder indices are neutral;
their integration yields an $O(1)$ factor.
The gain comes from non-ladder indices, and this
will justify the definition of the degree \eqref{def:d}.

More precisely, after an easy bookkeeping we notice that
in this way we gain roughly as many $\lambda^{2}$ factors
as many slopes and valleys we have (no gain from 
peaks or ladders).  Since the number of peaks
and valleys are the same,
$$
        p=v,
$$
and the peaks, valleys, slopes and ladders altogether sum up to  $n$,
$$
  p+v+\ell + s =n
$$
(see Remark (i) after Definition \ref{def:slope}),
we see that
$$
  v+s = n-p-\ell \ge \frac{1}{2}(n-\ell) = \frac{1}{2} d(\pi),
$$
since $n-2p-\ell = n-p-v-\ell = s\ge 0$.
Thus we gain
at least $\lambda^{d(\pi)}$. This would prove \eqref{star} with $\kappa =1$.
After various technical estimates that were neglected here,
 the actual value of $\kappa$ is reduced, but it still remains
positive and this completes the proof of Lemma \ref{lm:hard}.
 $\Box$

\bef\bec
\epsfig{file=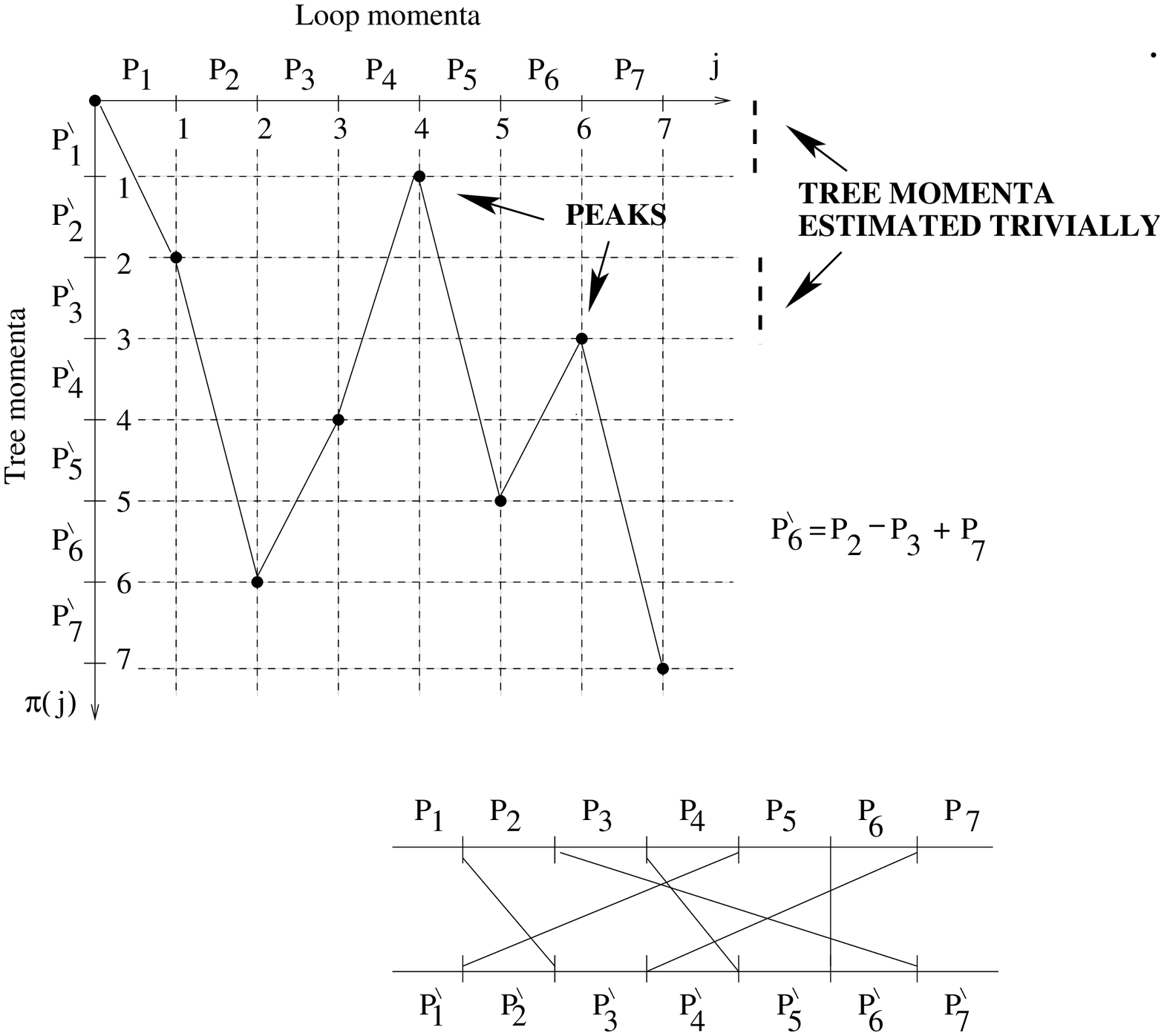, scale=0.6}
\eec
\caption{Valleys and peaks determine the order of integration}
\label{fig:val1}
\eef

\subsubsection{An example without ladder}\label{nolad}

As an example, we will show the integration procedure for the
graph on Fig.~\ref{fig:val1}. This is a permutation which
has no ladder index for simplicity, we will comment
on ladder indices afterwards.  The total integral is 
(for simplicity, we neglected the $\pm i\eta$ regularization
in the formulas below to highlight the structure better)
\begin{align}
   \lambda^{12}\int  &\rd\alpha \rd\beta \rd\bp
   \frac{1}{|\a - \om(p_1)|} \frac{1}{|\a - \om(p_2)|}
 \frac{1}{|\a - \om(p_3)|} \frac{1}{|\a - \om(p_4)|}
 \frac{1}{|\a - \om(p_5)|} \frac{1}{|\a - \om(p_6)|}
 \frac{1}{|\a - \om(p_7)|}
\non\\
    & \times\frac{1}{|\beta - \om(p_1)|} \frac{1}{|\beta - \om(p_1-p_4+p_5)|}
   \frac{1}{|\beta - \om(p_2-p_4+p_5)|} \frac{1}{|\beta - \om(p_2-p_4+p_5-
  p_6+p_7)|}
\non \\
&  \times\frac{1}{|\beta - \om(p_2-p_3+p_5-p_6+p_7)|} 
 \frac{1}{|\beta - \om(p_2-p_3+p_7)|} \frac{1}{|\beta - \om(p_7)|}.
\label{12}
\end{align}
Notice that every variable appears at least in three denominators,
for a graph of order $n$, typically every momentum appears in
$(const) n$ different denominators, so there is no way to start
the integration procedure (an integral with many
almost singular denominators is impossible to estimate directly 
with a sufficient precision).

Now we estimate  those $\beta$-factors  by $L^\infty$-norm 
(i.e. by $\lambda^{-2}$ according to \eqref{linfty})
whose corresponding primed momenta lied right above a peak;
in our case these are the factors
$$
   \frac{1}{|\beta - \om(p_1')|}\frac{1}{|\beta - \om(p_3')|}
$$
i.e. the first and the third $\beta$-factor in \eqref{12}. 
After this estimate, they will be removed from the integrand.
We will lose
a factor $(\lambda^{-2})^p=\lambda^{-4}$ recalling  
that $p=2$ is the number of peaks.
We are left with the integral
\begin{align}
    \lambda^{12} (\lambda^{-2})^p \int& \rd\alpha \rd\beta  \rd\bp
   \frac{1}{|\a - \om(p_1)|} \frac{1}{|\a - \om(p_2)|}
 \frac{1}{|\a - \om(p_3)|} \frac{1}{|\a - \om(p_4)|}
 \frac{1}{|\a - \om(p_5)|} \frac{1}{|\a - \om(p_6)|}
\non\\
    &\times \frac{1}{|\a - \om(p_7)|}\frac{1}{|\beta - \om(p_1-p_4+p_5)|}
    \frac{1}{|\beta - \om(p_2-p_4+p_5-
  p_6+p_7)|}
\non\\
&   \times\frac{1}{|\beta - \om(p_2-p_3+p_5-p_6+p_7)|} 
 \frac{1}{|\beta - \om(p_2-p_3+p_7)|} \frac{1}{|\beta - \om(p_7)|}.
\end{align}
Now the remaining factors can be integrated out by using the
following generalized version of \eqref{pointsing} (see Appendix~\ref{ineq})
\be
  \sup_{\a, \beta} \int \frac{\rd p}{|\a - \om(p) + i\eta|}
 \frac{1}{|\beta - \omega(p+u)-i\eta|} \leq \frac{|\log\eta|}{|u|+ \lambda^2}.
\label{pointsing2}
\ee

Suppose we can forget about the possible point singularity, i.e.
neglect the case when $|u|\ll 1$. Then the integral \eqref{pointsing2}
 is $O(1)$
modulo an irrelevant log factor.
Then the successive integration is done according to graph:
we will get rid of the factors $|\beta - \omega(p_1')|^{-1}$,
$|\beta - \omega(p_2')|^{-1}$, $|\beta - \omega(p_3')|^{-1}$, etc.,
in this order. The factors with $p_1'$ and $p_3'$ have already
been removed, so the first nontrivial integration will eliminate
\be
   \frac{1}{|\beta -\om(p_2')|} = \frac{1}{|\beta - \om(p_1-p_4+p_5)|}.
\label{145}
\ee
Which integration variable to use? Notice that the point
(1,2) was not a peak, that means that there is a momentum (in this
case $p_1$) such that $p_2'$ is the primed momenta with the largest
index that still depends on $p_1$ (the ``tower'' of $p_1$
ends at $p_2'$). This means that among all  the remaining $\beta$-factors
no other $\beta$-factor, except for \eqref{145},
 involves $p_1$! Thus only two
factors involve $p_1$ and not more, so we can integrate out
the $p_1$ variable by using \eqref{pointsing2}
$$
    \int dp_1 \frac{1}{|\a - \om(p_1)|}  \frac{1}{|\beta - \om(p_1-p_4+p_5)|}
    = O(1)
$$
(modulo logs and modulo the point singularity problem).

In the next step  we are then left with
\begin{align}
    \lambda^{12} (\lambda^{-2})^p \int & \rd\alpha \rd\beta \rd\bp
   \frac{1}{|\a - \om(p_2)|}
 \frac{1}{|\a - \om(p_3)|} \frac{1}{|\a - \om(p_4)|}
 \frac{1}{|\a - \om(p_5)|} \frac{1}{|\a - \om(p_6)|}
\non\\
&    \times \frac{1}{|\a - \om(p_7)|}
    \frac{1}{|\beta - \om(p_2-p_4+p_5- p_6+p_7)|}
\non\\
&   \times\frac{1}{|\beta - \om(p_2-p_3+p_5-p_6+p_7)|} 
 \frac{1}{|\beta - \om(p_2-p_3+p_7)|} \frac{1}{|\beta - \om(p_7)|}.
\non
\end{align}
Since we have already taken care of the $\beta$-denominators with $p_1', p_2'$,
the next one would be the $\beta$-denominator with $p_3'$,
but this was estimated trivially (and removed) at the beginning. So the next one 
to consider is
$$
   \frac{1}{|\beta -\om(p_4')|} = 
 \frac{1}{|\beta - \om(p_2-p_4+p_5- p_6+p_7)|}.
$$
Since $p_4'$ is not above a peak, there is
a $p$-momentum whose tower has the lowest point at
the level 4, namely $p_4$. From the graph we thus conclude
that $p_4$ appears only in this $\beta$-factor (and in one
$\alpha$-factor), so again it can be integrated out:
$$
   \int \rd p_4  \frac{1}{|\a - \om(p_4)|}
 \frac{1}{|\beta - \om(p_2-p_4+p_5- p_6+p_7)|} \leq O(1)
$$
(modulo log's and point singularity). 

We have then
\begin{align}
    \lambda^{12} (\lambda^{-2})^p \int & \rd\alpha \rd\beta \rd\bp
   \frac{1}{|\a - \om(p_2)|}
 \frac{1}{|\a - \om(p_3)|}
 \frac{1}{|\a - \om(p_5)|} \frac{1}{|\a - \om(p_6)|}
\non\\
&    \times \frac{1}{|\a - \om(p_7)|}
\frac{1}{|\beta - \om(p_2-p_3+p_5-p_6+p_7)|} 
 \frac{1}{|\beta - \om(p_2-p_3+p_7)|} \frac{1}{|\beta - \om(p_7)|}.
\non
\end{align}

Next, 
$$
    \frac{1}{|\beta -\om(p_5')|} = 
\frac{1}{|\beta - \om(p_2-p_3+p_5-p_6+p_7)|} 
$$
includes even two variables (namely $p_5$, $p_6$) that
do not appear in any other $\beta$-denominators (because the
towers of $p_5$ and $p_6$ end at the level 5, in other words
because right below the row of $p_5'$ there is a valley). We can freely choose
which one to integrate out, say we choose $p_5$, and perform
$$
  \int \rd p_5  \frac{1}{|\a - \om(p_5)|}
 \frac{1}{|\beta - \om(p_2-p_3+p_5-p_6+p_7)|} \leq O(1).
$$
We are left with
\begin{align}
    \lambda^{12} (\lambda^{-2})^p \int &\rd\alpha \rd\beta \rd\bp
   \frac{1}{|\a - \om(p_2)|}
 \frac{1}{|\a - \om(p_3)|} \frac{1}{|\a - \om(p_6)|}
\non\\
&    \times \frac{1}{|\a - \om(p_7)|}
 \frac{1}{|\beta - \om(p_2-p_3+p_7)|} \frac{1}{|\beta - \om(p_7)|}.
\end{align}
Finally
$$
    \frac{1}{|\beta -\om(p_6')|} = \frac{1}{|\beta - \om(p_2-p_3+p_7)|}
$$
can be integrated out either by $p_2$ or $p_3$ (both towers end
at the level of $p_6'$), e.g.
$$
   \int dp_2  \frac{1}{|\a - \om(p_2)|} \frac{1}{|\beta - \om(p_2-p_3+p_7)|}
  \leq O(1).
$$
The last $\beta$-factor is eliminated by the $\beta$ integration
by \eqref{log}
and then in the remaining integral,
$$
    \lambda^{12} (\lambda^{-2})^p \int \rd\alpha \rd\bp
   \frac{1}{|\a - \om(p_3)|} \frac{1}{|\a - \om(p_6)|}
    \frac{1}{|\a - \om(p_7)|},
$$
one can integrate out each remaining momenta one by one.
We have thus shown that the value of this permutation is
$$
   \mbox{Val}(\pi) \leq \lambda^{12}(\lambda^{-2})^p = \lambda^8
$$
modulo log's and point singularities.

In general, the above procedure gives
$$
    \lambda^{2n-2p}\leq \lambda^{n} = \lambda^{d(\pi)}
$$
if there are no ladders,  $\ell =0$.

\subsubsection{General algorithm including ladder indices and other fine points}

Finally, we show how to deal with ladder indices. The idea is that
first one has to integrate out the consecutive ladder indices
after taking a trivial Schwarz inequality to decouple
the $\alpha$ and $\beta$-factors of the consecutive ladder indices
and then proceed similarly to  \eqref{firstattempt}. 
It is important that only propagators with ladder
indices will be Schwarzed, for the rest of the integrand
we will use the successive integration procedure
explained in the previous section.  In this way
the ladder indices remain neutral for the final bookkeeping, 
in fact, the integral
$$
  \int \frac{\rd p}{|\alpha - \om(p) - i\eta|^2} \sim \lambda^{-2}
$$
exactly compensates the $\lambda^2$ prefactor carried by
the ladder index. Actually, one needs to take care that
not only the $\lambda$-powers but
even the constants cancel each other as well, since an error $C^n$
would not be affordable when $n$, being the typical number of the
collisions, is $\lambda^{-\kappa}$. Therefore the above bound 
will be improved to 
\be
   \int \frac{\lambda^2}{|\alpha - \om(p) - i\eta|^2} \rd p = 
  1 + O(\lambda^{1-12\kappa}),
\label{exact}
\ee
(see \eqref{exact1}),
and it is this point where the careful choice of the
remormalized propagator $\om(p)$ plays a crucial role.
This is also one error estimate which further restricts the value of $\kappa$.
After integrating out the ladder indices, we perform the successive estimates
 done in Section~\ref{nolad}.

\bigskip

There were two main issues  have been swept under the rug
(among several other less important ones...).
First, there are several additional log factors floating
around. In fact, they can be treated generously, since with this
procedure we gain a $\lambda^{(const.) d(\pi)}$
factor and the number of log factors is also comparable
with $d(\pi)$. However, here  the key point again is that by the ladder integration
in \eqref{exact} we do not lose any log factor; even not
a constant factor!

The second complication, the issue of the point singularities,
is much more serious and this accounts for the serious reduction 
of the value of $\kappa$ in the final theorem.
In higher dimensions, $d\ge 3$, one point singularity 
of the form $(|p|+\eta)^{-1}$ is integrable, but
it may happen that the {\it same} point singularity arises 
from different integrations of the form \eqref{pointsing2} 
along our algorithm. This would yield a high
multiplicity point singularity whose integral
is large.

\medskip

 First, we indicate with an example
that overlapping point singularities indeed do occur; they certainly
would occur in the ladders, had we not integrated out the ladders
separately. 

The structure of the integral for
a set of consecutive ladder indices, $\{ k, k+1, k+2, \ldots k+m\}$
 is as follows:
\begin{align}
 \Omega=  \int & \rd p_k \rd p_{k+1}\ldots \rd p_{k+m}\frac{1}{|\alpha - \om (p_k)|}
\frac{1}{|\alpha - \om (p_{k+1})|}\ldots \frac{1}{|\alpha - \om (p_{k+m})|}
\non\\
& \times \frac{1}{|\beta - \om (p_k+u)|}
\frac{1}{|\beta - \om (p_{k+1}+u)|}\ldots \frac{1}{|\beta - \om (p_{k+m}+u)|}.
\label{ladde}
\end{align}
Here we used that if the consecutive ladders are
the points $(k,s), (k+1, s+1), (k+2, s+2), \ldots$, then
the corresponding momenta are related as
$$
   p_k- p_s' = p_{k+1} - p_{s+1}' =  p_{k+2} - p_{s+2}' =\ldots
$$
i.e. one can write $p_{s+i}' = p_{k+i} + u$ with a fixed 
vector $u$ (that depends on all other momenta but not on $p_k, p_{k+1},
\ldots p_{k+m}$, e.g. $u=p_2-p_4+p_9$).

Using \eqref{pointsing2} successively, we obtain
$$
  \Omega \leq \Big(\frac{1}{|u|+\lambda^{2}}\Big)^{m+1},
$$
i.e. the {\bf same} point singularity arises from each
integration. We call this phenomenon {\it accumulation
of point singularities}.
Since $u$ is a linear combination of other momenta, that
need to be integrated out, at some point we would face with
\be
   \int  \Big(\frac{1}{|u|+\lambda^{2}}\Big)^{m+1} \rd u \sim 
   \Big(\lambda^{-2}\Big)^{m-2}.
\label{um}
\ee
Since $m+1$ consecutive ladder indices carry a factor $\lambda^{2(m+1)}$,
we see that from a consecutive ladder sequence we might gain only
$\lambda^6$, irrespective of the length of the sequence.
It turns out that even this calculation is a bit too optimistic;
the formula \eqref{um} is only rough caricature, also 
propagators depending on $u$ are present in the $u$ integration. In reality,
eventually, the ladder indices are integrated out to $O(1)$.

\medskip

Based upon this example, one may think that the accumulation of
point singularities occurs only in the presence of the consecutive ladders;
once they are taken care of separately (and not gaining but also not
losing from them), the remaining graph can be integated
out without any accumulation of point singularities.
We conjecture that this is indeed the case, but we could not
quite prove this. Instead, we estimate several more
$\beta$-factors (propagators with ``tree momenta'') by
the trivial $L^\infty$-bound to surely exclude this scenario. They are carefully chosen
(see the concept of ``uncovered slope indices'' in
Definition 10.3 of \cite{ESY2}) so that after their trivial
removal, along the successive integration algorithm
for the remaining propagators, indeed no accumulation of point
singularity occurs.

This completes the sketch of the proof of estimating the
non-repetition Feynman diagrams, i.e. the proof of
 Theorem~\ref{thm:only}.  
\qed

\section{Feynman graphs with repetitions}\label{sec:rep}
\setcounter{equation}{0}

In this short section we indicate how to prove Theorem~\ref{thm:negl}.
Recall that $\psi_s^{err}$ contains terms with many $(n\gg \lambda^{-\kappa})$
potentials and terms with (not immediate) recollisions. 

The estimate of the terms with a large number of collisions
is relatively easy; these are still non-repetition graphs,
so the integration procedure from Section~\ref{sec:genint} applies.
Permutations with degree $d(\pi)\ge C/\kappa$
have a  much smaller contribution than the required $o(t^{-2})$ error.
Permutations with low complexity contain macroscopically long
 ladders (i.e. ladders with length $cn$ with some positive constant $c$)
and these can be computed quite precisely. The precise calculation
reveals a factor $1/(cn)!$ due to the time-ordered integration
(see discussion around \eqref{timeorder11}) and the value
of such graphs can be estimated by
$$
    \frac{(\lambda^2 t)^n}{(cn)!} e^{-c'\lambda^2 t}
$$
The combinatorics
of such low complexity graphs is at most $n^{C/\kappa}$, so
their total contribution 
is negligible if $n\gg \lambda^2 t\sim \lambda^{-\kappa}$.

\medskip

The repetition terms from $\psi_s^{err}$ require a much more
laborous treatment. 
The different repetition patterns are shown on Fig.~\ref{fig:nest1}.
When one of the repetitions shows up
(apart from immediate repetition that were renormalized),
we  stop the expansion to reduce
complications (see \eqref{stoprule}).
Actually the stopping rule is a bit more involved, because
the repetition pattern has to collect sufficient
``repetitions'' to render that term sufficiently small
even after paying the price for the unitary estimate,
but we will no go into these details.
Finally, each term we compute by ``bare hand'', after applying
a certain graph surgery to reduce the number  of different cases.
One example of this reduction is shown on Fig.~\ref{fig:estrec3}
while Fig.~\ref{fig:rec} shows the result of an explicit estimate. 
The explicit estimates rely on the fact that the
repetition imposes a certain  momentum restriction
that reduces the  volume of the maximal overlap of singularities,
 reminiscent to the mechanism behind the crossing estimate, Section~\ref{sec:ex}.
Unfortunately, even after the graph surgery reduction, still
a considerably number of similar but not identical cases 
have to be estimated on a case by case basis.

\bef\bec
\epsfig{file=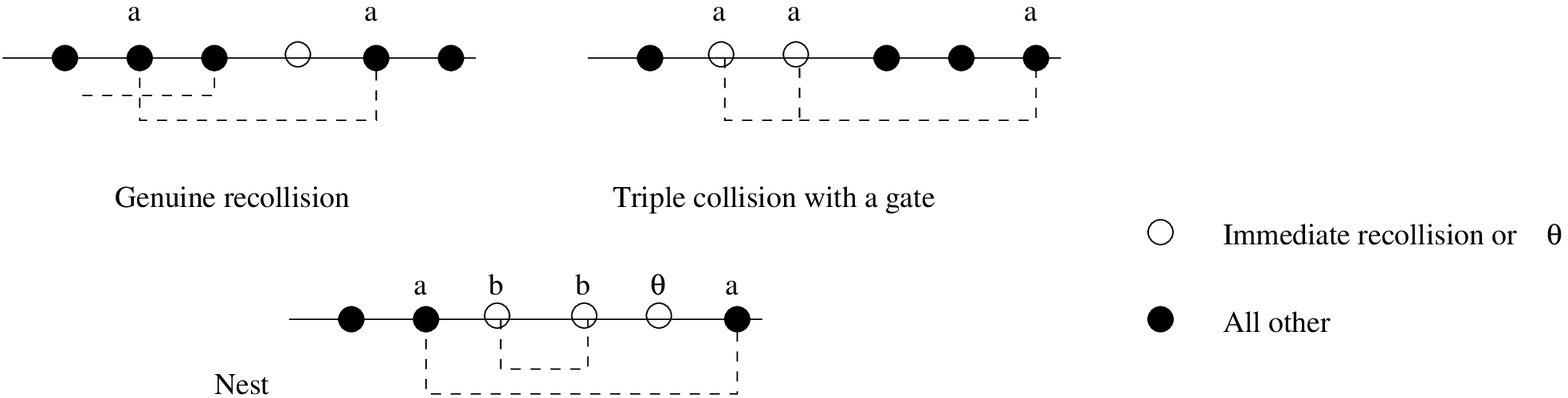, scale=0.80}
\eec
\caption{Various repetition patterns}
\label{fig:nest1}
\eef

\bef\bec
\epsfig{file=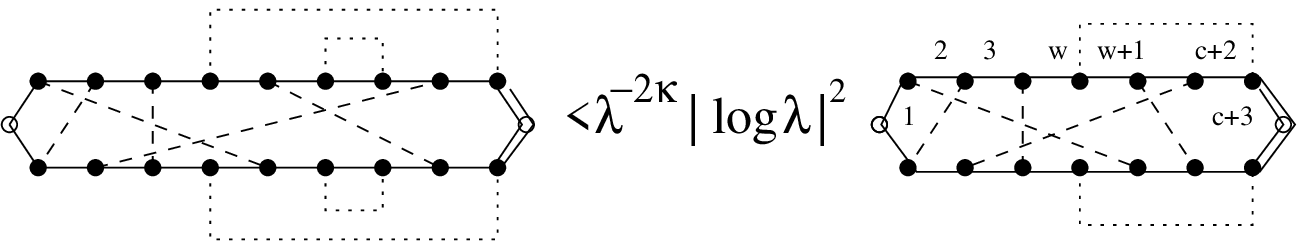, scale=1}
\eec
\caption{Removal of a gate}
\label{fig:estrec3}
\eef

\bef\bec
\epsfig{file=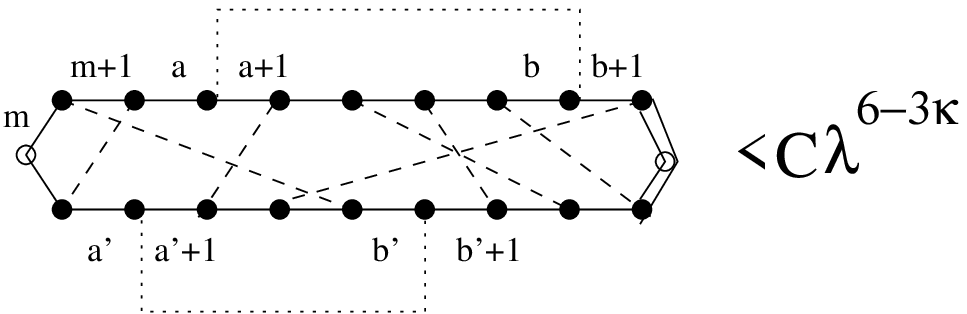, scale=1}
\eec
\caption{Two sided recollision}
\label{fig:rec}
\eef

\subsection{Higher order pairings: the lumps}\label{sec:replump}

We also mention how to deal with the higher order pairings
that we postponed from Section~\ref{sec:momspace}.
Recall that the stopping rule monitored
whether we have a (non-immediate) repetition in
the collision sequence $A=(\a_1, \a_2, \ldots, \a_n)$.
This procedure has a side effect when computing the expectation:
$$
   \bE \prod \ov{\wh V_{\alpha_j}(p_{j+1}-p_j) }\wh V_{\alpha_j}(q_{j+1}-q_j)
   = \sum_{\a_\ell\neq \a_k} \prod_j 
e^{i\alpha_j [p_{j+1}-p_j - (q_{j+1}-q_j)]}
$$
The non-repetition restriction $\a_\ell\neq \a_k$ destroys 
the precise delta function $\sum_\a e^{i\a p}=\delta(p)$.
This requires to employ the following Connected Graph Formula
that expresses the restricted sum as a linear combination
of momentum delta functions:

\begin{lemma} [Connected Graph Formula]
Let $\cA_n$ be the set of partitions of $\{ 1,\ldots, n\}$. There
exist explicit coefficients $c(k)$ such that 
$$\sum_{\a_\ell\neq \a_k}e^{iq_j\a_j}
=\sum_{{\bf A}\in \cA_n} \prod_{\nu}  
c(|A_\nu|) \delta \Big(\sum_{\ell\in A_\nu} q_\ell\Big)
\qquad {\bf A} = (A_1, A_2, \ldots),
$$
where the summation is over all partitions ${\bf A}$
of the set $\{ 1,\ldots, n\}$, and $A_1, A_2, \ldots, $
denote the elements of a fixed partition, in particular
they are disjoint sets and
$A_1\cup A_2\cup \ldots = \{ 1,\ldots, n\}$.
\end{lemma}

The appearence of nontrivial partitioning sets means that
instead of pairs we really have to consider ``hyperpairs'',
i.e. subsets (called {\it lumps}) of size more than 2.

Lumps have large combinatorics, but the  value of the corresponding
Feynman graph is small since in a large lump
 many more obstacle indices must coincide than in a pairing.
Nevertheless, their treatment would require setting
up a separate notation and run the algorithm 
of Section~\ref{sec:genint} for general lumps instead of pairs.

To avoid these complications and repeating a similar proof,
we invented a graph surgery that reduces lumps to pairs.
The idea is that lumps are artificially broken up into
pairs (i.e. permutations) and the corresponding Feynman graphs can
be compared. The break-up is not unique, e.g.
Fig.~\ref{fig:broke} shows  two possible  break-ups
of a lump of four elements into two pairs.

\bef\bec
\epsfig{file=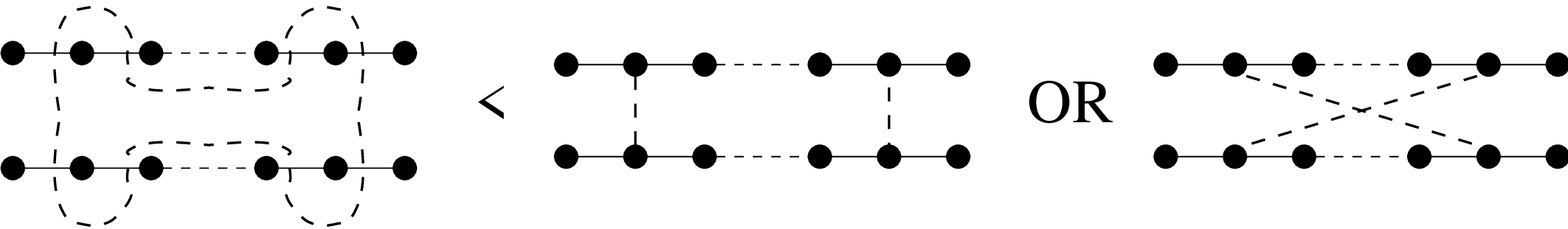, scale=0.80}
\eec
\caption{Breaking up lumps}
\label{fig:broke}
\eef

There are many break-ups possible, but the key idea is
that we choose  the break-up that gives the biggest $d(\pi)$.
The following combinatorial statement shows a lower estimate
on the maximal degree, demonstrating that every lump
can be broken up into an appropriate pairing with relatively
high degree.

\begin{proposition}
Let ${\bf B} = \{ B_1, B_2, \ldots\}$ be a partition of vertices.
$$
       s({\bf B}): = \frac{1}{2} \sum \{ \; |B_j| \; : \; |B_j|\ge 4\}
$$
Then there exists  a permutation
 $\pi$, compatible with ${\bf B}$ such that $d(\pi)
\ge \frac{1}{2}s({\bf B})$. 
\end{proposition}

Using this lemma we reduce the estimate of hyperpairs to the previous case
involving only pairs. This completes the very sketchy ideas
of the proof of Theorem~\ref{thm:negl}. \qed

\section{Computation of the main term}\label{sec:mainterm}
\setcounter{equation}{0}

In this section we sketch the computation of the main term, i.e.
the proof of Theorem~\ref{thm:mainterm}.
It is an explicit but nontrivial computation.
The calculation below is valid only in the discrete case 
as it uses the fact that for any incoming
velocity $u$, the collision kernel $\sigma(u,v)$
is constant in the outgoing velocity, i.e. the new
velocity is independent of the old one (this was the
reason why the diffusion coefficient could be directly
computed (see \eqref{gkdirect}) without computing
the velocity autocorrelation function via the Green-Kubo formula
\eqref{gk} as in the continuous case.

The computation of the main term for the continuous case
relates the diffusion coefficient to the underlying Boltzmann
process. This procedure is conceptually more general
but technically somewhat more involved (see Section 6 of \cite{ESY3}),
so here we present only the simpler method for the discrete model.

The Fourier transform of the rescaled Wigner transform $W(X/\e, V)$
is given by
$$
     \wh W_t(\e\xi, v) = \ov{\wh \psi_t}
\Big(v+\frac{\e\xi}{2}\Big) \wh \psi_t\Big(v-\frac{\e\xi}{2}\Big), 
$$
where recall that $\e= \lambda^{2+\kappa/2}$ is
the space rescaling and  $t=\lambda^{-2-\kappa}T$.

We want to test the Wigner transform against a macroscopic observable, 
i.e. compute
$$
\langle \cO, \bE \wh W_t\rangle  =
\langle \cO (\xi,v), \bE \wh W_t(\e\xi, v)\rangle 
= \int \rd v \rd\xi \; \cO (\xi,v)\; \bE \wh W_t(\e\xi, v).
$$
Recall from Lemma \ref{lm:cont}
that the Wigner transform enjoys the following continuity property:
$$
     \langle \cO, \bE \wh W_\psi\rangle- \langle \cO, \bE \wh W_\phi\rangle
     \leq C\sqrt{\bE \|\psi\|^2 + \bE \|\phi\|^2} \sqrt{ \bE \| \psi-\phi\|^2},
$$
in particular, by using \eqref{errsmall},
it is sufficient to compute the Wigner transform
of 
$$
   \sum_{n=0}^K \psi_{n,t}^{nr}.
$$
We can express this Wigner transform in terms of Feynman diagrams,
exactly as for the $L^2$-calculation. The estimate \eqref{onlyladderwigner}
in our key Theorem \ref{thm:only} implies, that only the ladder
diagrams matter (after renormalization).
Thus we have
\begin{align}\label{ow}
   \langle \cO ,\bE \wh W_t\rangle \approx &
 \sum_{k\leq K}  \lambda^{2k}
\int_\bR \rd \alpha \rd \beta
 \; e^{it(\alpha-\beta)+ 2 t \eta}  \\  &
\times \int \rd \xi \rd v  \cO(\xi, v) 
\ov{R_\eta\Big(\a, v +\frac{\e\xi}{2}\Big)}
   R_\eta\Big(\beta, v -\frac{\e\xi}{2}\Big)  \nonumber \\
 & \times \prod_{j=2}^k \Big[ \int \rd v_j
  \ov{R_\eta\Big(\a, v_j +\frac{\e\xi}{2}\Big)}
   R_\eta\Big(\beta, v_j -\frac{\e\xi}{2}\Big)\Big ]
\nonumber \\ &
 \times \int \rd v_1
  \ov{R_\eta\Big(\a, v_1 +\frac{\e\xi}{2}\Big)}
   R_\eta\Big(\beta, v_1 -\frac{\e\xi}{2}\Big)
\overline{\wh W_0}(\e\xi, v_1) \; , \nonumber
\end{align}
with the renormalized propagator
$$
        R_\eta(\a, v):= \frac{1}{\alpha - e(v)
-\lambda^2\theta(v) + i\eta} \; . 
$$
We perform each $\rd v_j$ integral.
The key technical lemma is the following:
\begin{lemma}\label{lm:keyt}
Let $f(p)\in C^1(\bR^d)$ , 
$a: = (\a +\beta)/2$, $\lambda^{2+ 4\kappa}\leq \eta\leq \lambda^{2+\kappa}$
and fix $r\in \bR^d$ with $|r|\le \lambda^{2+\kappa/4}$. Then we have 
\begin{align*}
\int  & \frac  {\lambda^2 f(v) }{\Big( \a - e(v-r)
      -\lambda^2\ov{\theta}(v-r)  - i\eta \Big)
     \Big(\beta - e(v+r)  -
       \lambda^2\theta(v+r)+i\eta \Big)} \; \rd v  \\
&     =   -2\pi i\int   \frac{\lambda^2 f(v)\; \delta(e(v)-a)}{ (\a-\beta)
 + 2 (\nabla  e)(v) \cdot r -
       2 i \lambda^2  {\cal I} (a)} \, \rd v + o(\lambda^{1/4})
\end{align*}
where
$$
   \cI(a): =  \mbox{Im} \int\frac{\rd v}{ a- e(v)-i0} 
= \int \delta(e(v)-a)\;\rd v\;,
$$
in particular, $\cI(e(p))=\mbox{Im} \,\theta(p)$.
\end{lemma}
The proof of Lemma \ref{lm:keyt} relies on the following (approximate) 
identity
$$
     \frac{1}{(\alpha - \ov g(v-r)-i0)(\beta - g(v+r) +i0)} 
    \approx \frac{1}{\alpha -\beta + g(v+r)-\ov g(v-r)}
$$
$$ 
\times \Big[ \frac{1}{\beta - g(v) +i0} - \frac{1}{\alpha - 
\ov g(v) -i0} \Big]
$$
and on careful Taylor expansions. \qed

\medskip

Accepting this lemma, 
we change variables $a=(\a + \beta)/2$,
 $b= (\alpha-\beta)/\lambda^2$ and choose $\eta\ll t^{-1}$
in \eqref{ow}. Then we get
\begin{align}
   \langle \cO ,\bE \wh W_t\rangle \approx \!\! &
 \sum_{k\leq K} \! \int \rd\xi \rd a\rd b
 \; e^{it\lambda^2 b}\Bigg(\prod_{j=1}^{k+1} \int\!
   \frac{-2\pi i  F^{(j)}(\xi, v_j) \; 
\delta(e(v_j)-a) }{ b + \lambda^{-2}\e (\nabla e)(v_j)\cdot \xi -
       2 i \cI (a)} \rd v_j \Bigg)  \nonumber
\end{align}
with $F^{(1)}:= \wh W_0$, $F^{(k+1)}:= \cO$, and  $F^{(j)}\equiv 1$,
$j\neq 1, k+1$. Here $W_0$ is the rescaled Wigner transform
of the initial state.

Let  $\rd\mu_a(v)$ be
the normalized surface measure of the level surface $\Sigma_a:=\{ e(v) = a\}$
defined via the co-area formula, i.e. the
integral of any function $h$ w.r.t.  $\rd\mu_a(v)$ is given by
$$
    \langle h \rangle_a:=\int h(v) \rd\mu_a(v)
 : =\frac{\pi}{\cI (a)} \int_{\Sigma_a}
 h(q) \frac{\rd m(q)}{|\nabla e(q)|}
$$
where $\rd m(q)$ is the Lebesgue measure on the surface $\Sigma_a$.
Often this measure is denoted by $\delta(e(v)-a)\rd v$, i.e.
$$
\int h(v) \rd\mu_a(v) = \frac{\pi}{\cI (a)}
     \int h(v) \delta(e(v)-a)\rd v.
$$

Let $H(v) : = \frac{\nabla e(v)}{ 2\cI(a)}$. Then we have
\begin{align}
  \langle \cO ,\bE \wh W_t\rangle \approx 2\cI(a) \sum_{k\leq K}  \int \rd\xi
\int_\bR \rd a\rd b
 \; e^{i2t\lambda^2 \cI(a) b}\Bigg(\prod_{j=1}^{k+1} \int
   \frac{- i  F^{(j)}(\xi, v_j)  }{ b + \lambda^{-2}\e H(v_j)\cdot \xi -
        i } \rd \mu_a(v_j) \Bigg).  \nonumber
\end{align}
We expand the denominator up to second order
\begin{align}\label{tay}
&\int
   \frac{- i  }{ b +\e\lambda^{-2}H (v)
     \cdot \xi - i }\; \rd\mu_a(v)  \\
& = \frac{-i }{ b  -i  } \int
     \Bigg[ 1 - \frac{\e\lambda^{-2}
       H(v)\cdot \xi}{b- i }
     +\frac{ \e^2\lambda^{-4} [H (v)\cdot \xi]^2}{ (b-  i)^2}
     + O\Big( (\e\lambda^{-2} |\xi|)^3 \Big)\Bigg]  \rd\mu_a(v). \nonumber
\end{align}
After summation over $k$,  and recalling that
$\xi=O(1)$ due the decay in the observable $\cO$,
the effect of the last (error) term  is $K(\e\lambda^{-2})^3 = 
\lambda^{-\kappa} (\lambda^{\kappa/2})^3=o(1)$, thus
we can keep only the first three terms on the right hand side of
\eqref{tay}.

The  linear term cancels out by symmetry: $H(v)=-H(-v)$. 
To compute the quadratic term, we define 
$$
    D(a):
   = 4 \cI (a)\int \rd\mu_a(v)\;  H(v) \otimes H(v),  \;
$$
which is exactly the diffusion matrix \eqref{diffm}. Thus
\begin{align}
   \langle \cO ,\bE \wh W_t\rangle \approx  \sum_{k\leq K} & \int \rd \xi
   \int_\bR \rd a \; 2\cI(a)
\int \wh W_0(\e\xi, v_1)
   \rd\mu_a(v_1)
\int \cO(\xi, v) \rd\mu_a(v)\nonumber \\
& \times \int_\bR \rd b  \; e^{2i\lambda^2\cI (a)tb}\;
\Big( \frac{-i }{ b-i }\Big)^{k+1} \Bigg[ 1
     + \frac{ \e^2\lambda^{-4}
 \langle \xi, D(a)\xi\rangle}{4 \cI(a)}
     \frac{ 1}{      (b-i )^2}\Bigg]^{k-1}. \nonumber
\end{align}
Setting
$$
   B^2:= \frac{ \e^2\lambda^{-4}
 \langle \xi, D(a)\xi\rangle}{4 \cI(a)},
$$
the arising geometric series can be summed up:
$$
  \sum_{k=0}^\infty \Big( \frac{-i}{ b-i}\Big)^{k+1}
  \Bigg[ 1 +\frac{B^2}{(b-i)^2}\Bigg]^{k+1}
   = (-i)\frac{(b-i)^2 + B^2}{ (b-i)^3 +i (b-i)^2 + iB^2}.
$$
We now perform the $\rd b$ integration 
analytically by using the formula (with $A:= 2\lambda^2 \cI(a)$):
$$
(-i) \int_\bR \rd b  \; e^{itAb}
\;\frac{(b-i)^2 + B^2}{ (b-i)^3 +i (b-i)^2 + iB^2} =2\pi e^{-tAB^2} + o(1)
$$
from the dominant residue $b=iB^2$. 
We can then compute
$$
    tAB^2 
 = \e^2\lambda^{-4-\kappa} \frac{T}{2} \langle \xi, D(a)\xi\rangle
   =\frac{T}{2} \langle \xi, D(a)\xi\rangle  .
$$
In particular, this formula shows that to
 get a nontrivial limit, $\e$ has to be chosen as
 $\e = O(\lambda^{-2-\kappa/2})$,
i.e. the diffusive scaling emerges from the calculation. Finally, we have
$$
  \langle \cO, \bE \wh W \rangle \approx \int \rd\xi \int_\bR \rd a \; \cI (a)
\Big( \int \cO(\xi, v) \rd\mu_a(v)\Big)
\langle \wh W_0\rangle_a  \exp\Big( - \frac{T}{2} \langle 
\xi, D(a)\xi\rangle \Big)
$$
where 
$$
f(T, \xi , a) : =\langle \wh W_0\rangle_a  \exp\Big( - \frac{T}{2} \langle 
\xi, D(a)\xi\rangle \Big)
$$ 
is the solution of the heat equation  \eqref{heate}
in Fourier space. This completes the sketch of the
calculation of the main term and hence the proof
of Theorem~\ref{thm:mainterm}.
$\;\;\Box$

\section{Conclusions}

As a conclusion, we informally summarize the main achievements.

\begin{itemize}

\item[(1)]
We rigorously proved diffusion from a Hamiltonian quantum 
dynamics in an environment of 
fixed  time independent random scatterers
in the weak coupling regime.
The quantum dynamics is described up to a time scale
$t\sim \lambda^{-2-\kappa}$ with some $\kappa>0$,
i.e. well beyond the kinetic time scale. The typical number
of collisions converges to infinity.

\item[(2)] We identified the quantum diffusion constant
(or matrix) and we showed that it coincides with 
the diffusion constant (matrix) obtained from 
the long time evolution of the Boltzmann equation.
This shows that the two-step limit (kinetic limit
of quantum dynamics followed by the scaling limit of the random walk)
yields the same result, up to the leading order,
as the one-step limit (diffusive limit of
quantum dynamics).

\item[(3)] We controlled the interferences of random waves in
a multiple scattering process 
with infinitely many collisions
with random obstacles in the extended states regime.

\item[(4)] In agreement with (2), we showed that
quantum interferences and memory effects do not become relevant
up to our scale. We remark that this is expected to hold for any $\kappa$ in $d=3$, but
 {\it not expected} to hold for $d=2$ [localization].

\item[(5)] As our main technical achievement,
 we classified and estimated Feynman graphs up to 
{\it all orders}.
 We gained 
an extra $\lambda$-power 
{\it per each non-ladder vertex}
compared with the usual power counting estimate
relying on the $L^\infty$ and $L^1$-bounds for the
propagators 
and on the tree and loop momenta.

\end{itemize}

\appendix
\section{Inequalities}\label{ineq}
\setcounter{equation}{0}

In this appendix we collect the precise bounds on
integrals of propagators that are used in the actual proofs.
Their proofs are more or less elementary calculations, however
the calculations are much easier for the continuous
dispersion relation. For details, see the Appendix B
of \cite{ESY3} for the continuous model and in Appendix A
of \cite{ESY4} for the discrete case.

We define the weighted Sobolev norm
$$
   \| f \|_{m,n} := \sum_{|\alpha|\leq n}
\| \langle x \rangle^m  \partial^\alpha f(x)\|_\infty
$$
with $\langle x \rangle : = (2+x^2)^{1/2}$ (here $\alpha$ is a 
multiindex).

\begin{lemma} [Continuous dispersion relation \cite{ESY3}] 
Let $e(p)=\frac{1}{2}p^2$.
Suppose that $\lambda^2 \ge \eta \ge \lambda^{2+ 4 \kappa}$ with
$\kappa \le 1/12$. Then we have,
\be
    \int \frac{|h(p-q)| \rd p}{|\alpha - \om(p)+ i\eta|}
    \leq \frac{C \| h\|_{2d,0}\, |\log\lambda| \; 
\log \langle \alpha\rangle}{\langle \alpha\rangle^{1/2}
 \langle |q| -\sqrt{2|\alpha|}\rangle } \; ,
\label{eq:logest}
\ee
and for $0\le a<1$
\begin{align}
  \int \frac{|h(p-q)| \rd p}{|\a -e(p) + i\eta|^{2-a}}
 \leq & \frac{C_a \| h \|_{2d, 0}\,\eta^{-2(1-a)}}{\langle \alpha\rangle^{a/2}
\langle |q| -\sqrt{2|\alpha|}\rangle } \; .
\label{eq:3aint}
\end{align}
For $a=0$ and with $h:= \wh B$, the following more precise estimate holds.
There exists a  constant $C_0$, depending only
on finitely many $\| B\|_{k,k}$ norms,  such that
\be
  \int \frac{ \lambda^{2}|\wh B(p-q)|^2 \; \rd p 
 }{|\alpha-\ov\om(p)-i\eta|^{2}}  \leq
    1+ C_0\lambda^{-12\kappa}\big[\lambda + |\a- \om(q)|^{1/2}\big] \, .
\label{eq:ladderint}
\ee
\end{lemma}
One can notice that an additional decaying factor $h(p-q)$
must be present in the estimates due to the possible (but irrelevant)
ultraviolet divergence. In the applications $h=\wh B$.
Moreover, an additional decaying factor in $\a$ was also
saved, this will help to eliminate the logarithmic divergence
of the integral of the type
$$
   \int_\bR \frac{\rd\a}{|\a - c +i\eta|} \sim |\log \eta|
$$
modulo the logarithmic divergence at infinity. 

The following statement estimates singularity overlaps:

\begin{lemma}\label{lemma:main} 
Let $\rd\mu(p)= {\bf 1} (|p|\leq \zeta)\rd p$, in applications
$\zeta\sim \lambda^{-\kappa}$.
For any $|q|\leq C\lambda^{-1}$
\begin{align}
  I_1:=\int \frac{\rd \mu( p)}{|\a - e(p) +i\eta|\,
  |\beta - e(p+q) +i\eta|}\, 
 & \leq  \frac{C\zeta^{d-3}|\log\eta|^2}
   {\tri q \tri }
\label{withoutp}
\\
  I_2:=\int \frac{\rd \mu( p)}{|\a - e(p) +i\eta|\,
  |\beta - e(p+q) +i\eta|}\, 
  \frac{1}{\tri p-r\tri } & \leq  \frac{C\eta^{-1/2}\zeta^{d-3}|\log\eta|^2}
   {\tri q \tri }
\label{withp}
\end{align}
uniformly in $r,\alpha, \beta$. Here $\tri q\tri: = |q|+\eta$.
\end{lemma}

To formulate the statement in the discrete case, we redefine
$$
  \tri q \tri : = \eta + \min\{ |q-\gamma|\; : \; \mbox{$\gamma$ is
a critical point of $e(p)$}\}
$$
for any momentum $q$ on the $d$-dimensional torus.
It is easy to see that there are $2^d$ critical points of
the discrete dispersion relation
$e(p)= \sum_{j=1}^d (1-\cos p^{(j)})$.

\begin{lemma} [Discrete dispersion relation \cite{ESY4}]
The following bounds hold for the dispersion relation
$e(p)= \sum_{j=1}^d (1-\cos p^{(j)})$
in $d\ge 3$ dimensions.
\begin{align}
    \int \frac{\rd p}{|\a - e(p) +i\eta|}\frac{1}{\tri p-r\tri }
  & \leq  c|\log\eta|^3 \;,
\label{1dee}
\\
  I:=\int \frac{\rd p}{|\a - e(p) +i\eta|}\frac{1}{|\beta - e(p+q) +i\eta|}
  &\leq  \frac{c\eta^{-3/4}|\log\eta|^3}
   {\tri q \tri }
\label{nopont}
\\
  I(r):=\int \frac{\rd p}{|\a - e(p) +i\eta|}\frac{1}{|\beta - e(p+q) +i\eta|}
  \frac{1}{\tri p-r\tri } &\leq  \frac{c\eta^{-7/8}|\log\eta|^3}
   {\tri q \tri }
\label{withp1}
\end{align}
uniformly in $r,\alpha, \beta$. 
With the (carefully) renormalized dispersion relation it also holds that
\be
 \sup_\alpha \int 
\frac{ \lambda^{2} \; \rd p  }{|\alpha-\ov\om(p)-i\eta|^{2}} \leq
    1+ C_0\lambda^{1-12\kappa}  
\label{exact1}
\ee
(compare with \eqref{eq:ladderint}).
\end{lemma}

Because the gain in the integrals \eqref{nopont}, \eqref{withp1}
compared with the trivial estimate $\eta^{-1}|\log \eta|$,
is quite weak (compare with the much stronger bounds
in Lemma \ref{lemma:main} for the continuous case),
we need one more inequality that contains four denominators.
This special combination occurs in the estimates of the
 recollision terms. The proof of this inequality is
considerably more involved than the above ones, see \cite{ES}.
The complication is due to the lack of convexity 
of the level sets of the dispersion relation (see Fig.~\ref{fig:disprel}).
We restrict our attention to the $d=3$ dimensional case, 
as it turns out, this complication is less serious in higher dimensions.

For any real number $\a$ we define
\be
\tri \alpha \tri :=  \min\{ |\alpha|, |\alpha -2|, |\a-3|, 
|\a -4|, |\a-6|\} \; 
\label{def:tria}
\ee
in the $d=3$ dimensional model.
The values $0,2,4,6$ are the critical values of $e(p)$.
The value $\a=3$ is special, for which the
level surface $\{ e(p) = 3\}$ has  a flat point.
 In general,
in $d\ge 3$ dimensions, $\tri \a\tri$ is the minimum of $|\a -d|$ and
of all $|\a - 2m|$, $0\leq m\leq d$.

\begin{lemma}\label{lemma:4dee}  [Four Denominator Lemma \cite{ES}]
For any $\Lambda>\eta$ there exists $C_\Lambda$ such that
for any $\alpha \in [0,6]$ with $\tri \a\tri \ge \Lambda$,
\be
\cI=  \int\frac{\rd p\rd q\rd r}{|\a -e(p)+i\eta||\a -e(q)+i\eta| 
|\a -e(r)+i\eta||\a -e(p-q+r+u)+i\eta|} \leq C_\Lambda
|\log\eta|^{14}
\label{14}
\ee
uniformly in $u$.  
\end{lemma}

The key idea behind the proof of Lemma \ref{lemma:4dee} is to study the decay
property of the Fourier transform 
of the level sets $\Sigma_a:=\{ p\; :\; e(p)=a\}$, i.e. the quantity
$$
  \wh\mu_a(\xi) = \int_{\Sigma_a} e^{ip\xi}\rd p:= \int_{\Sigma_a}
  \frac{e^{ip\xi}}{|\nabla e(p)|} \rd m_a(p),
$$
where $m_a$ is the uniform (euclidean) surface measure on $\Sigma_a$.
Defining
$$
   I(\xi)= \int \frac{e^{ip\xi}\rd p}{|\a - e(p)+i\eta|}, 
$$
we get by the coarea formula that
$$
    I(\xi) = \int_0^6 \frac{\rd a}{|\a-a+i\eta|} \wh\mu_a(\xi).
$$
From the convolution structure of $\cI$, we have
$$
   \cI = \int I(\xi)^4 e^{-iu\xi}\rd \xi \leq \int |I(\xi)|^4 \rd \xi
   \leq \Big( \int_0^6 \frac{\rd a}{|\a-a+i\eta|}\Big)^4
  \sup_a \int |\wh \mu_a(\xi)|^4 \rd \xi .
$$
The first factor on the right hand side is of order $|\log \eta|^4$, so
the key question is the decay of the Fourier transform, $\wh\mu_a(\xi)$,
 of the level set, for large $\xi$. It is well known (and
follows easily from stationary phase calculation) that for
surfaces $\Sigma\subset\bR^3$ whose Gauss curvature is strictly separated away
from zero (e.g. strictly convex surfaces), the Fourier transform
$$
          \wh\mu_\Sigma(\xi): = \int_\Sigma e^{ip\xi} \, \rd p
$$
decays as
$$
   |\wh\mu_\Sigma(\xi)|\leq \frac{C}{|\xi|}
$$
for large $\xi$. Thus for such surfaces the $L^4$-norm of the Fourier
transform is finite. For surfaces, where only one principal curvature
is non-vanishing, the decay is weaker, $|\xi|^{-1/2}$. Our surface $\Sigma_a$
even has a flat point for $a=3$. A detailed analysis shows
that although the decay is not sufficient uniformly in all
direction $\xi$, the exceptional directions and their neighborhoods
are relatively small so that the $L^4$-norm is still finite.
The precise statement is more involved, here we flash up the main
result. Let $K=K(p)$ denote the Gauss curvature of the level set $\Sigma_a$
and let $\nu(p)$ denote the outward normal direction at $p\in \Sigma_a$.
\begin{lemma}[Decay of the Fourier transform of $\Sigma_a$, \cite{ES}]
For $\nu\in S^2$ and $r>0$ we have
$$
   \wh \mu_a(r\nu)\lesssim \frac{1}{r} + \frac{1}{r^{3/4}|D(\nu)|^{1/2}+1},
$$
where $D(\nu)=\min_j|\nu(p_j)-\nu|$ where
 $p_j$'s are finitely many points
on the curve $\Gamma= \{ K=0\}\cap \Sigma_a$,
 at which the neutral direction of the Gauss
map is parallel with $\Gamma$.
\end{lemma}
This lemma provides sufficient decay for the $L^4$-norm to be
only logarithmically divergent, which can be combined with
a truncation argument to obtain \eqref{14}.

\bigskip
We stated all above estimates (with the exception of
the precise \eqref{eq:ladderint} and \eqref{exact1})
for the bare dispersion relation $e(p)$.
The modification of these estimates for the renormalized
dispersion relation $\om(p)$ 
is straightforward if we are prepared to lose an
additional $\eta^{-\kappa}$ factor by using 
$$
  \frac{1}{|\a - \om(p) +i\eta|} \leq \frac{1}{|\a - e(p) +i\eta|}+
   \frac{c\lambda^2}{|\a - e(p) +i\eta||\a - \om(p) +i\eta|}
  \leq \frac{1+c\lambda^2\eta^{-1}}{|\a - e(p) +i\eta|} \; .
$$

\section{Power counting for integration of Feynman diagrams}
\label{app:B}

\begin{lemma}\label{lm:grint}
Let $\Gamma$ be a connected  oriented 
graph with set of vertices  $V(\Gamma)$ and
set of edges $E(\Gamma)$. Let  $N=|V(\Gamma)|$ be the number
of  vertices and $K=|E(\Gamma)|$ be the number of edges. Let
$p_e\in \bR^d$ 
denote a (momentum) variable associated with the edge $e\in E(\Gamma)$
and $\bp$ denotes the collection $\{ p_e\;:\;e\in E(\Gamma)\}$.
Since the graph is connected, $K\ge N-1$.
Let $R: \bR^d\to \bC$ denote a function with finite $L^1$ and $L^\infty$ 
norms, $\| R\|_1$ and $\|R\|_\infty$.
For any vertex $w\in V(\Gamma)$, let
$$
    \Omega_{w} : = \int \Delta_w(\bp)\prod_{e\in E(\Gamma)} R(p_e) \rd p_e,
$$
where 
\be
\Delta_w(\bp)=\prod_{v\in V(\Gamma) \atop v\neq w} \delta\big(\sum_{e\; :\; 
 e\sim v} \pm p_e\big)
\label{deltaww}
\ee
is a product of delta functions expressing Kirchoff laws at all 
vertices but $w$. Here $e\sim v$ denotes the fact that the edge $e$ is
adjacent to the vertex $v$ and $\pm$ indicates that the
momenta have to be summed up according to the orientation
of the edges with respect to the vertex. More precisely, 
if the edge $e$ is outgoing with respect to the vertex $v$,
then the sign is minus, otherwise it is plus. 

Then the integral $\Omega_w$ is independent of the choice of the special vertex
$w$ and
 the following bound holds
$$
 |\Omega_w|\leq \|R\|_1^{K-N+1}\| R\|_\infty^{N-1}.
$$
\end{lemma}

{\it Proof.}  First notice that the arguments of the delta
functions for  {\it all} $v\in V(\Gamma)$,
$$
\sum_{e\; :\;  e\sim v} \pm p_e,
$$
sum up to zero (every $p_e$, $e\in E(\Gamma)$,
 appears exactly twice, once with
plus once with minus). This is the reason why one delta
function had to be excluded from the product \eqref{deltaww},
otherwise they would  not be independent.
It is trivial linear algebra to see that, independently of
the choice of $w$, all $\Delta_w$
determine the same linear subspace in the space of momenta,
$(\bR^{d})^{E(\Gamma)}= \bR^{dK}$.

Let $T$ be a spanning tree in $\Gamma$
and let $\cT = E(T)$ denote the edges in $\Gamma$ that
belong to $T$. Let $\cL = E(\Gamma)\setminus \cT$ denote
the remaning edges, called ``loops''.
 Momenta
associated with $T$, $\{ p_e\; :\; e\in \cT\}$, are called
tree-momenta, the rest are called loop-momenta.
Under Kirchoff law at the vertices, all tree momenta
can be expressed in terms of linear combinations (with
coefficients $\pm 1$) of the loop momenta:
\be
      p_e = \sum_{a\in \cL}  \sigma_{e,a} p_a, \qquad  \forall e\in \cT,
\label{kir}
\ee
where the coefficients $\sigma_{e,a}$ can be determined as follows. Consider the graph
$T\cup \{ a\}$, i.e. adjoin the edge $a$ to the spanning tree.
This creates a single loop $L_a$ in the graph that involves some
elements of $\cT$. We set $\sigma_{e,a}=1$ if $e\in L_a$ and
the orientation of $e$ and $a$ within this loop coincide. 
We set $\sigma_{e,a}=-1$ if
the orientation of $e$ and $a$ are opposite, and 
finally $\sigma_{e,a}=0$ if $e\not\in L_a$.

It is easy to check that the $N-1$ linear relations \eqref{kir}
are equivalent to the ones determined by the product \eqref{deltaww} of 
delta functions. For, if all tree momenta are defined
as linear combinations of the loop momenta given by \eqref{kir},
then the Kirchoff law is trivially satisfied. To check this, notice
that when
summing up the edge momenta at each vertex, only those loop
momenta $p_a$ appear whose loop $L_a$ contains $v$.
If $v\in L_a$, then $p_a$ appears exactly twice in the sum,
once with plus and once with minus, since the momenta $p_a$ 
flowing along the loop $L_a$ once enters
and once exits at the vertex $v$.
Thus the relations \eqref{kir} imply the relations in \eqref{deltaww},
and both sets of relations have the same cardinality, $N-1$.
Since the relations \eqref{kir} are obviously independent
(the tree momenta are all different), we obtain that both sets
of relations determine the same subspace (of codimension $d(N-1)$)
of $\bR^{dK}$.

Thus we can rewrite
$$
  \Omega=\Omega_w= \int \Bigg( \prod_{e\in E(\Gamma)} R(p_e) \rd p_e\Bigg)
  \prod_{e\in \cT} \delta\big( p_e - \sum_{a\in \cL}  \sigma_{e,a} p_a\big);
$$
in particular, $\Omega_w$ is independent of $w$.

Now we estimate all factors $R(p_e)$ with $e\in \cT$ by $L^\infty$-norm,
then we integrate out all tree momenta and we are left with the $L^1$-norms
of the loop momenta:
$$
   |\Omega|\leq \| R\|_\infty^{N-1} 
\int \Big( \prod_{e\in \cL} |R(p_e)| \rd p_e\Big)
  \Big(\prod_{e\in \cT} 
\delta\big( p_e - \sum_{a\in \cL}  \sigma_{e,a} p_a\big)
\rd p_e\Big) = \| R\|_\infty^{N-1}\| R\|_1^{K-N+1}
$$
completing the proof of Lemma \ref{lm:grint}. $\Box$

\end{document}